\documentclass[preprint2,trackchanges]{aastex63}
\usepackage{natbib}
\usepackage{graphics}
\usepackage{xcolor}
\newcommand\kms{km s$^{-1}$}
\shorttitle{Star and Cluster Formation  }
\shortauthors{Megeath, Gutermuth \& Kounkel}

\begin{document}


\vskip -0.5 in

\title{Low Mass Stars as Tracers of Star and Cluster Formation}


\author{S. T. Megeath}
\affil{Ritter Astrophsyical Research Center, Dept. of Physics and Astronomy, Univ. of Toledo, Toledo, OH 43606, USA}
\email{s.megeath@utoledo.edu}
%\and
\author{R. A. Gutermuth }%\altaffilmark{2}}
\affil{Dept. of Astronomy, University of Massachusetts, Amherst, MA 01003, USA.}
%\and
\author{M. A. Kounkel}% \altaffilmark{3}}
\affil{Dept. of Physics and Astronomy, Vanderbilt University, Nashville, TN 37235, USA}

\begin{abstract}
We review the use of young low mass stars and protostars, or young stellar objects (YSOs), as tracers of star formation. Observations of molecular clouds at visible, infrared, radio and X-ray wavelengths can identify and characterize the YSOs populating these clouds, with the ability to detect deeply embedded objects and all evolutionary stages. Surveys with the Spitzer, Herschel, XMM-Newton and Chandra space telescopes have measured the spatial distribution of YSOs within a number of nearby ($< 2.5$~kpc) molecular clouds, showing surface densities varying by more than three orders of magnitude. These surveys have been used to measure the spatially varying star formation rates and efficiencies within clouds, and when combined with maps of the molecular gas, have led to the discovery of star-forming relations within clouds. YSO surveys can also characterize the structures, ages, and star formation histories of embedded clusters, and they illuminate the relationship of the clusters to the networks of filaments, hubs and ridges in the  molecular clouds from which they form. Measurements of the proper motions and radial velocities of YSOs  trace the evolving kinematics of clusters from the deeply embedded phases through gas dispersal, providing insights into the factors that shape the formation of bound clusters. On 100 pc scales that encompass entire star-forming complexes, Gaia is mapping the young associations of stars that have dispersed their natal gas and exist alongside molecular clouds. These surveys reveal the complex structures and motions in associations, and show evidence for supernova driven expansions. Remnants of these associations have now been identified by Gaia, showing that traces of star-forming structures  can persist for a few hundred million years.  
\end{abstract}



%% and \citet commands to identify citations. The citations are

%% their objects with \objectname{} or \object{}. Each macro takes the
%% differs from what is to be printed in the paper. The text appearing 
%% Otherwise, LaTeX will issue an error. 

\section{Introduction}

Over the past thirty years, there has been a rapid advance in our understanding of the evolution of baryonic matter from the Big Bang to galaxies containing richly structured interstellar mediums. One of the chief products of this evolution are low mass ($\le 1$~M$_{\odot}$) stars, which are the dominant form of stellar mass produced in the baryonic cycles of galaxies. The dominance of low mass stars is a consequence of their lifetimes and their status as the primary product of  star formation. The initial mass function of star forming regions and clusters  peaks around 0.25~M$_{\odot}$ and decreases with a Salpeter power-law form for masses in excess of 1~M$_{\odot}$ \citep{1955ApJ...121..161S,2010ARA&A..48..339B}. Approximately 40\% of the initial stellar mass and 80\% of all stars formed have masses between 0.08-1~M$_{\odot}$.

The goal of this review is to better establish low mass stars as tracers of the star formation process over the spatial scales of clusters, associations and molecular cloud complexes. It builds on an unparalleled, two decade era of multi-wavelength surveys with space-based telescopes as well as ground-based and airborne observations. These near-field star formation studies are producing a deeper understanding of  star and cluster formation in local regions of our galaxy and setting the stage for future studies probing more distant and diverse environments in our galaxy and others.

Molecular clouds were first efficiently surveyed for embedded low mass stars with near-IR detector arrays in the 1990s \citep[e.g.][]{1992ApJ...393L..25L,1993prpl.conf..429Z}. Surveys with the  Spitzer Space Telescope,  Herschel Space Observatory, XMM-Newton, and Chandra Space Telescope have since provided the means to identify and characterize low mass young stellar objects (hereafter: YSOs) in  young clusters and molecular clouds \citep[e.g.][]{2007prpl.conf..361A,2009ApJS..181..321E,2015ApJ...802...60K,2013ApJ...767...36S}. Gaia has measured the parallaxes and proper motions of the less embedded stars in clusters and molecular clouds, as well as the more evolved associations of young  stars outside the clouds \citep[e.g.][]{2018AJ....156...84K,2019ApJ...870...32K}. 

%Follow up studies of the YSOs inhabiting the clouds are now providing a detailed accounting of the infall, accretion and feedback processes and how they are modified by the natal environment. 

Despite their numbers, the  faintness of young low mass stars limits our ability to efficiently identify and study them beyond 2.5~kpc. Existing surveys of the molecular clouds, associations and clusters within 500 pc of the Sun have produced the most complete censuses of YSOs. These are complemented by surveys of clouds and clusters between 500 pc and 1 kpc which include more high mass star forming regions; these are needed to study how environment and feedback from massive stars shapes star and cluster formation. Observations of regions between 1 and 2.5~kpc include complexes comparable to those studied in other galaxies. The Cygnus-X region in particular provides a relatively nearby (1.4 kpc) example of the massive star forming complexes found in other galaxies \citep{2010ApJ...720..679B,2012A&A...539A..79R,2014AJ....148...11K,2020ApJ...896...60P}. Surveys of the nearest clouds, however, are essential for understanding biases in studies of $> 500$~pc regions.

The James Webb Space Telescope and forthcoming extremely large telescopes will extend surveys for young low to intermediate mass stars to more distant regions of our galaxy as well as nearby galaxies such as the LMC. Studies of more distant galaxies will continue to depend on high mass stars, partially resolved massive star clusters, and the integrated light of older low mass stars as tracers of star formation \citep[e.g.][]{2010ARA&A..48..339B,2019ARA&A..57..227K}. Although observations beyond the nearest 2.5~kpc are needed for a representative view of star formation in the universe,  their interpretation requires insights from local studies.

{\color{black}
\section{Questions and Scope}
\label{sec:scope}}

We will overview how surveys of low mass YSOs  are addressing the following questions about the star formation process:

\begin{enumerate}

\item  What are the rates, densities and efficiencies of star formation in molecular clouds, and how do they vary with environment? 

%PDFS, SFR, SFE determination

\item What are the differences between diffuse and clustered star formation, and how do they depend on cloud properties?

%Star forming relations. Hubs and filaments

\item What is the duration of cluster formation, and during this time, what processes shape their structure and kinematics? 

%star formation history, kinematics of cluster, feedback.

\item How does star formation in cloud complexes produce a mixture of bound clusters and unbound associations, and how long do these assemblages persist in the galactic disk after star formation ends?

%bound cluster formation. Associations and clusters.

\end{enumerate}

%%As observational capabilities advance and build upon each other, the results of these surveys are advancing from 2D maps of the distribution of YSOs to 6D measurements of position and velocity. The surveys are also characterizing the properties of individual sources: evolutionary stage, mass, age and luminosity.  

%Our goal is to review both the observational methods and the advances in understanding they are bringing. We describe the use of 2D maps of YSOs to study their spatial distributions, measure star formation rates and efficiencies, and extract clusters, as well as the more recent use of 6D maps to study 3D spatial structure and kinematics. Although our focus is YSOs, we also discuss the dense gas structures and molecular cores from which they form.

This review focuses on observational studies, and we include theoretical approaches only where they directly pertain to the interpretation of observational results or provide necessary context for the observations. Reviews of theory can be found in \citet{2014prpl.conf..243K,2014prpl.conf..291L,2019ARA&A..57..227K,2020SSRv..216...64K,2020SSRv..216...76B}. A rigorous dialogue between observations and theory is beyond the scope of this review. Instead, our goal is to build a foundation for such a dialogue.  

We will concentrate on nearby ($< 1$~kpc) clouds and complexes with particular emphasis given to the Orion region. Although this sample is not representative of the diverse star forming environments found in our galaxy and others, it can provide a detailed understanding of star formation over a considerable range of environmental conditions. Studies of these nearby regions include extensive observations of specific examples that complement results drawn from large samples, and this review will contain insights drawn from both samples and examples. 

Ultimately, insights from these near-field studies will need to be extrapolated to the full range of gas densities, radiation fields, metallicities and tidal fields present in galaxies. This will require a detailed physical understanding of nearby regions and input from the rich samples of more distant clusters and clouds inhabiting diverse environments in our galaxy and others.   

%We do not overview more distant regions with different metallicities, higher radiation fields, or strong tides; these are issues that will be addressed by the next generation of telescopes with observations of the galactic center or nearby dwarf galaxies.

%Ultimately, insights from this near-field sample can be extrapolated to the full range of gas densities, radiation fields, and potentially even metallicities and tidal fields present in galaxies. 

We will not discuss the IMF, although fundamental questions remain. Does the IMF vary, and if so, what environmental factors control its form? Is there primordial mass segregation within  clouds and clusters? For a discussion of how surveys of low mass stars are helping resolve these questions, 
we refer the reader to \citet{2010ARA&A..48..339B}, \citet{2011ApJ...727...64K,2012ApJ...745..131K}, \citet{2013ApJ...764..114H}, and \citet{2018AJ....156..271L}.

In Sec.~\ref{sec:techniques}, we overview current observational methods for identifying and characterizing YSOs. Sec~\ref{sec:spatial} introduces methods for measuring the spatial distribution of YSOs and discusses analyses of YSO surface densities in star forming regions. The determination of star formation rates, efficiencies, and relations in molecular clouds is the topic of Sec.~\ref{sec:sfr&e}. In Sec~\ref{sec:clusters}, the demographics, properties and evolution of clusters are discussed.  
%Evidence for primordial mass segregation in clouds and clusters is presented in Sec.~\ref{sec:primordial}.  
Finally, the evolution of clusters and association brings us from cloud to galactic scales in Sec.~\ref{sec:assoc}.  The results are summarized in Sec~\ref{sec:summary} and the appendices elaborate on several analyses described in the text.
 
\section{Techniques for YSO Identification}
\label{sec:techniques}

Tracing star formation with low mass YSOs requires the capability to efficiently detect and identify young stars and protostars over large swaths of the sky. Space-based infrared and X-ray observatories, such as Chandra, XMM-Newton, Spitzer, WISE and Herschel, have identified and characterized low mass YSOs across fields many square degrees  in extent with modest levels of contamination. The censuses obtained with these telescopes have been augmented with  ground-based visible, near-IR and radio data and airborne IR data. More recently, astrometry with  Gaia  is providing a new means for identifying members through parallaxes or proper motions, although the wavelengths used by Gaia are not well suited for detecting embedded stars. In this section, we discuss the identification and classification of YSOs, emphasizing the strengths and weaknesses of each technique. 

\begin{figure*}
\epsscale{1.15}
\plotone{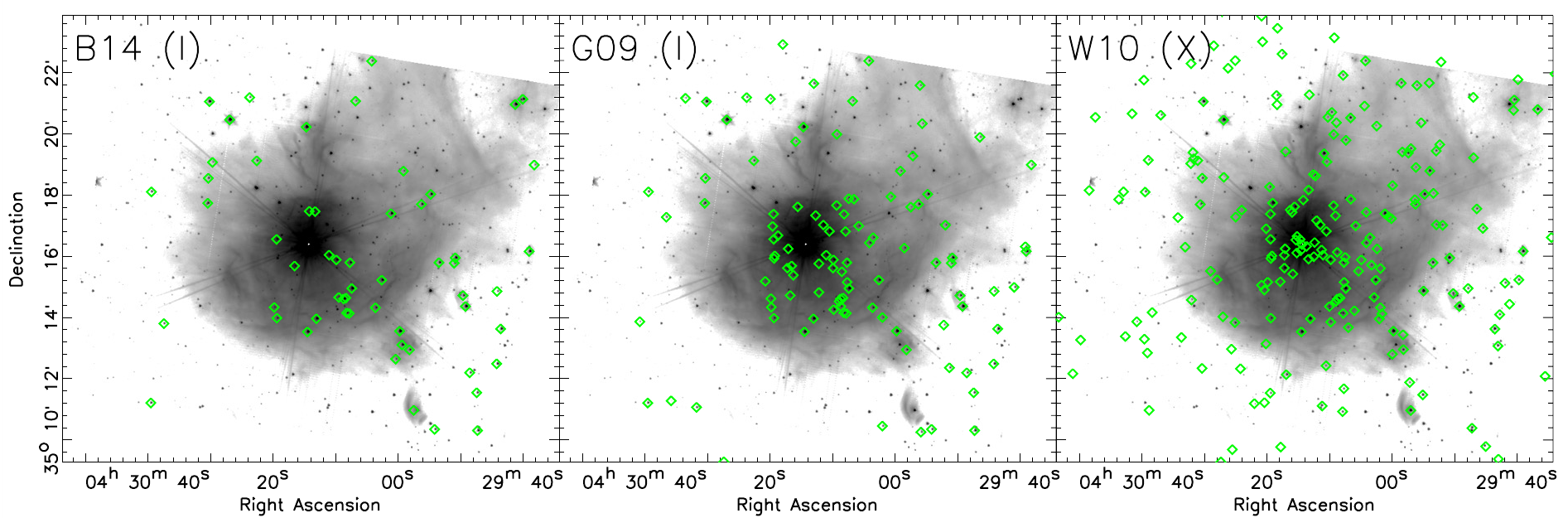}
\caption{A comparison of different methods for identifying YSOs. These data show the YSOs (diamonds) identified  toward the  LkH$\alpha$101 young cluster using {\it Spitzer}+2MASS data  \citep[B14,][]{2014ApJ...786...37B}, {\it Spitzer}+2MASS data with   alternative identification criteria \citep[G09,][]{2009ApJS..184...18G}, and a combination of {\it Spitzer}, 2MASS and Chandra data \citep[W10,][]{2010ApJ...715..671W}.  The  8~$\mu$m image in the background shows the bright IR nebulosity toward this region.  While the {\it Spitzer} data is only identifying dusty YSOs, the Chandra data also finds young stars without disks or with optically thin disks.  
\label{fig:lkha101}}
\end{figure*}

\subsection{Mid-IR Imaging}

The sensitivity of  Spitzer at wavelengths longward of $3$~$\mu$m fueled a revolutionary advance in mapping the distribution of YSOs with dusty disks and infalling envelopes \citep[e.g.][]{2004ApJS..154..363A,2004ApJS..154..367M,2004ApJS..154..374G,2004ApJS..154..379M,2004ApJ...617.1177W,2007prpl.conf..361A,2007ApJ...663.1149H}. At these wavelengths, the emission from circumstellar dust outshines the stellar photospheres and is easily detected by Spitzer. Dusty YSOs can be identified and classified though a variety of methods: fitting model SEDs to the photometry \citep[][]{2007ApJS..169..328R,2013ApJS..209...31P}, using multiple color and magnitude criteria \citep{2007ApJS..171..447R,2009ApJS..184...18G,2014AJ....148...11K,2016AJ....151....5M}, or a combination of those methods \citep{2006ApJ...644..307H,2008ApJ...680..495H}. Fig.~\ref{fig:lkha101} compares the application of different approaches on the LkH$\alpha$101 cluster.

Spitzer and 2MASS provide photometry in ten wavelength bands from 1.2 to 160~$\mu$m. Surveys for dusty YSOs rely primarily on the 1.2 to 24~$\mu$m data due to the low angular resolution of the Spitzer 70 and 160~$\mu$m imaging. To maximize the completeness of the sample of YSOs, these methods typically do not require detections in all eight bands, and  multiple criteria utilizing different combinations of the bands are often used to identify young stars \citep{2009ApJS..184...18G,2012AJ....144..192M}. In regions with bright nebulosity, criteria using a combination of ground-based $H$ and $K$-band and Spitzer 3.6 and 4.5~$\mu$m band photometry enhance completeness. In these bands, the contrast between YSOs and nebulosity is the highest. Other criteria are required to detect sources that are deeply embedded or whose disks are primarily detected at  wavelengths $> 4.5$~$\mu$m \citep[e.g.][]{2007ApJ...669..493W,2008ApJ...674..336G}. While the 2MASS point source catalog provides all-sky coverage in the $J$, $H$ and $Ks$-bands,  deeper NIR data bring higher sensitivities and angular resolution that can enhance  completeness \citep[e.g.][]{2004ApJS..154..374G,2008ApJ...675..491A,2012ApJ...752..127M,2013ApJS..209...31P,2015ApJ...809...87W,2019A&A...622A.149G}.

There are limitations to these approaches.  Reddening in the Spitzer bands, as given by the molecular cloud extinction laws of \citet{2007ApJ...663.1069F} and \citet{2009ApJ...690..496C}, can affect the classification of YSOs. Highly reddened pre-main sequence (pre-ms) stars with disks can be misclassified as protostars \citep{2012AJ....144...31K,2015ApJS..220...11D}. Some YSO classification schemes are constructed around reducing such biases in YSO characterization \citep[e.g.][]{2008ApJ...674..336G,2009ApJS..184...18G}.
Another limitation is that young stars without optically thick disks or infalling envelopes cannot be reliably identified, particularly if the stars are not detected in the {\it Spitzer} 24~$\mu$m band. This includes stars with no disks, optically thin disks, or disks with large inner holes \citep{2010AJ....140..266W,2012ApJ...750..157C,2013ApJ...762..100C,2015ApJS..220...11D}.

Photometry covering the 3.6 to 24~$\mu$m range can also be used to estimate the luminosities of protostars. Determining the SED slopes and integrated fluxes of protostars over this wavelength range,  \citet{2012AJ....144...31K} and \citet{2013AJ....145...94D} found empirical relationships between the slopes and fluxes and the bolometric luminosities of protostars. The relationships can be used when far-IR data is not available. 

\begin{figure*}[t!]
%\epsscale{2.3}
%\vskip -0.17 in
\plottwo{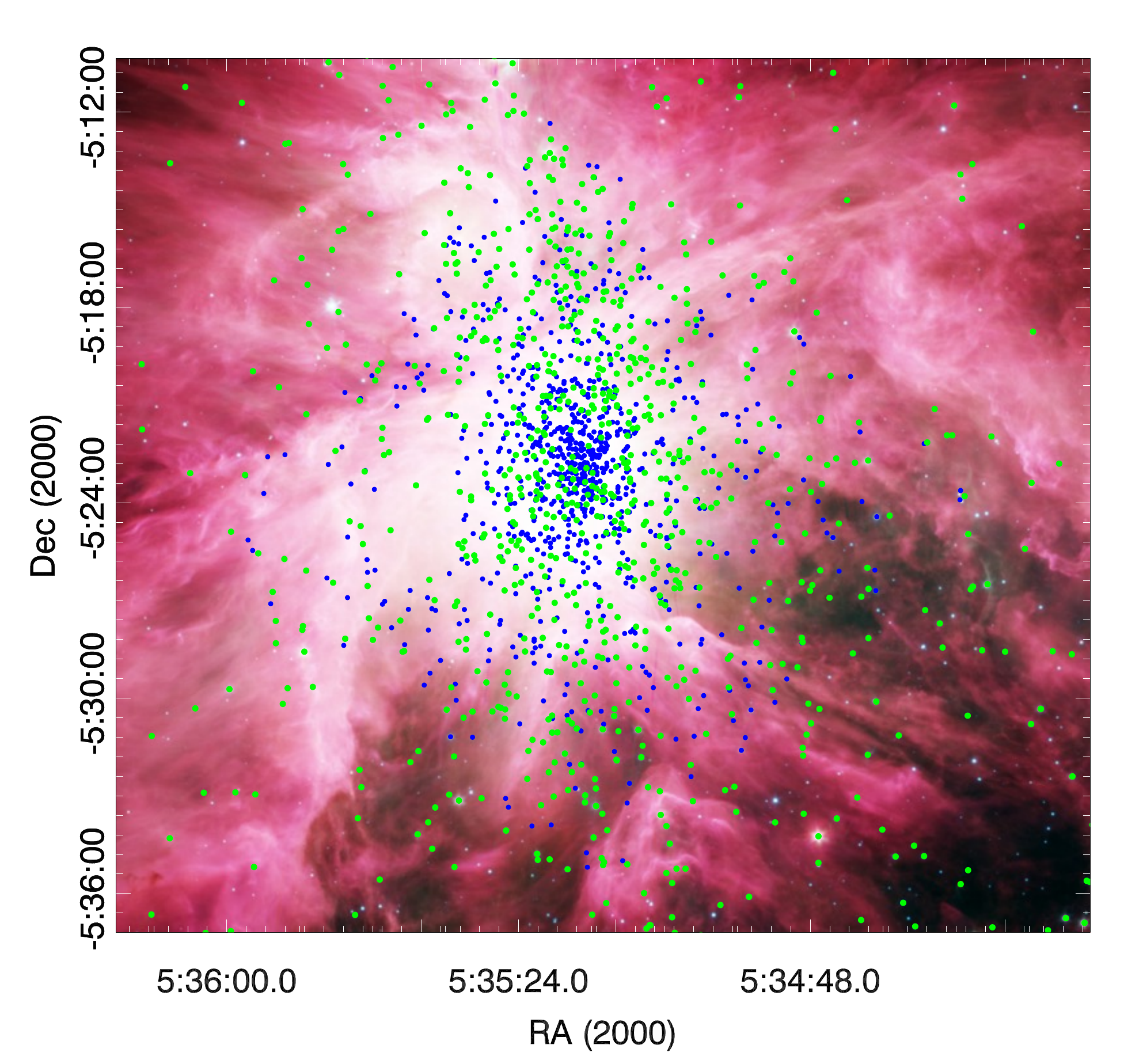}{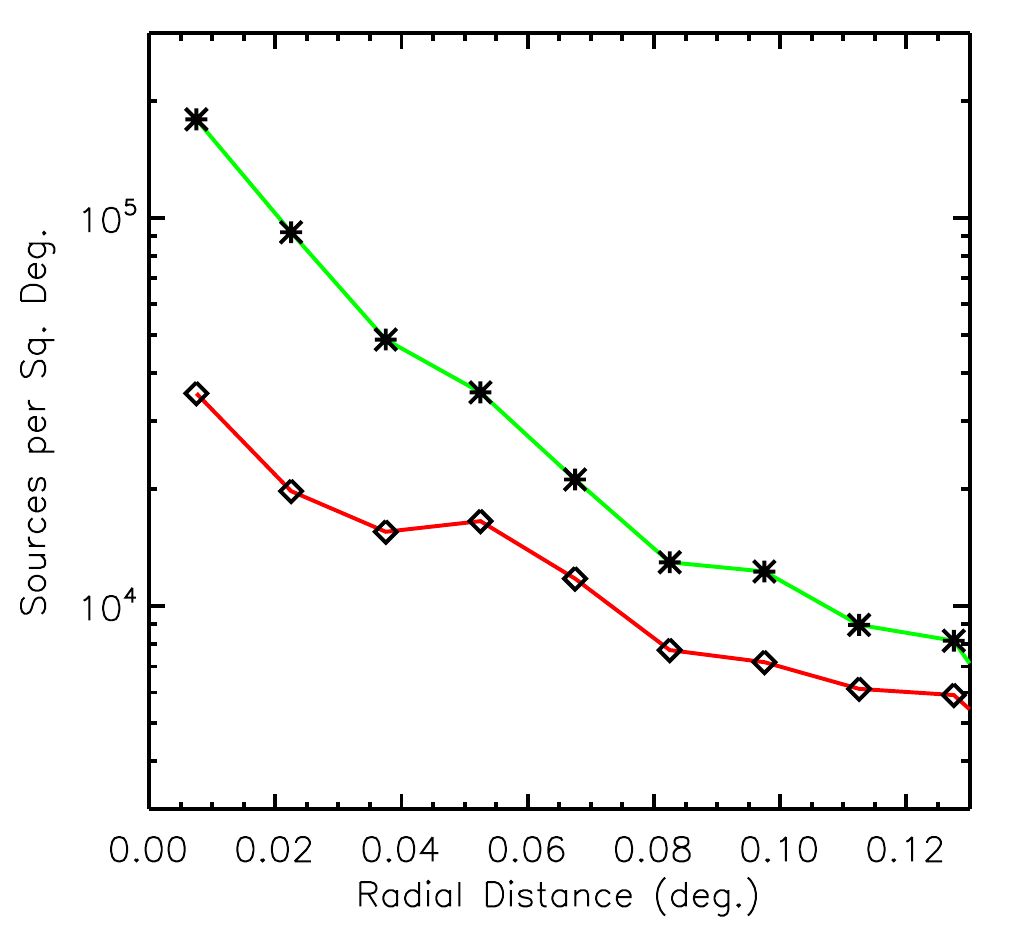}
\caption{ A demonstration of how X-ray identified YSOs can be used to correct for biases in {\it Spitzer}  surveys towards bright nebulae \citep{2016AJ....151....5M}. {\bf Left:} comparison of Chandra and Spitzer identified YSOs in the center of the Orion Nebula Cluster. The background figure is the Spitzer/IRAC 3.6, 4.5 and 8~$\mu$m image of the nebula. This region is characterized by bright, spatially varying  nebulosity which limits the detection of the mid-IR sources. The green dots are the IR-excess sources identified by Spitzer, the blue dots are X-ray sources with near-IR analogs found by the COUP survey and not identified by Spitzer. These include both diskless young stars and stars where confusion with the bright mid-IR nebulosity precludes their detections. {\bf Right:} the surface density of YSOs vs radius from the center of the nebula.  The red curve is for YSOs identified by {\it Spitzer} while the green curve is that augmented by the COUP data. }
\label{fig:oncdetection}
\end{figure*}

The primary sources of contamination are  galaxies, particularly AGN, that are not spatially resolved by Spitzer. To minimize their numbers, color and magnitude criteria were applied to remove likely contaminants  \citep{2009ApJS..184...18G,2014AJ....148...11K} or to assign a probability that a source is a background galaxy \citep{2007ApJ...663.1149H}. These methods  utilized data from wide-field extragalactic surveys to define the colors, magnitudes and densities of galaxies \citep{2007ApJ...663.1149H,2009ApJS..184...18G}. Nearby reference fields were also used to estimate the level of contamination \citep{2012AJ....144..192M}. For regions observed against the galactic plane, contamination by AGB stars must also be considered  \citep{2008AJ....136.2413R}. Most recently, \citet{2021A&C....3600470C} and \citet{2021ApJS..254...33K}  applied machine and statistical learning techniques to separate YSOs from contamination, using YSO catalogs from nearby clouds as training sets. 

WISE  brought the ability to identify dusty YSOs over the entire sky, although with lower sensitivities and angular resolutions. Given the bright nebulosity in the mid-IR, the WISE data is susceptible to source confusion, particularly in the 22~$\mu$m wavelength band \citep[e.g.][]{2015AJ....149...64G}. Different schemes have been suggested for identifying sources including SED slopes \citep{2019A&A...622A.149G} and color criteria \citep{2014ApJ...791..131K,2016ApJ...827...96F}. The adopted criteria can be adjusted and optimized for the level of background contamination and source confusion \citep[e.g.][]{2017A&A...608L...2P}. Toward the L1641 region of the Orion~A cloud, which lacks bright nebulosity, \citet{2019A&A...622A.149G} recovered 59\% of the Spitzer identified dusty YSOs with the WISE data. Recently, \citet{2019MNRAS.487.2522M} combined WISE and Gaia data to find YSOs using machine learning.

Spitzer surveys of nearby star forming regions also suffer from a spatially varying incompleteness, particularly in areas containing bright nebulosity \citep{2009ApJS..184...18G}. \citet{2016AJ....151....5M} found that the sensitivity to point sources decreased as the median absolute deviation (MAD) of the signal in a field increased. The increase in the MAD is typically due to  presence of highly structured mid-IR nebulosity, which is commonly found in star forming regions, although other stars can contribute to the MAD in crowded fields. They estimated that the fraction of detected stars dropped to about 10\% in the brightest parts of the Orion Nebula compared to regions with only faint nebulosity (Fig.~\ref{fig:oncdetection}). Since the brightest nebulosity  is found in the rich clusters that typically contain OB stars, the detection rate of YSOs is systematically lower in fields with high YSO densities. This bias can be corrected by using X-ray data or by using the detection rates of fake stars in the mid-IR images, as described by \citet[][also see Fig.~\ref{fig:oncdetection}]{2016AJ....151....5M}.

\subsection{X-ray Imaging}

X-ray surveys with the Chandra Space Telescope and XMM-Newton are also producing censuses of YSOs in clusters as well as entire molecular clouds \citep[e.g.][]{2013ApJ...768...99P,2015ApJ...802...60K}, while the ongoing X-ray eROSITA survey will  obtain a homogeneous selection of young stars across much of the sky. These observatories primarily detect X-rays from  magnetically driven activity in the coronae of young stars \citep[e.g.][]{1999ARA&A..37..363F}. The deepest X-ray survey of a star forming region to date is that of the Chandra Orion Ultradeep Project (COUP), which obtained a 9.7 day exposure of the central region of the ONC \citep{2005ApJS..160..379F}. These data detect X-ray emission from 90\% of the member stars. These stars have X-ray to bolometric luminosity ratios 1000 times that of the Sun and show fractional X-ray luminosities as large as $\log{L_x/L_{bol}} \sim -3$ \citep{2005ApJS..160..401P}. The emission is highly variable, and flares can increase the X-ray flux by as much as a factor of 100 for tens of hours \citep{2005ApJS..160..423W}. 

X-rays in the energy regime probed by Chandra and XMM-Newton (0.5-8~kev) are relatively unaffected by interstellar absorption and can be used to detect embedded populations. In addition, X-ray emission is a signature of young stars that does not require the presence of  disks or envelopes, and therefore can identify diskless stars. X-ray surveys are also not affected by the bright nebulosity found in the mid to far-IR. Although a complementary near-IR detection is usually required to confirm an X-ray source as a YSO \citep{2005ApJS..160..353G}, the comparatively faint nebulosity at near-IR wavelengths results in surveys which are less biased by nebulosity than mid-IR surveys (Fig.~\ref{fig:oncdetection}).

The disadvantage of X-ray observations is incompleteness. Although COUP detected almost 90\% of the YSOs in the Orion Nebula \citep{2016AJ....151....5M}, surveys of 200-500 pc regions with more typical integration times of many tens of kiloseconds often detect 30-50\% of the Spitzer-identified YSOs \citep[e.g.][]{2010AJ....140..266W,2013ApJ...768...99P}. The undetected sources are less active or lower mass stars that are fainter at X-ray wavelengths and may only be detected during flares \citep{2005ApJS..160..401P}. For this reason, X-ray surveys, particularly with Chandra, have focused primarily on clusters  \citep[e.g.][]{2011ApJ...743..166W,2014ApJ...787..107K}. In comparison, cloud surveys, such as the XMM-Newton survey of Orion A, lack the sensitivity of the cluster observations \citep{2013ApJ...768...99P}. 
%Combined IR and X-ray surveys have been used to mitigate incompleteness in the two data \citep{2012MNRAS.426.2917G,2016AJ....151....5M}, although the incompleteness in detecting diskless YSOs remains.

The X-ray flux is dependent on the bolometric luminosities or surface areas of  stars, although with an order of magnitude of scatter \citep{2005ApJS..160..401P,2010AJ....140..266W}. This scatter, and the rapid evolving luminosities of pre-ms stars, makes corrections for incompleteness difficult.  Current efforts to account for incompleteness use an empirical X-ray luminosity function, or XLF. \citet{2015ApJ...802...60K} adopt an XLF measured from the COUP data to derive a correction for less complete X-ray data. They find this approach produces similar corrections to those derived from an adoptive IMF.
%An additional issue is whether the disks reduce the X-ray luminosity \citep{2005ApJS..160..401P,2007A&A...468..425T}; this may explain why the disk fraction obtained by X-ray detetected sources are lower than those obtained using IR source counts \citep{2008ApJ...674..316H,2010AJ....140..266W}.

X-ray imaging can also detect extragalactic sources such as AGN.  In cases where mid-IR data available, these contaminants can be identified by their faintness in the IR \citep{2013ApJ...768...99P}.  Given their faintness in the IR, it is typically assumed that X-ray sources with visible or IR counterparts are YSOs \citep{2005ApJS..160..353G}. Sources without visible or IR counterparts are dominated by galaxies, although a significant fraction may be deeply embedded YSOs \citep{2005ApJS..160..353G}. X-ray active foreground stars may also contaminate X-ray samples; these are expected to be small in number \citep{2006ApJS..163..306G,2012ApJ...750..125A}.

\subsection{Far-IR Imaging}

The Herschel Space Observatory brought the capability for far-IR surveys of star-forming regions. The 70~$\mu$m band of the PACS camera onboard Herschel has a similar angular resolution to Spitzer's 24~$\mu$m band and is used to identify internally heated protostars. Together with PACS 100 and 160~$\mu$m data, and sub-millimeter data from ground-based telescopes or  SPIRE  on Herschel, the assembled SEDs can be used to characterize and classify YSOs \citep{2016ApJS..224....5F,2020ApJ...905..119F}. In particular, the far-IR observations are crucial for determining the luminosities and luminosity evolution of protostars \citep{2008ApJS..179..249D,2017ApJ...840...69F}. Toward the Orion clouds, Herschel observations also identified  18 protostars characterized by bright emission at 70~$\mu$m and faint or undetected emission in the mid-IR bands probed by Spitzer. These sources consequently have extreme 24-70~$\mu$m colors and are referred to as PACs Bright Red Sources or PBRS \citep{2013ApJ...767...36S}. Of these 18 sources, eleven were not identified as protostars by Spitzer and eight were not detected by  Spitzer at 24~$\mu$m. This demonstrates the presence of a small population of protostars  missed by Spitzer surveys. This small population contains the youngest known protostars in Orion \citep{2013ApJ...767...36S,2015ApJ...798..128T,2020ApJ...890..129K}. 
%\citet{2013ApJ...767...36S} also found several other candidate sources which  they did not classify as PBRS on the basis of their l70~$\mu$m to 24~$\mu$m color threshold, but were not identified by Spitzer. This suggests that there remains a small population of dusty YSOs which have not been identified with current criteria.

\begin{figure*}[h]
\epsscale{1.1}
\plotone{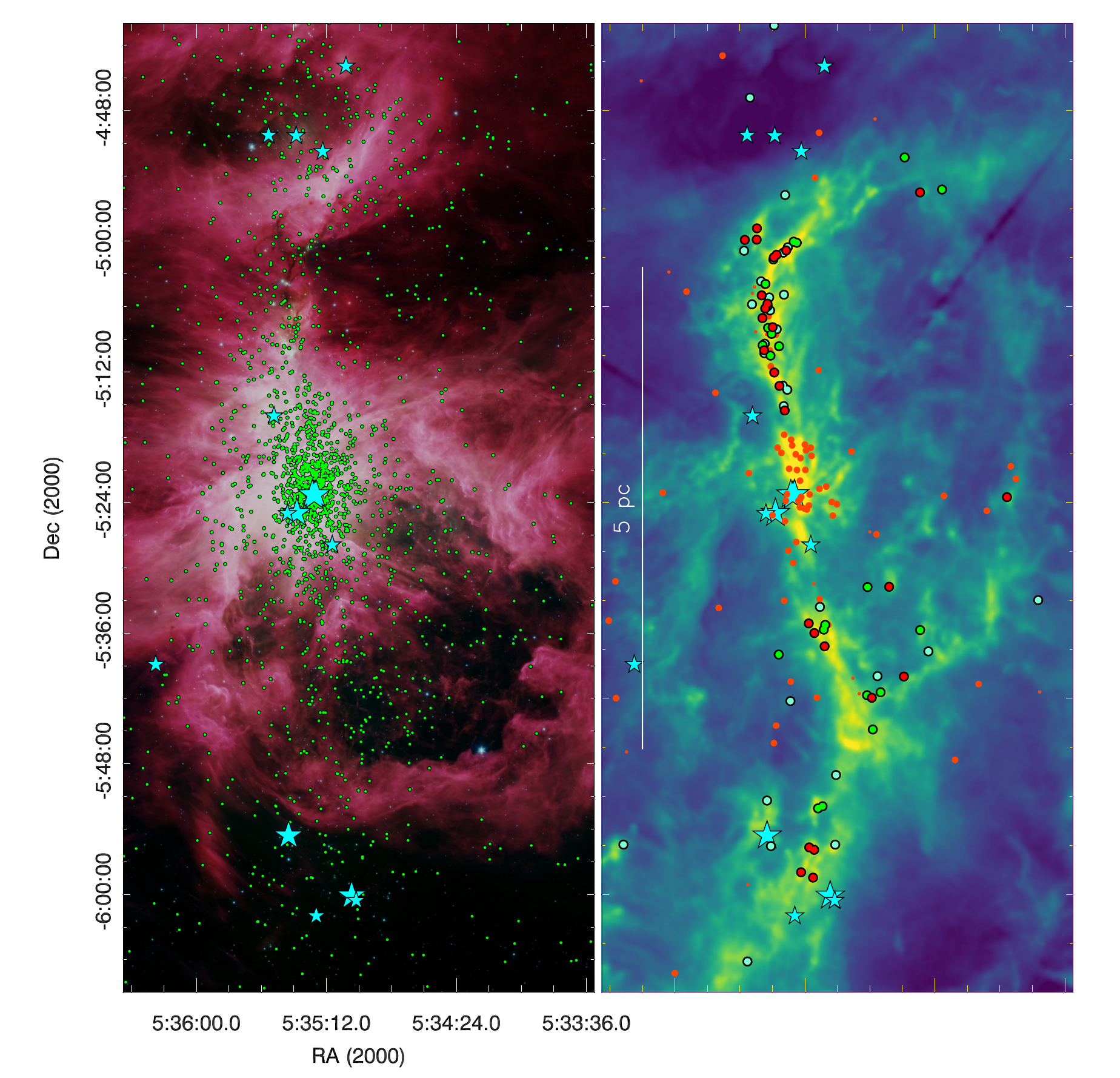}
\caption{The spatial distribution of YSOs in the integral-shaped filament (ISF) which hosts the Orion Nebula Cluster (ONC). The left panel shows the Spitzer 3.6 (blue), 4.5 (green) and 8~$\mu$m  (red) image with the pre-ms stars with disks displayed as green dots, these data have been corrected for incompleteness using Chandra data \citep[Fig.~\ref{fig:oncdetection},][]{2012AJ....144..192M,2016AJ....151....5M}. The right panel is a $N(H_2)$ column density map of the ISF constructed from  Herschel and Planck data  \citep{2016A&A...590A...2S}.  The markers with black circles show the location of protostars from the Herschel Orion Protostar Survey \citep[HOPS,][]{2016ApJS..224....5F}, with red colors denoting Class~0 protostars, green Class~I protostars, and turquose flat spectrum protostars. Orange dots without black circles are protostar candidates  identified by Spitzer that are not in the HOPS catalog. These Spitzer-only sources include protostars toward the bright Orion Nebula which were not targeted by HOPS, but those displaced from the ISF are likely to be extragalactic contaminants \citep{2012AJ....144..192M,2016ApJ...825...91L}. The O and B stars are the large and small turquoise star symbols, respectively \citep{1994A&A...289..101B}; the bubbles created by these stars are apparent in the left panel \citep{2020A&A...639A...2P}.}
\label{fig:isf}
\end{figure*}

\subsection{Radio Interferometry}
\label{sec:radio}

YSOs emit radio emission that can be used to identify young low-mass stars without any bias due to extinction. This emission is produced by several different mechanisms.  At millimeter wavelengths, typically 0.8-8~mm, radio observations detect primarily thermal emission from disks \citep[e.g.][]{2016ApJ...828...46A}. Surveys at these wavelengths have been used to characterize disks around YSOs \citep[e.g.][]{2018ApJ...860...77E,2020A&A...640A..27V}, often through targeted surveys of previously identified YSOs \citep[e.g]{2020ApJ...890..130T,2021ApJ...913..123G}. At these wavelengths, dense cores can also be detected in modest ($\sim 1000$~AU) spatial resolution surveys \citep{2017A&A...600A.141K}; these  include unstable pre-stellar cores which are the earliest, detectable stage of low mass star formation \citep{2000prpl.conf...59A}. IR observations are  needed to distinguish starless cores from protostars. 

At wavelengths $> 8$~mm, emission from ionized gas often dominates. Thermal free-free emission is commonly detected towards YSOs and is thought to originate from shock-ionized gas in outflow jets \citep{2018A&ARv..26....3A}.  In a VLA C-band (4-6~cm) survey of Perseus protostars, \citet{2018ApJS..238...19T} detected 61\% of Class~0 protostars, 53\% of Class~1 protostars and 75\% of pre-ms stars with disks. They ascribe the detections primarily to free-free emission in jets. Free-free emission is also  detected towards disks ionized by UV radiation from neighboring OB stars \citep{1987ApJ...321..516C}. This requires YSOs to be near massive stars. 

Non-thermal gyrosynchrotron emission can be detected from the active stellar coronae found around young stars \citep{1999ARA&A..37..363F}. VLA surveys of nearby molecular clouds find evidence for gyrosynchrotron emission towards almost all evolutionary classes of YSOs, from Class I protostars through diskless pre-ms stars, although there is some evidence that later evolutionary classes are more likely to exhibit such emission \citep{2013ApJ...775...63D,2015ApJ...805....9O,2016ApJ...818..116P,2014ApJ...790...49K}. A deep 5 GHz survey of the ONC found that 38\% of the X-ray or IR sources had compact radio sources \citep{2016ApJ...822...93F}. 
%The same survey found that 52\%of the radio sources had no IR or X-ray counterpart.  
Subsequent VLBA imaging detected 22\% of the compact radio sources found by the VLA; their detection at this angular resolution required the emission to be non-thermal in nature \citep{2021ApJ...906...23F}. Less than a third of these sources are detected in more than a single epoch due to the variability of the non-thermal emission.   Multi-epoch VLBA and VLA astrometry of the YSOs have been used to identify companions \citep{2017ApJ...834..142K}, determine parallaxes \citep{2008ApJ...675L..29L,2017ApJ...834..142K,2018ApJ...853...99D} and measure the motions of YSOs in young clusters \citep{2017ApJ...834..139D}. 

%number of VLA sources in Serpens: 2015ApJ...805....9O - undetected sources.
%number of VLA sources in Ophichus: 2013ApJ...775...63D - undetected source but most could be galaxies
%number of VLA sources in Perseus: 2016ApJ...818..116P
%vla proper motions in ONC 2017ApJ...834..139D
%distaance to Serpens Aquila 2017ApJ...834..143O put in
%VLBA distance and kinematics to perseus clouds, most info comes from gaia (leave out): 2018ApJ...865...73O
%VLBA distance and kinematics of taurus: 2018ApJ...859...33G put in distance
%distance to Ophiuchus 2017ApJ...834..141O
%oph A senstivive 3 cm
%lkha VLBA parallax: 2018ApJ...853...99D put in distance

A significant number of radio sources do not have IR or X-ray counterparts yet share the characteristics of known YSOs \citep{2016ApJ...818..116P}. In the \citet{2021ApJ...906...23F} VLBA survey of the Orion Nebula, 28\% of the sources lacked an IR or X-ray counterparts, much larger than the expected level of extragalactic contamination. These might be YSOs hidden by extinction.  In the OMC2 region of Orion, several YSOs have been identified by their cm and mm emission that are too deeply embedded to detect at IR or X-ray wavelengths \citep{2017ApJ...840...36O,2019ApJ...886....6T}. 

%Although radio surveys are currently not as efficient as surveys at other wavelengths, 
Radio surveys detect deeply embedded YSOs undetected at IR or X-ray wavelengths. They can  identify diskless pre-ms stars with active coronae, and for these sources,  provide precision astrometry. With ALMA, and forthcoming facilities such as the SKA and ngVLA, radio surveys will emerge as an essential tool for studying low-mass YSOs.

%Similar to the VLA case, 28\% of the sources lacked IR or X-ray counterparts.  The presence of  this much larger than the expected level of extragalactic contamination and may result from outflow knots or compact structures in the Orion Nebula.
%123/557
%These data have been extensivley used to study jets and photoionization of disks, and to a lesser degree, the coronae around young stars.  
%The emission at these wavelengths is not affected by extinction, and it has also been used to detect YSOs that are too deeply embedded to observe in IR or radio wavelengths.  OMC2 FIR4.  Orion Nebula.  

\subsection{Visible to Near-IR Spectroscopy}

Spectroscopic observations of candidate YSOs at optical and near-IR wavelengths use either photospheric absorption lines or the presence of accretion-driven emission lines as indicators of youth. At visible wavelengths, Li I absorption (6708 \AA) is an unequivocal confirmation of stellar youth in convective stars, as this element depletes rapidly prior to the star reaching the main sequence \citep[e.g.,][]{1997AJ....113..740B}. Lithium absorption is an effective diagnostic for populations with ages up to 20--30 Myr \citep[e.g.,][]{2014A&A...563A..94J,2016A&A...596A..29M}. 

The H$\alpha$ line (6562.8 \AA), both independently and in conjunction with other nearby emission lines (e.g.\ He I, O I, N II, Ca II), is an indicator of accretion from a disk and therefore another signature of youth \citep{2003ApJ...582.1109W,2005ApJ...626..498M}. Since  accretion is fed by disks, the sample of YSOs identified by accretion lines should be identical to the sample of young stars with dusty disks found by {\it Spitzer} \citep{2009AJ....137.4777W}. Around 15\% of  stars, however, show mismatches in their classifications by spectroscopic and mid-IR criteria; this may be due to the presence of disks without strong H$\alpha$, the presence of  chromospheric H$\alpha$ lines toward active stars without disks, or perhaps the inability to detect  disks due to nebulosity or weak mid-IR emission \citep{2008AJ....135..966F,2017AJ....154...29K}. In particular, chromospheric emission lines may provide a new and currently under-utilized means for identifying  young stars without disks \citep{2002AJ....123..304H, 2008ApJ...675..491A,2019ApJ...871...46K}.

Spectroscopic observations of YSOs are limited to samples of sources targeted on the basis of existing data and thus are typically incomplete. Some regions, however, do have significant coverage. The most notable example is the Orion~A cloud, where observations from several studies have  amassed a sample in excess of 3000 YSOs \citep{2005AJ....129..363S,2008ApJ...676.1109F,2009A&A...504..461F,2012ApJ...752...59H,2013ApJS..207....5F,2016ApJ...821....8K,2017AJ....153..188F}.  In this sample, 50\% of the sources confirmed as young stars by visible spectra  do not have X-ray counterparts, consistent with the completeness of the X-ray detections estimated from known YSOs with IR excesses \citep{2013ApJ...768...99P}, and 15\% were not identified in surveys for IR excesses or X-rays.  

%\citep{2005ApJS..160..353G,2013ApJ...768...99P,2014ApJ...787..107K,2014ApJ...787..109G,2014ApJ...787..108G}. 

%Orion A has been a popular region of the spectroscopic follow up due to a high density of potential targets. Another notable but much more diffuse region with extensive spectroscopic coverage is the Orion OB1a and OB1b. Spectra of 2000 stars over the area of $\sim$150 deg$^2$ have been observed over 20 years \citep{2019AJ....157...85B}.

Near-IR spectroscopy can characterize embedded stars undetected at visible wavelengths \citep{2009AJ....137.4777W}. While near-IR spectra have comparatively fewer useful lines that can be used as an unequivocal signature of youth, large surveys such as APOGEE are obtaining near-IR spectra of hundreds of embedded stars in nearby star-forming regions. Considerable effort has been invested in using these spectra for confirming and characterizing young stars. In particular, well-calibrated $\log g$ measurements of stars can be used to separate low mass YSOs from field dwarfs or red giants, as well as to estimate their ages \citep{2020AJ....159..182O}. Lower spectral resolution ($\lambda/\Delta \lambda \sim 300$) 1-2.5~$\mu$m spectra are also effective at identifying young, mid-M type stars by their surface gravities \citep{2008ApJ...685..313P,2017AJ....153...46L}. Near-IR accretion lines can also be detected, particular in the hydrogen line series \citep{2017A&A...600A..20A}. Although they have not been used as a primary diagnostic for identifying young stars, they are a promising means for future surveys.   

%Unlike spaced-based IR and X-ray surveys,
Spectroscopic surveys at near-IR wavelengths are currently expanding rapidly. While the initial (SDSS-III) APOGEE survey targeted only known members of young clusters \citep{2015ApJ...799..136F,2015ApJ...807...27C,2016ApJ...818...59D},  SDSS-IV APOGEE  implemented a broader selection criteria that was prone to contamination, but  offered a more comprehensive coverage of several star-forming regions \citep{2018AJ....156...84K,2019AJ....157..196K}. The forthcoming SDSS-V survey will obtain near-IR spectra of young stars across the entire Galaxy.

%catalog of sources to a mix of whatever targeting criteria each individual study has utilized as well as to what could be observed in a particular timeframe with a particular instrument. Thus, it is impossible to obtain a ``complete'' sample with this method.

\subsection{Visible and near-IR photometry}

New facilities for wide-field mapping from ground-based telescopes have opened up opportunities to survey clouds at visible and near-IR wavelengths \citep[e.g.][]{2005ApJ...632..397G,2015A&A...581A.140S,2016A&A...587A.153M,2017A&A...604A..22B,2019MNRAS.486.1718S}. Near-IR (1-2.5~$\mu$m) observations can detect deeply embedded stars and brown dwarfs and are less affected by nebulosity than visible or mid-IR data; however, at these wavelengths, it is difficult to distinguish YSOs from field stars. A common approach is to count YSOs statistically by subtracting out estimates of the field star contamination. In the center of the ONC, contamination is low and can be corrected by using nearby reference fields and modeling that takes into account the effect of extinction on the surface density of background stars 
\citep{1994AJ....108.1382M,2000ApJ...540..236H,2002ApJ...573..366M}. More generally, finding excesses in the surface densities of stars toward molecular clouds is an effective means for finding and characterizing embedded clusters, but cannot identify lower density populations \citep{2000AJ....120.3139C,2017A&A...608A..13L}.

Although visible-light observations are hampered by extinction, color-magnitude diagrams (CMDs) and variability at these wavelengths have been used to select candidate YSOs for spectroscopic followup \citep{2003ApJ...593.1093L,2008ApJ...675..491A,2017AJ....153...46L,2005AJ....129..907B,2019AJ....157...85B}.  CMDs using standard filters, as well as non-standard filters selected to give effective temperatures \citep{2012ApJ...748...14D}, have also been used to identify YSOs by locating their isochrones in the HR diagram \citep{2017A&A...604A..22B}.  Recently, \citet{2021AJ....162..282M} applied machine learning to 2MASS photometry and Gaia photometry and parallaxes to search for pre-ms stars within 5 kpc, independent of the presences of IR excesses. 

%start here on read
\subsection{Kinematic \& Astrometric Data}
\label{sec:kin_id}

Kinematical studies bring the capability to identify and characterize assemblages of young stars sharing a common origin. Parallaxes can further be used to identify populations at a common distance. Newly-born YSOs inherit their kinematics from their progenitor molecular cloud (see Sec.~\ref{sec:kinematics}). As they age, these populations will begin to disperse, but in bulk they will persist in comoving groups for tens if not hundreds of Myr (see Sec.~\ref{sec:assoc}). The launch of Hipparcos led to the first large scale surveys of stellar kinematics in several nearby star forming regions and OB associations. These surveys identified members of clusters and associations and shed light on their 3D structure \citep{1999AJ....117..354D}. Due to their low resolution and sensitivity, the surveys studied primarily the high mass stars in the closest associations.

For individual star forming regions, multiple studies have measured  proper motions through long temporal baseline, high angular resolution, ground-based and space-based imaging. These data typically span on the order of a decade between observations. Such observations were conducted in various portions of the spectrum, including optical, near-IR, and radio \citep[e.g.,][]{1988AJ.....95.1755J,2005A&A...438..769D,2006A&A...460..499B,2011AJ....142...71C,2015ApJ...815....2W,2016MNRAS.460.2593W,2016ApJ...833...95D,2017A&A...597A..90D,2017ApJ...834..139D,2019AJ....157..109K}. Many of these studies were able to reach precisions in proper motions on the order of 1 mas yr$^{-1}$, progressively improving over time. For young populations with sufficiently distinct proper motions relative to the field (e.g, nearby moving groups, Sco Cen OB association, Taurus, and Ophiuchus), these studies can identify probable members based on their proper motions alone. 
In other populations that are located further away or towards the galactic anticenter, proper motions are used to identify contaminants to existing YSO catalogs by their discrepant motions \citep[e.g.,][]{1988AJ.....95.1755J}. 

Twenty years after Hipparcos, Gaia is delivering several orders of magnitude improvements in both the precision of parallaxes and proper motions and the number of sources observed. The Gaia catalogs are revolutionizing our ability to identify low mass members of star-forming complexes through systematic searches for overdensities of stars with coherent kinematics. Additionally, with known distances to the individual stars it is now possible to construct  reliable Hertzsprung-Russell diagrams. In these diagrams, low mass YSOs  lie above the main sequence, separating them from the foreground and background populations, and allowing a precise measurement of stellar ages.

Since the Gaia DR2 release, a number of star-forming complexes and OB associations have been analyzed \citep[e.g.,][]{2018AJ....156...84K, 2018A&A...619A.106G, 2018A&A...620A.172Z,2018A&A...617L...4R, 2018ApJ...869L..33O, damiani2018, 2018AJ....156..271L,2018A&A...618A..93C, 2019ApJ...870...32K}. Autonomous searches of these regions revealed previously undiscovered coherent populations, many with ages of a few dozen Myr \citep{2019AJ....158..122K}; these populations trace the dynamical evolution of stars after they disperse their natal molecular gas.

Radial velocity measurements with visible light and near-IR spectrometers can also identify young stars since the velocity dispersions of stars in clusters and clouds are typically much less than the  range of radial velocities due to galactic motions \citep{2001AJ....121.2124D,2019ApJ...871...46K}. The radial velocities of YSOs often follow that of the molecular gas  \citep{2005AJ....129..363S,2006ApJ...648.1090F,2009ApJ...697.1103T,2016A&A...589A..80H}, although  substantial deviations from the gas velocities are sometimes apparent \citep{2016A&A...590A...2S,2017ApJ...845..105D}. Measurements of the velocity dispersion from radial velocities must account for the inflation of the dispersion by binarity \citep{2012A&A...547A..35C,2019ApJ...871...46K}.

\section{The 2D Distribution of YSOs}
\label{sec:spatial}

%Surveys of molecular clouds map the 2D spatial distribution of YSOs. Early surveys with near-IR arrays  established the importance of clusters for low mass star formation \citep[e.g.][]{ 1991ApJ...371..171L,1992ApJ...393L..25L,2000AJ....120.3139C}. These surveys, however, did not reliably distinguish young stars from background stars and instead mapped the surface density of all stars and then subtracted out the estimated contribution from background stars. Statistics of the background contribution limited their ability to map low surface density populations of YSOs. Surveys that could efficiently map the distribution of YSOs over the range of surface densities  present in molecular clouds required the deployment of the Chandra, XMM, WISE, and  Spitzer space telescopes.

{\color{black}

Ground-based surveys with near-IR arrays made the first maps of the 2D distribution of YSOs in molecular clouds \citep[e.g.][]{ 1991ApJ...371..171L,1992ApJ...393L..25L,2000AJ....120.3139C}. Although the ground-based surveys identified clusters of young stars, the detection of lower density populations awaited the deployment of IR and X-ray space telescopes. 

Their capabilities are demonstrated by the map of the Orion Nebular Cluster, or ONC, in Fig.~\ref{fig:isf}.  The dense peak of the ONC found in previous near-IR and visible imaging is apparent in the center of the plot \citep{1998ApJ...492..540H}, but so is the elongated population of YSOs that surrounds the cluster and is aligned with the filamentary cloud. The protostars closely follow the filament while the more evolved pre-ms stars show a wider distribution. 

The resulting 2D distribution of YSOs have led to a full characterization of the range of stellar densities, star formation efficiencies, and star formation rates in clouds.  This resulted in one of the most important discoveries, the presence of star forming relations in individual molecular clouds.}

\begin{figure}[t!]
%\epsscale{1.1}
%\plotone{figures/figure_yso_pdf_pasp.pdf}
%\includegraphics[width=0.42\textwidth,trim=0 0 0 0, clip]{figures/figure_yso_pdf_pasp.pdf}
\includegraphics[width=0.42\textwidth]{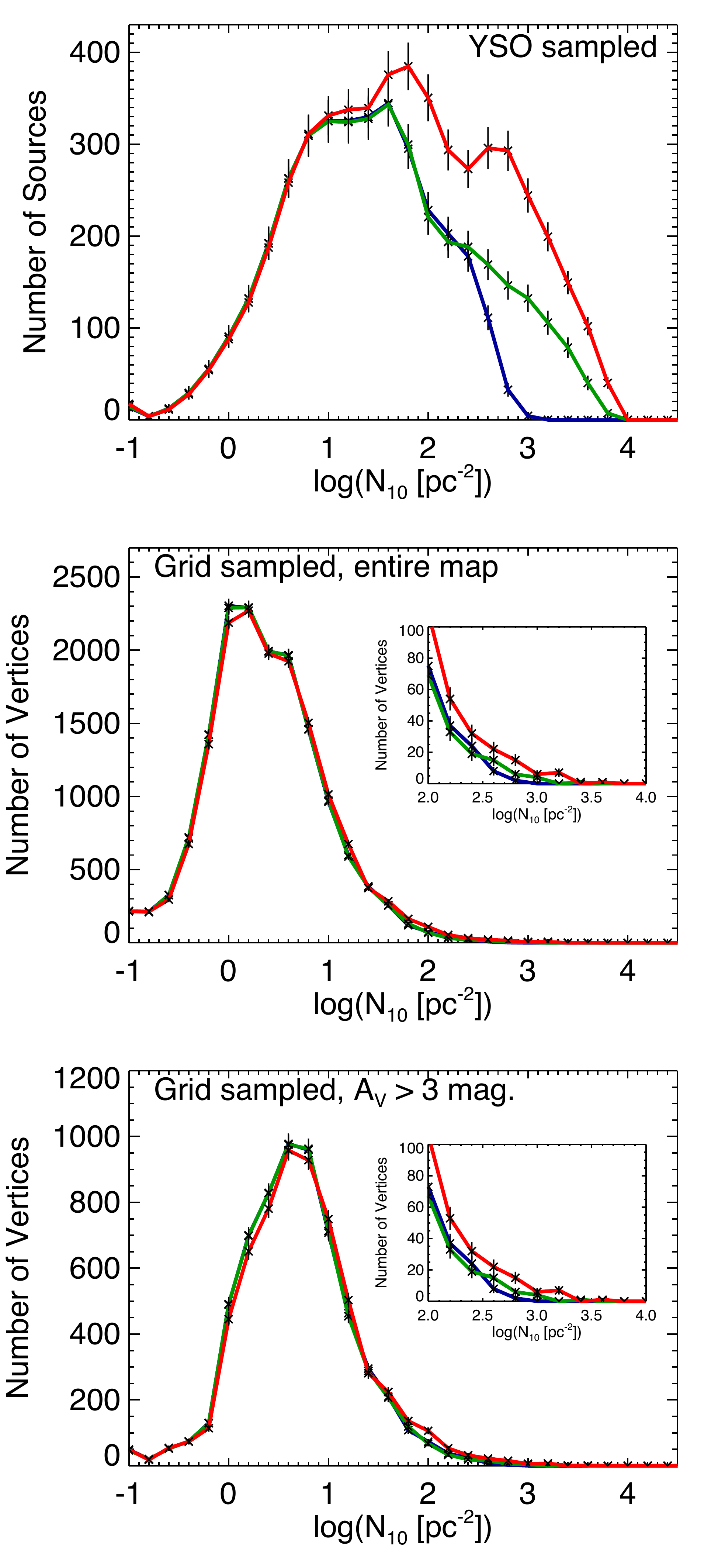}
%\vskip -0.1 in
\caption{Nearest neighbor surface density PDFs for  Orion~A showing the effect of sampling and A$_v$ cutoff. The blue histograms give the densities for the dusty YSOs determined by  Spitzer data alone, the green give the densities from Spitzer data corrected with Chandra data, and the red gives the Spitzer and Chandra data corrected for incompleteness \citep{2016AJ....151....5M}.}
\label{fig:yso-area}
\end{figure}

\subsection{Stellar Surface Density PDFs}
\label{sec:PDF}

The range of YSO surface densities can be characterized by histograms of the surface density, similar to the gas column density probability density functions (PDFs) used to study molecular clouds \citep[e.g.][]{2009A&A...508L..35K,2013ApJ...766L..17S,2015A&A...577L...6S,2015A&A...576L...1L,2016MNRAS.461...22P}. The range of densities necessitates the use of adaptive methods for measuring the surface density. PDFs can be generated by sampling the surface densities  centered on each YSO, where there is one density value for each object (YSO sampled PDFs), or by sampling the density over a uniform grid covering the mapped region \citep[grid or area sampled PDFs,][]{2011ApJ...739...84G}. The nearest neighbor density provide an adaptive measurement, 

\begin{equation}
N_{n} = \frac{n-1}{\pi r_n^2},~\sigma_{n} = \frac{N_n}{\sqrt{n-2}},
\end{equation}

\noindent
where $N_n$ is the surface density and $\sigma_n$ is its uncertainty. Here, $r_n$ is the distance to the nth nearest neighbor from either a YSO or a grid point for the YSO  and grid sampled distributions, respectively \citep{1985ApJ...298...80C,2016AJ....151....5M}. Other methods for obtaining grid sampled surface densities include smoothing the YSO distribution by a gaussian kernel \citep{1993AJ....105.1927G},  smoothing by a kernel with an adaptive width \citep{2000AJ....120.3139C}, and Voronoi tesselation \citep{2014ApJ...787..107K}.

\begin{figure}[t!]
\vskip -0.1 in
\includegraphics[width=0.4\textwidth]{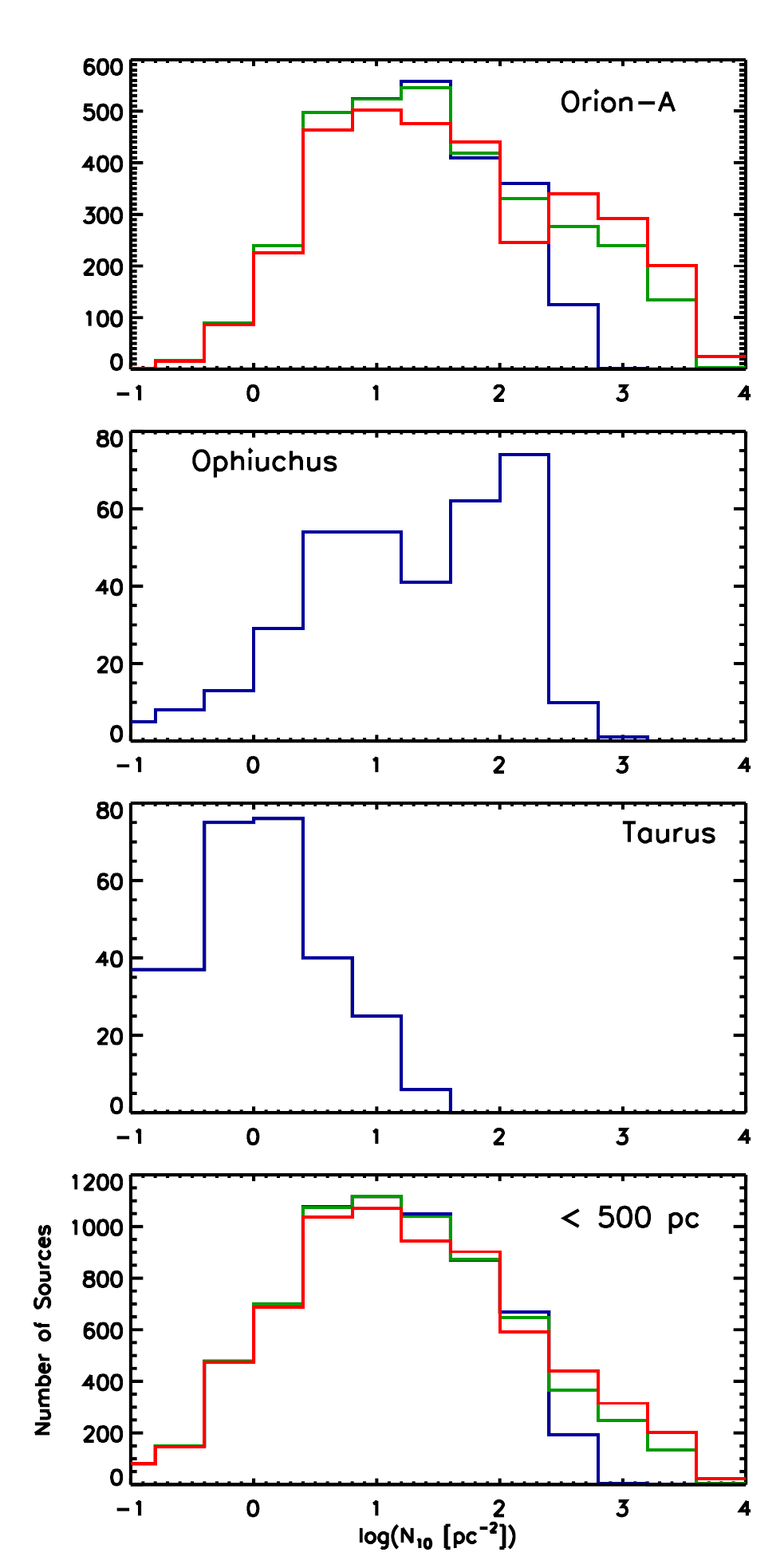}
\caption{YSO sampled PDFs for the spatial distribution of dusty YSOs in Orion~A, Ophiuchus, and Taurus molecular clouds.  The different colors show the PDFs resulting from the three different completeness corrections applied to the Orion~A data, as described in Fig.~\ref{fig:yso-area}. }
\label{fig:PDF}
\end{figure}

The  grid and YSO sampled PDFs of the Orion~A cloud are shown in Fig.~\ref{fig:yso-area}. Since low density regions with small numbers of YSOs dominate the projected area of molecular clouds, the grid sampled PDFs peak at lower surface densities than the YSO sampled PDF. The grid sampled PDFs also depend on the selected area; for example, by reducing the sampled region to the area of the cloud where $A_V > 3$, the peak of the PDF shifts to higher densities. In comparison, as long as most of the YSOs are within the survey, YSO sampled PDFs are insensitive to the boundaries of the survey field. Using estimates of the fractions of undetected YSOs, the PDFs can be corrected for incompleteness \citep{2016AJ....151....5M}.

\begin{figure*}
\epsscale{1.05}
\plottwo{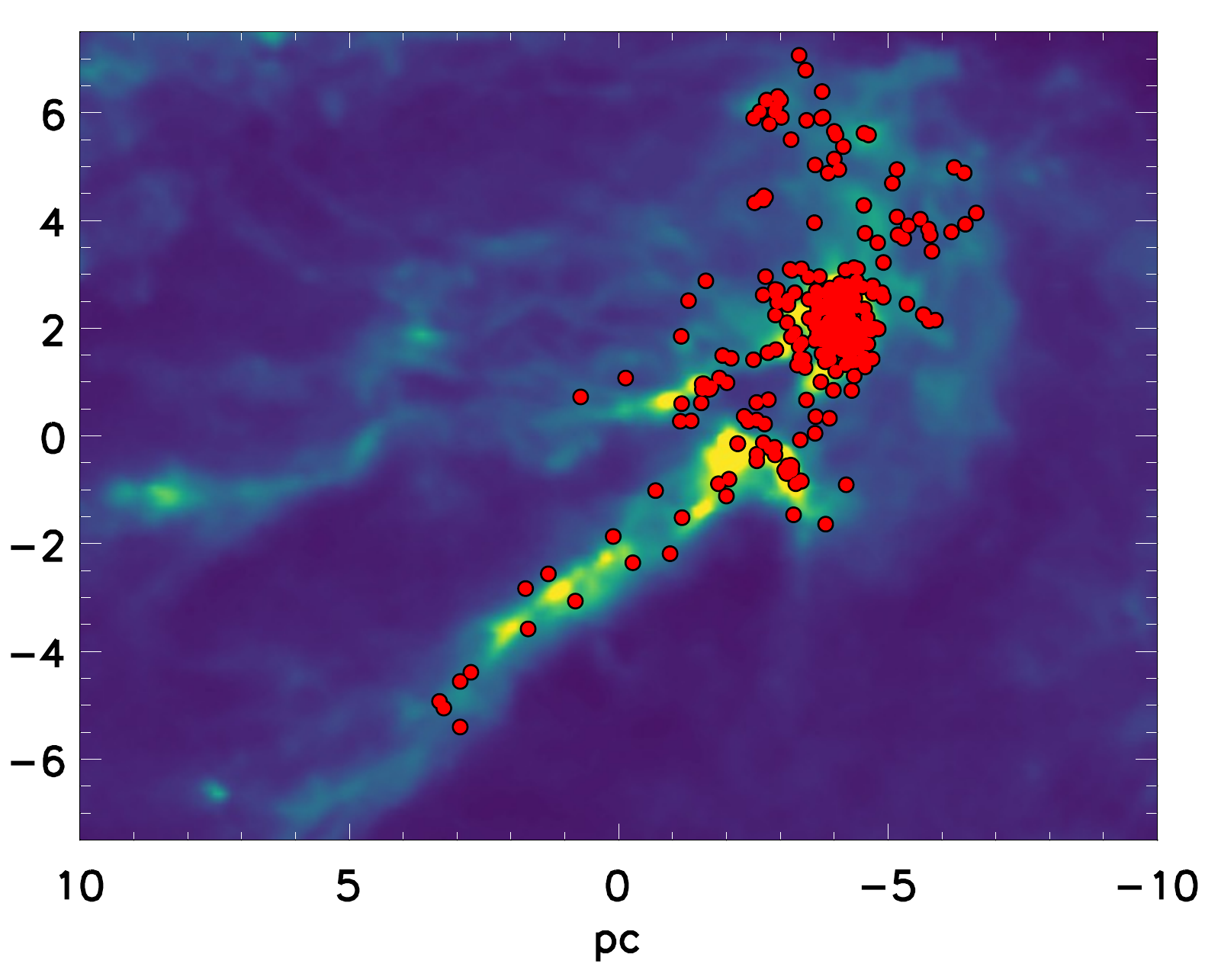}{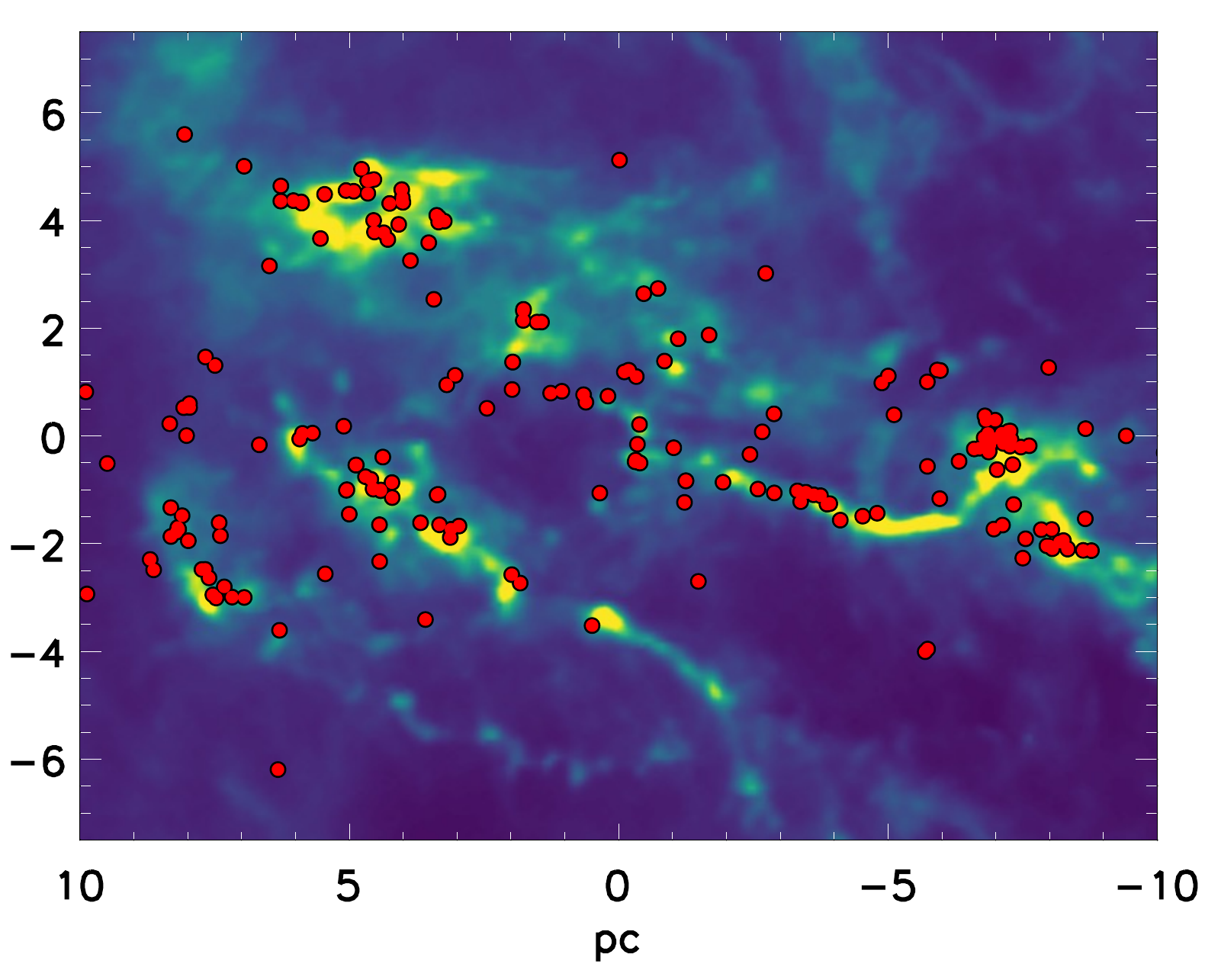}
\caption{The distribution of YSOs in the Ophiuchus and Taurus molecular clouds. Left: Planck-derived extinction map of Ophiuchus  with the YSOs from the SESNA processing of the c2d survey \citep[Gutermuth in prep.,][]{2009ApJS..181..321E}. Right: Planck-derived extinction map of  Taurus  with the YSOs from \citet{1995ApJS..101..117K}. The coordinate scale is in parsecs at the distances of 130 pc and 140 pc for Taurus and Ophiuchus, respectively. The maps are from the Planck Legacy Archive \citep{2014A&A...571A..11P}.} \label{fig:oph_taurus}
\end{figure*}

%The stellar surface PDFs can be constructed either by measuring the densities over a grid of points within a molecular cloud (grid base) or from positions centered on known YSOs (YSO based). In the literature, the densities of the stellar PDFs are YSO-based, giving the range of densities found around known YSOs.   Stellar surface density PDFs can be generated by sampling the densities centered on each YSO (i.e. there is one value for each YSO, hereafter local surface density), or by sampling the density at points of a uniform grid covering the mapped region (hereafter grid surface density). The differences of these two techniques are shown in Figure 1 where we show the PDF for the Orion~A cloud used a grid based and local based methods. These show the grid based histogram shifted to lower surface densities; this reflects that low density regions with small small numbers of stars dominate the projected area of molecular clouds. For this reason, the grid based PDFs depend on the extent of the grid. Most of the Spitzer surveys are limited to regions with $A_V > 3$; reducing or increasing this value will push the peak to lower and higher densities, respectively. In comparison, as long as most of the young stars are within the surveyed regions, local surface density PDFs are not strongly dependent on the extent of that region. 

The YSO sampled PDFs for three nearby clouds, Taurus, Ophiuchus and Orion~A, are shown in Fig.~\ref{fig:PDF}. These show a several order of magnitude spread in the surface densities of YSOs, from a few stars per pc$^{2}$ to almost 10,000 per pc$^{2}$. The combined PDF of all the clouds within 500 pc covered by the c2d, Gould Belt and Orion surveys is also shown \citep[using data from the SESNA program for the c2d and Gould belt regions, Gutermuth et al.~in prep.,][]{2009ApJS..181..321E,2012AJ....144..192M,2015ApJS..220...11D}. The combined surface density distribution of the local star forming regions was first examined by \citet{2010MNRAS.409L..54B}. They found that the combined PDF has a lognormal form with a peak of 22 stars pc$^{-2}$ and a dispersion of $\sigma_{log_{10}\Sigma} = 0.85$. A caveat to their approach was that, due to the incompleteness in the Orion Nebula, they excluded the  inner regions of the ONC. When included, the distribution deviates from a lognormal distribution with a tail at high densities \citep[see Fig.~\ref{fig:PDF} and][]{2015ApJ...802...60K}.  
%reread here

\citet{2010MNRAS.409L..54B} interpreted the lack of a discontinuity in the composite PDF as evidence against distinct clustered and distributed modes of star formation operating in these nearby clouds, with only a quarter of the stars forming in dense clusters. {\color{black} \citet{2012A&A...545A.122P} and \citet{2012MNRAS.426L..11G} noted that such an interpretation was not unique. They showed that a superposition of multiple evolving clusters with different densities could reproduce the Bressert~et~al.\ PDF. Indeed, the PDFs of individual clouds show disparate PDF shapes, some dominated by high density clusters and others  by low density distributed populations (Fig.~\ref{fig:PDF}). Thus, the Bressert PDF does result from the superposition of multiple disparate PDFs.}  

\citet{2016AJ....151....5M} compared the PDFs of eight nearby molecular clouds.  The  clouds split into two distinct groups;  the five clouds with clusters and median densities $> 25$~YSO~pc$^{-2}$  and the three clouds without clusters and median densities below $< 10$~YSO~pc$^{-2}$. Examples of these two groups are the Ophiuchus and Taurus molecular clouds (Figs.~\ref{fig:PDF} and \ref{fig:oph_taurus}); these clouds are at similar distances, have comparable masses, and contain similar numbers of dusty YSOs. Yet, in Ophiuchus most of the YSOs are concentrated in a single cluster while in Taurus, the YSOs are distributed throughout the cloud. This indicates the degree of clustering is not simply a function of the mass of a cloud or the number of YSOs. Instead, the density structure of the molecular cloud appears to be the most relevant factor (see Sec.~\ref{sec:gas_env}). 

{\color{black} While some clouds contain both clusters as well as diffusely distributed stars following the filamentary gas structure, others only have diffuse star populations. Based on these two distinct types of PDFs, \citet{2016AJ....151....5M} argued that 10~YSO~pc$^{-2}$ is an appropriate density threshold for identifying stars in clusters.}

\section{Star Formation Rate and Instantaneous Efficiency}
\label{sec:sfr&e}

The rate and efficiency are fundamental statistics that can characterize star formation on the scales of clouds and galaxies. The number and density of low mass YSOs, discussed in the previous section, provide a new means for measuring the rate and efficiency at which stars form throughout molecular clouds. These have been used by different authors to determine star-formation relations in the clouds.  

\subsection{The Star Formation Rate (SFR)}
\label{sec:sfr}

The simplest estimate of the SFR is given by 

\begin{equation}
    SFR~({\rm M}_{\odot}~{\rm Myr}^{-1})= m_{\star} \frac{n_{\rm YSO}}{t_{\rm YSO}},
    \label{eqn:sfr_number}
\end{equation}

\noindent
where $n_{\rm YSO}$ is the number of YSOs, $t_{\rm YSO}$ is the lifetime of the YSOs in Myr, and $ m_{\star}$ is the average mass for a typical IMF ($0.5$~M$_{\odot}$).  The value of $n_{\rm YSO}$ is either the number of protostars \citep[e.g.][]{2010ApJ...723.1019H,2013A&A...559A..90L} or the total number of dusty YSOs \citep[e.g.][]{2011ApJ...739...84G}, including both protostars and pre-main sequence (pre-ms) stars with disks. 

\begin{figure}[t!]
\includegraphics[width=0.4\textwidth]{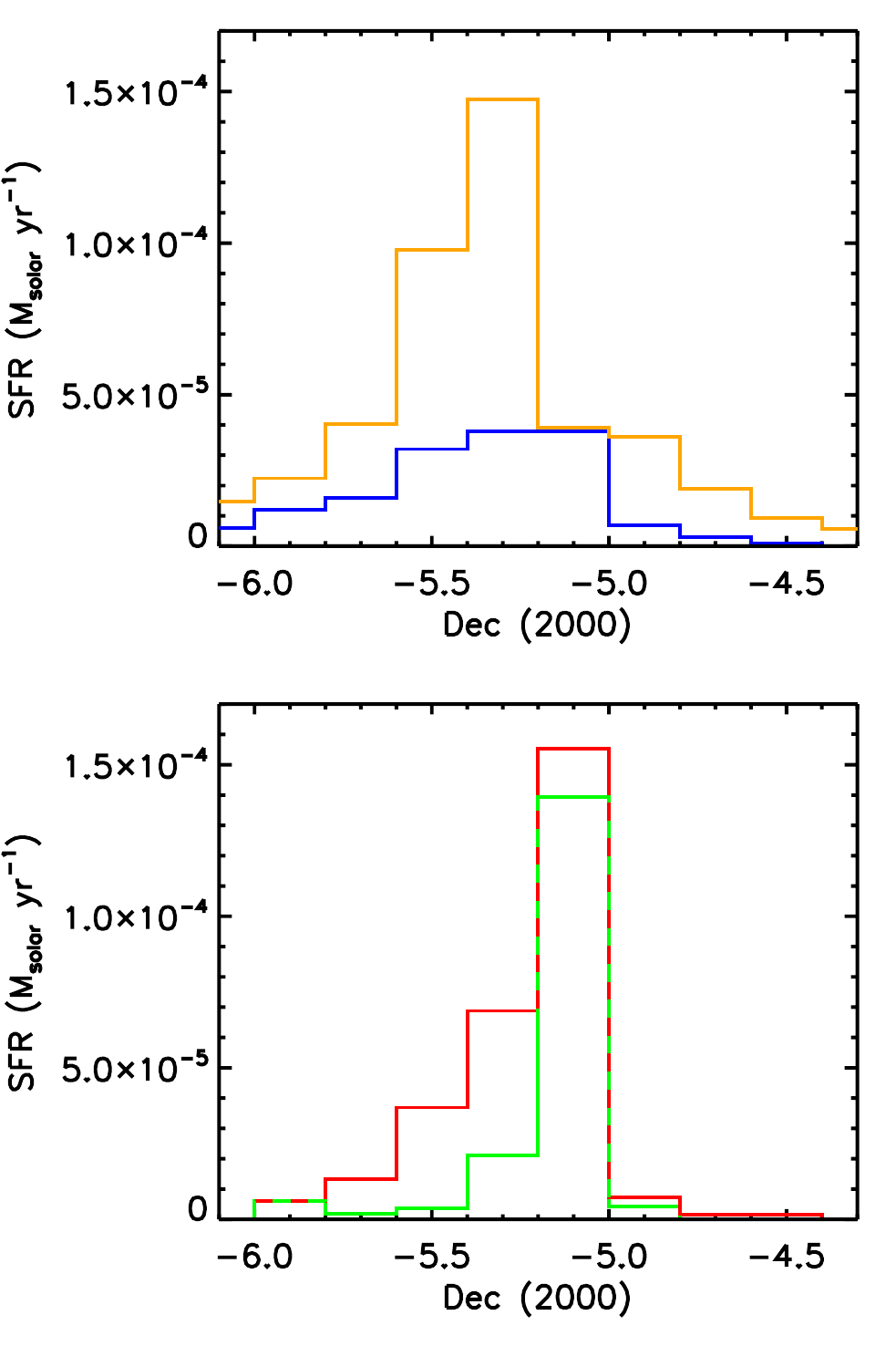}
\caption{The SFR in the ISF (Fig.~\ref{fig:isf}).  Each bin shows the total rate in that dec.~range.  {\bf Top:} the yellow histogram is the SFR calculated from the total number of YSOs and the blue line is that calculated from the number of protostars, both using Eqn.~\ref{eqn:sfr_number}. {\bf Bottom:} the SFR calculated using the luminosities of the protostars and Eqn.~\ref{eqn:sfr_lum}; the green histogram has only the HOPS protostars  while the red histogram includes a correction for protostars not observed by Herschel. See Appendix~\ref{sec:appendix_SFR}.}
\label{fig:sfr}
\end{figure}

When $n_{\rm YSO}$ is the number of protostars, $t_{\rm YSO}$ is the lifetime of protostars, $\approx 0.5$~Myr \citep{2014prpl.conf..195D}. If $n_{\rm YSO}$ is the total number of dusty YSOs, then a lifetime for pre-ms stars with disks is required. The half-life of the disks is typically adopted for this lifetime, $\approx 2$~Myr \citep{2008ApJ...686.1195H,2009AIPC.1158....3M,2009ApJS..181..321E}. The combined lifetime for protostars and pre-ms stars with disks, 2.5~Myr, can then be used for $t_{\rm YSO}$ \citep{2020ApJ...896...60P}. 

One caveat to this approach is that $t_{\rm YSO}$ may be affected by the local environment. The lifetime of protostars likely varies with their birth environment due to the dependence of the free fall times of the protostellar envelopes on the local gas properties. Evidence for this dependence is found by studies of protostellar luminosities, which find that protostars located in regions with high YSO densities (and therefore, high gas densities; see Sec.~\ref{sec:sf_law}) exhibit systematically higher luminosities \citep{2012AJ....144...31K,2014AJ....148...11K, 2014ApJ...782L...1E}. The higher luminosities likely reflect higher accretion rates, and therefore shorter protostellar lifetimes, in regions with high YSO and gas densities. This can result from shorter free fall times in the denser gas. 

{\color{black}
It is not clear whether the half-life of disks varies significantly with environment. The outer regions of disks are truncated by UV radiation from the surrounding environment, with the degree of truncation depending on the intensity of the radiation field \citep{2004ApJ...611..360A,2010ApJ...725..430M,2016MNRAS.457.3593F,2017MNRAS.468L.108H,2018ApJ...860...77E,2019MNRAS.488.1462P,2020A&A...640A..27V}. The effect of photoevaporation by UV on the inner regions of disks detected by Spitzer, however, is uncertain. Studies of clusters in the nearest 2~kpc have obtained mixed results on whether the  lifetimes of the inner disks are shortened by the UV environment \citep{2007ApJ...660.1532B,2012ApJ...750..125A,2020ApJ...889...50Y}.  Thus, the SFR calculated from the total number of YSOs may be less dependent on environment than the SFR determined from protostars alone.  }

An alternative method for measuring the SFR uses the combined luminosities of protostars to calculate a total instantaneous accretion rate. This accretion rate is proportional to the total luminosity of the protostars. Adopting a typical mass to radius relationship, the SFR is given by the equation 

\begin{equation}
    SFR~({\rm M}_{\odot}~{\rm yr}^{-1}) = \frac{1}{G} \left(\frac{r_{\star}}{m_{\star}}\right)\frac{1}{\eta} \sum_i L_{\rm proto}^i ,
    \label{eqn:sfr_lum}
\end{equation}

\noindent
where $L^i_{\rm proto}$ is the luminosity of the $ith$ protostar, $m_{\star}/r_{\star}$ is the mass-radius ratio, and $\eta$ is the fraction of the accretion energy radiated into space. Since in most cases there are no direct measurements of $\eta$ or the mass-radius ratios, values from models of protostars must be adopted (see Appendix~\ref{sec:appendix_SFR} for details). 

In Fig.~\ref{fig:sfr}, we show the star formation rate calculated three different ways: using Eqn.~\ref{eqn:sfr_number} with the total number of YSOs in a given declination bin ($t_{\rm YSO} = 2.5$~Myr), using Eqn.~\ref{eqn:sfr_number} with the number of protostars in that bin ($t_{\rm YSO} = 0.5$~Myr), and using Eqn.~\ref{eqn:sfr_lum} with the total accretion luminosity for all the protostars in the bin.  We find the answers can differ significantly.  If we use Eqn.~\ref{eqn:sfr_number} and the total number of YSOs to measure the SFR, we find the peak SFR occurs in the Orion Nebula. If we instead use the total number of protostars, the peak is to the north of the nebula. Although there are biases against detecting protostars in the Orion Nebula, this analysis suggests that the star formation is currently concentrated north of the Orion Nebula, and the location of star formation has shifted northward. Furthermore, comparison to the SFR from  Eqn.~\ref{eqn:sfr_lum} suggests that using the standard protostellar lifetime (0.5~Myr) results in an underestimate of the current star formation rate. This implies that the protostars north of the nebula are accreting at a higher than typical rate. The inconsistencies between these methods motivate future studies. Methods for measuring the SFR with protostars require a deeper understanding of protostellar lifetimes and accretion. Furthermore, resolving differences between methods which measure the time averaged rate and those that measure the instantaneous rate require an understanding of the time dependency of the SFR.

\subsection{The Star Formation Efficiency (SFE)}
\label{sec:SFE}

The instantaneous SFE is the fraction of the  total mass that is found in stars, 

\begin{equation}
    SFE = \frac{M_{\star}}{M_{\rm cloud}+M_{\star}}= \frac{ m_{ \star} n_{\scalebox{.5}{YSO}} }{M_{\rm cloud}+m_{ \star} n_{\scalebox{.5}{YSO}}},
\end{equation}

\noindent
where $M_{\star}$ is the current total mass in stars, $M_{\rm cloud}$ is the current molecular gas mass of the cloud, $m_{\star}$ is the mean stellar mass for a standard IMF, typically 0.5~M$_{\odot}$, and $n_{\rm YSO}$ is the number of YSOs. By directly counting the stars that constitute the bulk of the stellar population in both mass and number, this provides a direct and statistically robust measurement of the efficiency.  For more distant regions, the efficiency is determined by indirectly measuring the number of massive stars \citep[e.g.][]{2016ApJ...833..229L}. Although these approaches produce larger, more representative samples, they require more assumptions, are sensitive to statistical variations in the properties of the most massive stars, and cannot measure SFEs of smaller clouds without massive stars. 

The determination of the SFE requires a measurement of the cloud mass, which is derived from near-IR extinction maps, molecular line maps, or thermal dust emission maps \citep[see overview in][]{2006AJ....131.2921R}. This mass does not include gas dispersed by winds, outflows and radiation, and the measured SFE can be increased by the clearing of gas by feedback.

\begin{figure}[!t]
\includegraphics[width=0.48\textwidth]{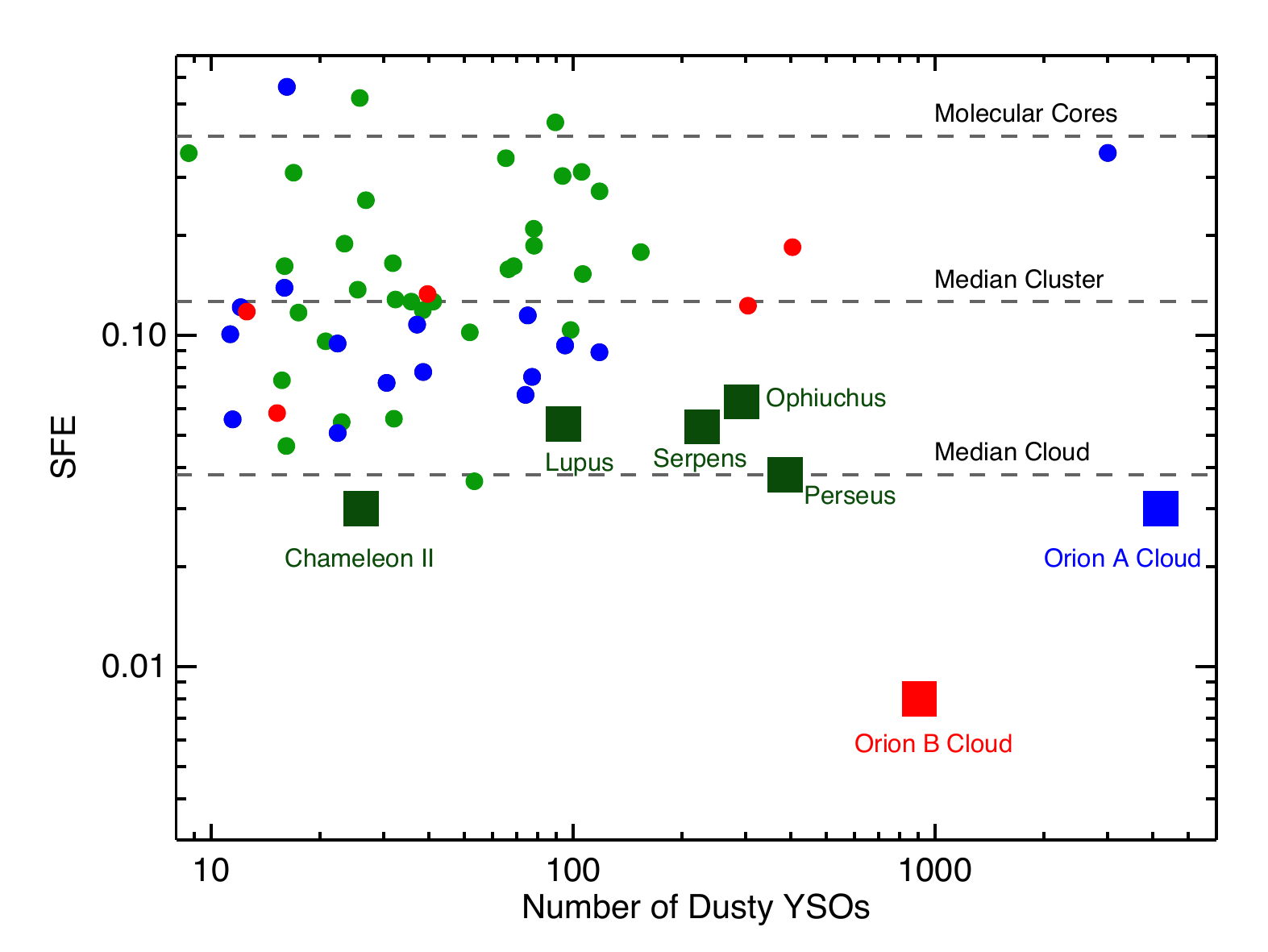}
\caption{The instantaneous SFEs for clusters and clouds surveyed by {\it Spitzer}. Circular markers are clusters while square markers are clouds. Blue and red markers are from the Orion~A and B clouds, respectively \citep{2016AJ....151....5M}, the light green makers are clusters from \citet{2009ApJS..184...18G}, and the dark green markers are clouds from \citet{2009ApJS..181..321E}. For the \citet{2009ApJS..184...18G} data, the cloud masses are determined by integrating 2MASS $A_V$ maps over the convex hulls determined for each cluster.  The Orion cluster gas masses are also determined by integrating the 2MASS $A_V$ maps over the region subtended by the cluster \citep{2016AJ....151....5M}. For the cluster data and the Orion clouds, the number of dusty YSOs has been divided by 0.75 to correct for diskless pre-ms stars \citep{2016AJ....151....5M}.  The dense core SFE is that estimated by \citet{2015A&A...584A..91K}.}
\label{fig:sfe}
\end{figure}

We show the SFEs for entire clouds and for clusters in Fig.~\ref{fig:sfe}. For the molecular clouds, the number of YSOs is that for the entire cloud and the mass is the total cloud gas mass.  The adopted gas mass varies with the size of the survey field and  the chosen gas tracer (e.g. C$^{18}$O vs $^{13}$CO) since most of the mass is in the extended, low density gas \citep[e.g.][]{2015A&A...575A..79S,2016MNRAS.461...22P}. The SFEs of clusters require cluster boundaries to determine the number of YSOs (see Sec.~\ref{sec:clusters}). The gas masses have been determined by the virial masses of clumps coincident with the clusters \citep{1992ApJ...393L..25L} or by the masses obtained by integrating column density maps over the regions subtended by the clusters \citep[e.g.][]{2016AJ....151....5M}.  

Fig.~\ref{fig:sfe} shows a picture apparent since the first IR cloud surveys \citep{1992ApJ...393L..25L}.  Molecular clouds have a median SFE of only 0.038, with Orion~B showing an unusually low SFE (0.0055).  Clusters show a median SFE of 0.13, although with significant scatter.   This scatter may arise from some clusters being at the onset of SF, and therefore gas rich, while other clusters are depleted of gas by star formation and the associated feedback.  In the case of the highest mass cluster, the ONC, the high SFE may result from an overestimate in $n_{\scalebox{.5}{YSO}}$ from completeness corrections  and an underestimate of the gas mass by the  $A_V$ maps due to the inability of 2MASS to penetrate the high extinction in parts of this cluster \citep{2016AJ....151....5M}.  Finally, estimates for individual molecular cores give a value of around 0.40 for the SFE \citep[e.g.][]{2015A&A...584A..91K}. Thus, there is a factor of three increase in SFE  from cloud to cluster scales, and another such increase from cluster to core scales.

{\color{black} The almost order of magnitude spread in the SFEs of the molecular clouds in Fig.~\ref{fig:sfe} may be due to cloud evolution. The observed SFE of a cloud will increase with time, rising first as gas is converted into stars, and then later due to the dispersal of the molecular gas. Thus, the instantaneous SFE will first underestimate and later overestimate the integrated SFE of a cloud, i.e. the total fraction of gas converted into stars over the lifetime of a cloud. This can lead to a large spread in the values of the observed SFEs for an ensemble of clouds even if the integrated efficiencies are similar \citep{2019MNRAS.488.1501G}. }

\begin{figure*}
\epsscale{1.3}
%\hskip -0.5 in 
%\plotone{figures/stargasplots_fits_stm.png}
%trim={5cm 0 0 0}
\includegraphics[width = 1.\textwidth]{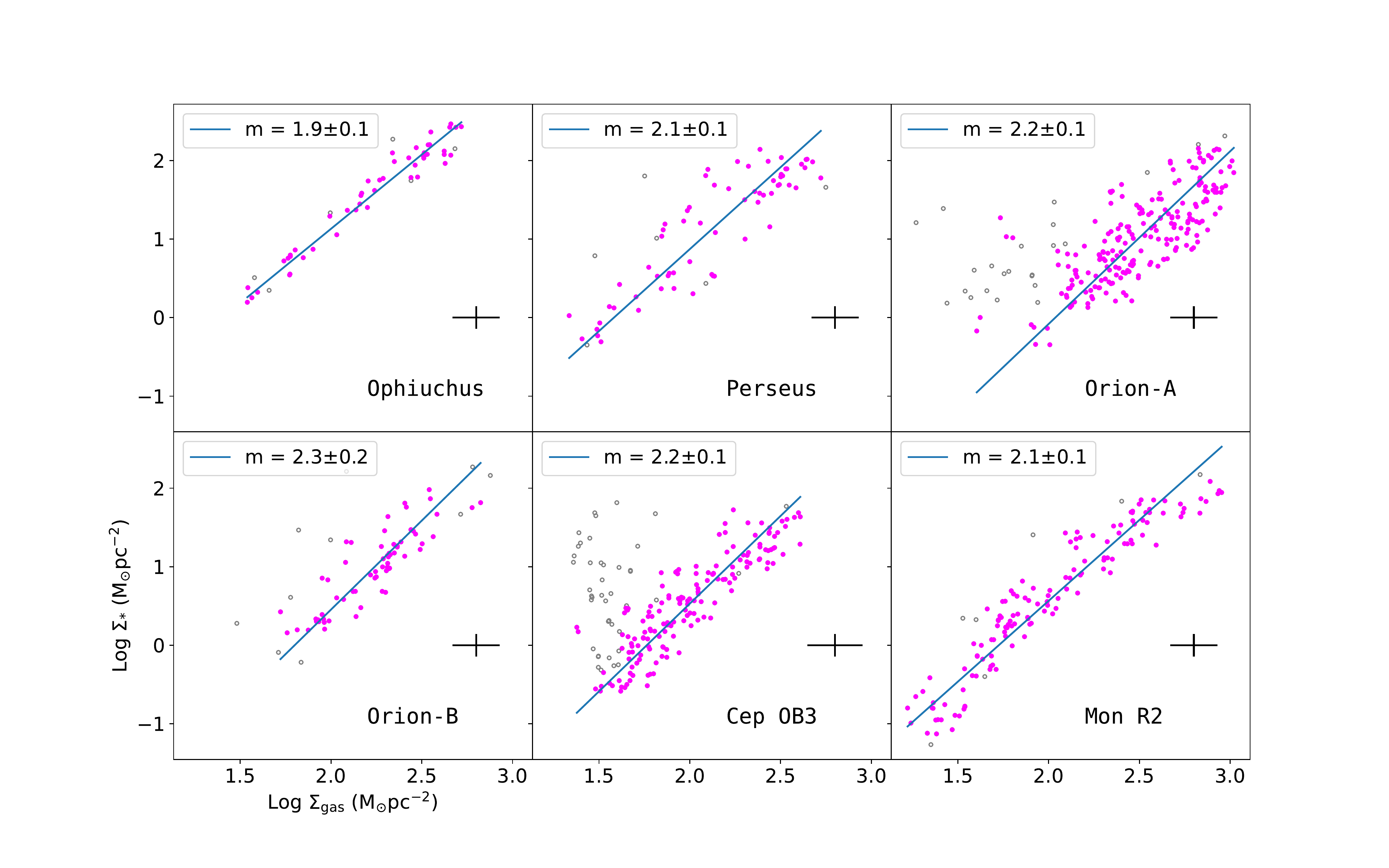}
\caption{The YSO surface densities  vs gas column densities  for six molecular clouds surveyed with Spitzer and Herschel.  These plots, adapted from \citet{2020ApJ...896...60P}, show  densities measured at the positions of protostars in regions where the CII/CI (i.e. pre-main sequence star/protostar) ratios are $< 30$; this  eliminates regions where gas dispersal has altered the correlation.  The value of $m$  are given for each cloud.}
\label{fig:sflaw}
\end{figure*}

\subsection{SF Relations in Molecular Clouds}
\label{sec:sf_law}

%The progressive increase in the SFE from cloud to core scales has motivated examinations of how the surface density of YSOs varies with the gas column density. 

Spitzer  YSO   have uncovered  star formation  (sf-) relations  within  individual molecular  clouds;  arguably one of the  most  important and unexpected results from these surveys. These relations were motivated by observations showing that the  the local densities of YSOs and SFRs in molecular clouds vary by three orders of magnitude (Figs.~\ref{fig:PDF} \& \ref{fig:sfr}). {\color{black}These variations must result, at least in part, from variations in the gas environment from which the stars form.}

%, yet the question remained: which property or properties are responsible.

%for regulating star formation?  Herschel and ground-based surveys of clouds show that the gas kinetic temperature in clouds varies by a factor of two, but that the  volume  and  column gas densities vary  by  more than an order of magnitude \citep{2014Sci...344..183K,2016MNRAS.461...22P,2017ApJ...843...63F}. This suggests  that  gas  density  is  one of the  factors  controlling the SFR and SFE. Turbulence  and magnetic fields may also play roles directly, or indirectly by shaping the gas density structure.

A growing body of work indicates that gas density is the primary environmental factor. Comparisons of $n_{\scalebox{.5}{YSO}}$ to the  masses of clouds suggested that star formation occurred primarily in regions where the gas column and volume densities exceeded a threshold \citep[$A_V \ge 7$ or $n(H) \ge 10^4$~cm$^{-3}$,][]{2010ApJ...724..687L,2010ApJ...723.1019H,2014Sci...344..183K}.  Subsequent work comparing  the spatially varying surface density of YSOs, $\Sigma_{\scalebox{.5}{YSO}}$, to the corresponding column density of gas, $\Sigma_{\rm gas}$, revealed a star-gas surface density correlation,

\begin{equation}
    \Sigma_{\rm YSO}(x,y) = \kappa \Sigma_{\rm gas}(x,y)^m,
\end{equation}

\noindent where, $x, y$ is the position in a cloud and $m \approx 2$.  \citep{2011ApJ...739...84G,2012ApJ...752..127M,2013ApJ...778..133L,2013A&A...559A..90L,2014ApJ...794..124R}. This correlation is interpreted as a sf-relation  within  clouds,

\begin{equation}
    \frac{\partial\Sigma_{\rm YSO}(x,y,t)}{\partial t} = \kappa \Sigma_{\rm gas}(x,y,t)^m,
\end{equation}

\noindent
where $\kappa$ is a constant \citep{2011ApJ...739...84G}. Here $\Sigma_{\rm YSO}$ is calculated from the density of dusty YSOs \citep{2011ApJ...739...84G,2020ApJ...896...60P} or from the  density of protostars \citep{2013A&A...559A..90L}. 

The sf-relation is apparent  when the stellar and gas surface densities are smoothed over 0.2 to 1~pc scales \citep{2011ApJ...739...84G}. Observations sampling $< 0.2$~pc spatial scales begin to resolve individual dense cores. \citet{2019MNRAS.483..407S} found that dense cores are present even in regions where the smoothed gas surface density is low. They also found that the number of cores per area correlates with the column density of gas smoothed over 0.2 to 1~pc scales. Since these cores will collapse into individual stars (or stellar systems), this correlation  likely gives rise to the observed sf-relation.  

%\citet{2013ApJ...778..133L} argued that the density thresholds inferred for nearby clouds were a result of this star formation law and steep drops in the molecular clouds PDFs at high gas column densities. 

Initial measurements showed that $m$ varied between $1.5-4$ across an assortment of clouds. One of the primary uncertainties in this intra-cloud sf-relation is the effect of evolution and gas dispersal on the star-gas correlation.  \citet{2020ApJ...896...60P} mitigated this uncertainty by using the  pre-ms star/protostar ratio as a proxy of age to exclude more evolved regions in a cloud. Using Herschel 160-500~$\mu$m data to  map the column densities of gas in  clouds, they found star-gas surface density correlations with $m=1.8-2.3$ in twelve nearby  clouds with gas masses ranging from 3000 to $1.8 \times 10^6$~M$_{\odot}$ (Fig.~\ref{fig:sflaw}). This value of $m$ implies that the local instantaneous SFE increases approximately linearly with $\Sigma_{\rm gas}$ \citep{2011ApJ...739...84G}, and thus provides a basis for understanding variations in the SFE from cloud to cluster scales.  

The scaling factor $\kappa$ of the star-gas correlation varies from cloud to cloud \citep{2013ApJ...778..133L,2020ApJ...896...60P}. Thus, this SF-relation only applies within a given molecular cloud and is not directly relatable to star-formation relations found on galactic scales \citep{2012ARA&A..50..531K,2012ApJ...745..190L}. 

\subsection{The Efficiency per Free Fall Time}
\label{sec:sf_ff}

The SFR  can also be expressed as the fraction of molecular gas that is converted into stars per free fall time, or efficiency per free fall time, 

\begin{equation}
  \epsilon_{ff} = \frac{\dot M_{\star} \tau_{ff}}{M_{gas}} \approx SFE \left(\frac{\tau_{ff}}{t_{SF}}\right),
\end{equation}

\noindent
where $\tau_{ff}$ is the free fall time for the gas and $t_{SF}$ is the interval over which star formation has occurred. Since $\tau_{ff} \propto \rho^{-0.5}$, a typical gas density $\rho$ must be determined. The efficiency has been calculated for dense clumps and entire molecular clouds, giving typical values of a few percent \citep[e.g.][]{2007ApJ...654..304K,2016ApJ...833..229L}.

%2011ApJ...729..133M}. 

%The substantial scatter in $\epsilon_{ff}$ for clouds, as determined using massive star tracers, suggested that $\epsilon_{ff}$ increased with time \citep{2016ApJ...833..229L}. 

\citet{2021ApJ...912L..19P} calclulated the effciency per free fall time for the clouds in their 2020 study, using the gas mass within successive gas column density contours to estimate the density \citep[see also][]{2014ApJ...782..114E}.  Using the number of YSOs within the surface density contours, they found that the intra-cloud sf-relation can be expressed as

\begin{equation}
    \frac{\partial\Sigma_{\rm YSO}(x,y,t)}{\partial t} = \epsilon_{ff} \frac{ \Sigma_{\rm gas}(x,y,t)}{t_{ff}},
\end{equation}

\noindent 
where $\Sigma_{\rm YSO}$ is the number of protostars in a gas surface density contour divided by the projected area of that contour, $\tau_{ff}$ is calculated from the density estimated within a contour, and $\epsilon_{ff}$ is the star formation efficiency per free fall time. Due to its linear nature, $\Sigma_{\rm gas}$ and $\Sigma_{\rm YSO}$ can be recast in terms of volume densities.

{\color{black} They found that for clouds within 1.5~kpc of the Sun, $\epsilon_{ff} \approx 0.026$; this value decreases to  $\epsilon_{ff} \approx 0.01$ if corrections for the cloud structure are applied to the free fall time \citep{2021arXiv210904665H}.  The variation in efficiency between clouds are small: $\sigma[log_{10}({\epsilon_{ff}})] = 0.2$.  Since this relation is found in clouds both with and without high mass stars, \citet{2021ApJ...912L..19P} suggest that star formation must be regulated, at least in part, on local scales by processes such as MHD turbulence, magnetic fields and protostellar outflows {\color{black} \citep[e.g.][]{2005ApJ...630..250K,2011ApJ...743L..29H,2011ApJ...730...40P,2012ApJ...761..156F,2015MNRAS.450.4035F,2021MNRAS.502.3646G}}.}

\begin{figure}[t!]
\epsscale{1.1}
\center{\plotone{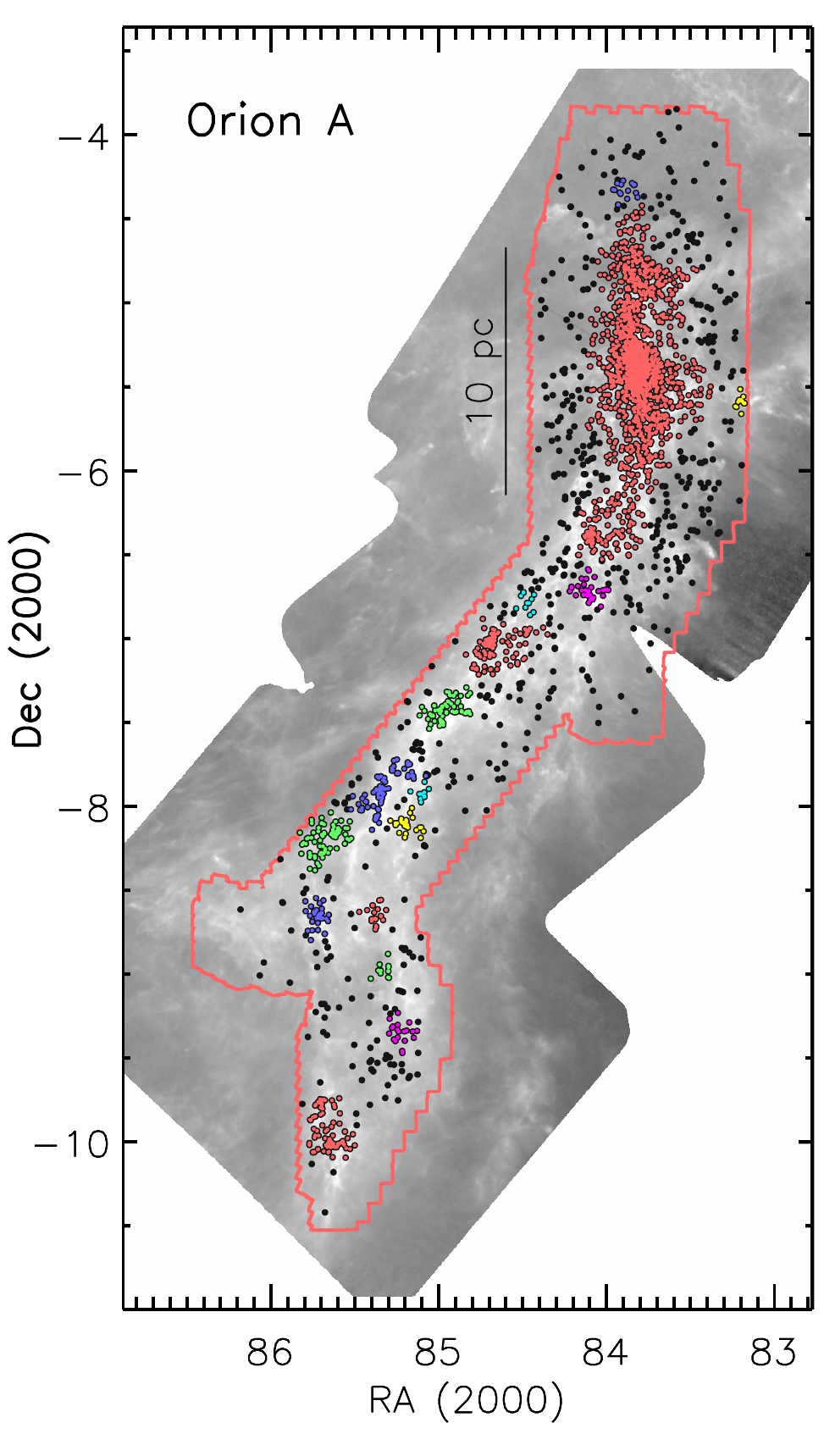}}
\caption{Distribution of dusty YSOs in the Orion A molecular cloud. The black dots are isolated YSOs, the colored dots show the different clusters and groups extracted by \citet{2016AJ....151....5M}, and the outline traces the extent of the Spitzer fields. The background image is the $N(H_2)$ map from \citet{2015A&A...577L...6S}. The ONC is the large cluster at the northern tip of the cloud marked in red.}
\label{fig:oriona_cluster}
\end{figure}

\section{Characterizing Embedded Clusters}
\label{sec:clusters}

Near-IR surveys of molecular clouds in the 1990s showed that most low mass stars formed in embedded clusters \citep[e.g.][]{1991ApJ...371..171L,1995AJ....109..709A,2000AJ....120.3139C,2008hsf1.book..621A}. Spitzer and Chandra, with their ability to identify isolated YSOs, showed that  these clusters  are typically density peaks in populations of YSOs that extend across clouds \citep[Figs~\ref{fig:oph_taurus} \& \ref{fig:oriona_cluster}, ][]{2007prpl.conf..361A}. 

Multiple approaches have been developed for isolating clusters. The ONC provides an example where many of these approaches have been applied to a single cluster. Both \citet{2000AJ....120.3139C} and \citet{2016AJ....151....5M} use non-parametric approaches for isolating the ONC, searching for contiguous regions of high YSO surface density that exceed a threshold density.  The resulting cluster included most of the integral shape filament (ISF), contains $\sim 2000-3000$ YSOs, and spans a 10 pc interval along the ISF (see Figs~\ref{fig:isf} \& \ref{fig:oriona_cluster}).

Parametric approaches have also proven fruitful. \citet{1998AJ....116.1816H},  \citet{2014ApJ...795...55D}, and  \citet{2018MNRAS.473.4890S} fit a King model, a $r^{-2.4}$ power-law, and Plummer sphere to the ONC, respectively.  They focused on the central density peak of the cluster,  which contains 1000~M$_{\odot}$ in the inner 0.7 pc \citep[Fig.~\ref{fig:isf},][]{2018MNRAS.473.4890S}.  \citet{2014ApJ...787..107K} decomposed the ONC into four isothermal ellipsoids. The primary ellipsoid has a half width of $0.31 \times 0.16$~pc and contains 834 members; within the boundaries of this ellipsoid are two ellipsoids tracing smaller density peaks. A fourth, highly elongated ellipsoid ($0.23 \times 0.04$~pc) is located to the north of the other three ellipsoids and extends along the ISF. These disparate solutions show that results from  different approaches must be compared with care.

\subsection{Cluster Demographics}
\label{sec:demo}

%Due to the ambiguities in defining and isolating embedded clusters, there have been few studies of the demographics of clusters in nearby clouds. 

Adopting a method for identifying clusters, we can assess  whether most stars form in large clusters, smaller clusters or groups, or isolation. Using the Spitzer catalog of dusty YSOs in the Orion clouds, and adopting a threshold density of 10 YSOs~pc$^{-2}$ for clusters (see Sec.~\ref{sec:PDF}),   \citet{2016AJ....151....5M} found 17 groups with 10-100 members, three clusters with 100-1000 members, and one cluster, the ONC, with $\sim 2000$ members (Fig.~\ref{fig:oriona_cluster}). Similar to the pre-Spitzer result of \citet{2000AJ....120.3139C}, they found the largest clusters in a cloud contained more stars than all the smaller clusters combined. 

Binning the number of YSOs  into logarithimic decades of cluster size, they found that in both clouds, the bin containing the most massive clusters contained  $\ge 60\%$ of the YSOs, while each of the lower bins contained $\le 20\%$ of the YSOs. Again, for the Orion~A and B clouds considered together, the ONC, which is the only $\ge 1000$ member cluster in the clouds, contained $ \sim 50$\% of the YSOs. Those in assemblages with $< 10$ members were considered isolated; these constituted $\sim 20\%$ of the YSOs. Thus the vast majority of stars form in clusters or groups. 

{\color{black} This result  for local clouds is in tension with extragalactic studies that show for bound clusters, the total mass of stars  per logarithmic decade of cluster size is approximately constant \citep[e.g][]{2009A&A...494..539L,2012ApJ...752...96F}.  The lack of agreement is not surprising since many embedded clusters may dissolve or lose members due to gas dispersal as they become bound clusters (see Sec.~\ref{sec:gas_dispersal}). Furthermore, extragalactic samples include clusters formed from many different clouds within a galactic disk. Finally, the clusters detected in other galaxies are more massive than those typically found near the Sun.} Nonetheless, the analysis of \citet{2016AJ....151....5M} should be extended to a larger sample of clouds to ensure that the Orion results are representative of clouds within 2 kpc of the Sun. 

%%%add DaRio model%%%

%The nearest large ($> 1000$ member), embedded cluster is the Orion Nebula Cluster (ONC) which is found in the integral shaped filament (ISF) of the Orion A cloud (Bally,Amy,Arce). Many observations has focused on the on the Orion Nebula region which contains the densest part of the cluster (e.g. hillenbrand \& hartmann, coup, dario, etc); however, this is the center of a highly active 10~pc star forming region which does not have an analog within 1~kpc. The axis of this region is the ISF, which is a molecular filament extending 7~pc. Stutz \& Gould find that the filament has a relatively mass per length, with a value of ?? per pcsq. This is much higher than the critical density for a thermally supported filament. The northern part of th filament is exceptionally active in SF with XX known protostars. This OMC 2-3 region is notable for the density and luminosity of the protostars. The filament here is formed into fibers, each with sub-sonic linewiths and mas to line ratios of XX. 

%To the north is the NGC 1977. Furthernorth. This suggests that this was a straight filament that was bent. Alternatively models of collapsing clouds claim that this can come from rotation.

\begin{figure*}[ht]
\epsscale{1.1}
\plotone{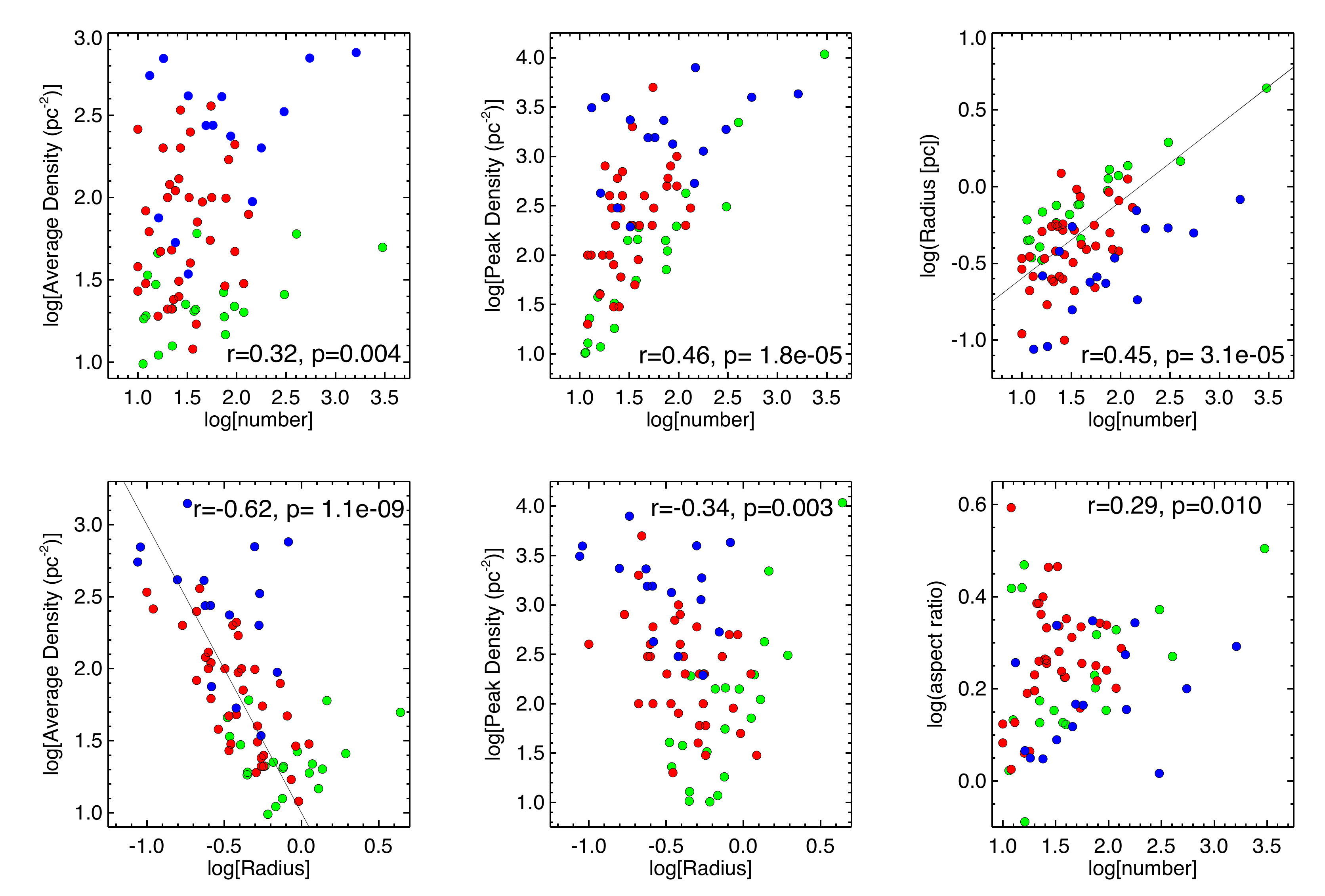}
\caption{Cluster properties for Spitzer and Chandra surveys of clusters; green dots are clusters and groups in Orion from \citet{2016AJ....151....5M}, red dots are cluster cores from \citet{2009ApJS..184...18G}, and the blue dots are sub-clusters from \citet{2014ApJ...787..107K}. While there is no overlap between the \citet{2016AJ....151....5M} and \citet{2009ApJS..184...18G} sample, the ONC is found in both the \citet{2016AJ....151....5M} and \citet{2014ApJ...787..107K}. The line in the radius vs number plot (upper right panel) is for a constant density of 64~pc$^{-2}$ while the line in the density vs radius plot (lower left panel) is for the density $= 10/{\rm radius}^2$.}
\label{fig:cluster_prop}
\end{figure*}

\subsection{Cluster 2D Structure}
\label{sec:structure}

Once a cluster has been extracted, its structural properties can be ascertained using the 2D spatial distribution of YSOs. 
%In this section, we overview the literature on cluster properties, highlighting methods, salient results and the incomplete nature of our current understanding. 
Figure~\ref{fig:cluster_prop} displays the properties of clusters within 1 kpc taken from \citet{2009ApJS..184...18G},  \citet{2015ApJ...812..131K} and \citet{2016AJ....151....5M}. See Appendix C for a discussion on how these properties are determined. {\color{black} As discussed below, the trends in Figure~\ref{fig:cluster_prop} imply that the properties of the embedded clusters are largely set by their formation in their parental gas structures. }  

The average surface densities of clusters depends on the extraction method. The clusters of \citet{2016AJ....151....5M} have densities near the adopted surface density threshold, the isothermal ellipsoids of \citet{2014ApJ...787..107K,2015ApJ...812..131K} show higher densities, and the cluster cores of \citet{2009ApJS..184...18G} fall in between. These systematic differences are also apparent in the average surface density vs radius plot, which shows an inverse correlation between density and radius. {\color{black} The isothermal ellipsoids and cluster cores often trace compact, high density structures hierarchically nested within large, lower density clusters \citep{2014ApJ...787..107K}.  In a sample of clusters out to distances of several kiloparsecs, \citet{2015ApJ...812..131K} found that the volume densities of ellipsoids decreased as the number of members increased. They suggested that the small, dense sub-groups merged and dynamically relaxed into larger, lower density clusters.} The observed inverse correlation between density and radius likely reflects this process.

%In contrast, clusters identified by a surface density thresholds suggest a trend of higher average surface densities with increasing radius.

In contrast, there is a strong correlation between the peak density of a cluster and the number of members. \citet{2016AJ....151....5M} find a correlation of $N_{peak} \propto n_{\scalebox{.5}{YSO}}^{1.2\pm0.1}$ where $N_{peak}$ is the peak density and $n_{\scalebox{.5}{YSO}}$ is the number of members. \citet{2016AJ....151....5M} also note that the mass of the most massive member also correlates with this density. 
%This indicates that larger clusters contain regions of higher gas and stellar densities which may also form more massive stars. 
There is no correlation of the peak density with radius. If the radius was determined by expansion, the peak density should decrease with increasing radius. These relationships thus indicate that the radii and densities of embedded clusters are determined primarily during their formation, and not by the subsequent evolution of the clusters.  

A strong correlation is also found between cluster radius and the number of members.  If we assign each member an average mass, a mass vs radius relationship of  $M \propto R^{2}$ is consistent with the correlation in Fig.~\ref{fig:cluster_prop}.  An analysis by \citet{2016A&A...586A..68P} found a slightly shallower  relationship for clusters of $M \propto R^{1.7}$.   {\color{black} A similar relationship is found for cluster-forming molecular clumps, suggesting that the correlation is inherited from the properties of the parental dense gas structures \citep{2016A&A...586A..68P}.}

Finally, we find that all clusters with more than 60 members are elongated with aspect ratios ranging from 1.4 to 3, the most extreme example being the ONC. Smaller clusters show a wide range of aspect ratios; much of this scatter may be due to the larger uncertainties for these clusters.  An alternative measure of the asymmetry is the azimuthal asymmetry parameter (AAP) of \citet{2005ApJ...632..397G}, which uses variations in the number of sources in pie shaped wedges centered on a cluster's center to establish azimuthal asymmetry. \citet{2016AJ....151....5M} use the AAP to show that all the Orion clusters with more than 100 members are asymmetric. Smaller clusters and groups may also be asymmetric, but lack the number of sources needed to have AAP values that deviate significantly from a symmetric cluster. {\color{black}  The elongation is inherited from the morphology of the parental dense gas structures (Sec.~\ref{sec:gas_env})}.

\begin{figure}[t!]
\epsscale{1.2}
%\plotone{figures/pasp_region8n9_cropped_v3.png}
\plotone{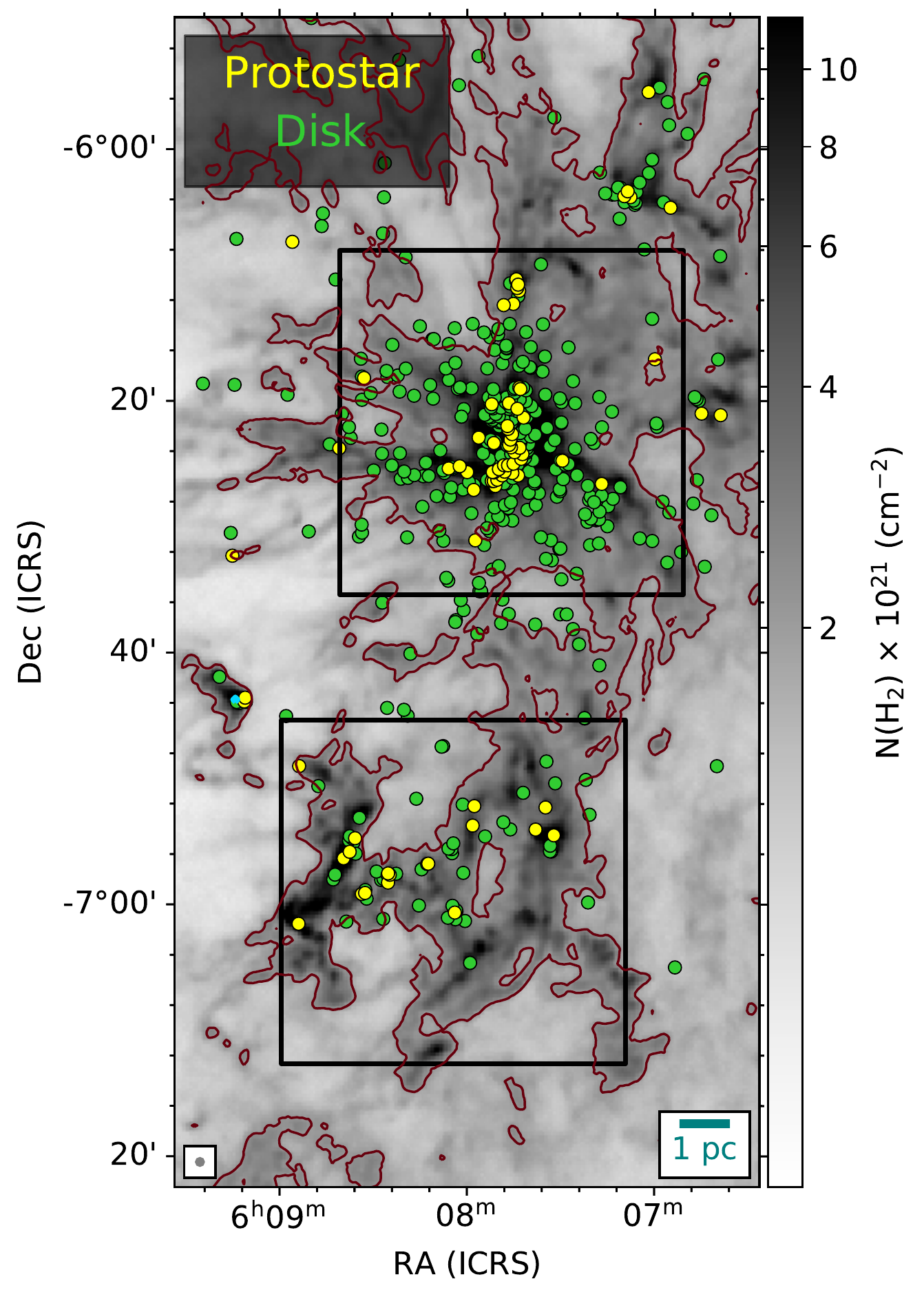}
\caption{The distribution of YSOs and gas in  the Mon~R2 molecular cloud. The YSOs are from the SESNA survey (Gutermuth in prep.) and the gas is derived from Herschel dust continuum imaging of the cloud. The boxes outline two regions studied by \citet{2016MNRAS.461...22P}; the upper region contains the rich Mon~R2 cluster and the lower contains a diffuse star forming region.}
%The right shows the PDFs in two different environments in the cloud Region 8 contains the  MonR2 cluster, while Regions 9 contains more distributed star formation sites. The gas PDF of the MonR2 cluster is well approximated by a power law from $log(N(H_2))$ from 21.75 to 23 while the region PDF shows a broken power-law shape, a steeper law with little gas above 22.5.  }
\label{fig:monr2}
\end{figure}

\subsection{The Gas Environment of Clusters}
\label{sec:gas_env}

In Sec.~\ref{sec:sfr&e}, we found that the structures of clusters appear to be strongly influenced by their formation as opposed to their subsequent evolution. A deep connection between gas environment and cluster properties is also implied by the sf-relation discussed in Sec.~\ref{sec:sf_law}. This relation shows that a continuum of stellar densities, rates and efficiencies are present in molecular clouds, and that clustered and diffuse star formation are the extremes of this continuum. Regions of a cloud with high column densities of gas form stars rapidly and produce embedded clusters with high SFEs. In the low column density regions that dominate the surface area of a cloud, a diffuse population of stars is formed at a low SFE. Given the typical gas column density PDFs of molecular clouds \citep[e.g.][]{2013ApJ...778..133L}, and the concentration of high density gas in massive structures, most YSOs form in embedded clusters as discussed in Sec.~\ref{sec:demo}. 
%2000AJ....120.3139C,2009ApJS..181..321E,2016AJ....151....5M}.

Despite this continuum, there are distinct differences in the structure of the dense gas found in clusters and in areas of diffuse star formation. Fig.~\ref{fig:monr2}  compares two regions of the Mon~R2 cloud. The northern region contains the Mon~R2 cluster. Here, the dense gas is organized into a hub with filaments extending outward radially, i.e.\ a hub-filament system \citep{2009ApJ...700.1609M}. The YSOs are  concentrated in the central hub, which is elongated. In contrast, the low column density southern region shows a network of filaments with widths of $\sim 0.1$~pc; these are similar to the filaments found in Herschel observations of many nearby clouds \citep[e.g.][]{2019A&A...621A..42A}. The concentration of gas into filaments reconciles the low average density of this region with the requirement for pockets of the dense gas needed to form stars; accordingly, the observed protostars are found in the filaments. Due to its stronger gravity, much higher gas densities are attained in the hub, resulting in the high density of YSOs \citep{2016MNRAS.461...22P}. 
%The density of the gas is much higher in the hub; its gas column density PDF shows a power-law tail extending to high column densities (100~A$_v$). In contrast the gas column density PDF of region 9 shows a broken power-law form which drops steeply at high densities \citep{2016MNRAS.461...22P}.  
Both the high and low density regions have a mixture of protostars and pre-ms stars with disks, indicating that star formation has been sustained over several generations. 

%L1495: 39 members, Oph 147 memters
Embedded clusters are found in hubs, like that in Mon~R2, or ridges. 
%Hubs span a range of masses: from the 200 ~M$_{\odot}$ L1495 hub in Taurus in \citep{2019ApJ...871..134S} to the 1000~M$_{\odot}$ hub in the Mon~R2 cloud \citep[masses from][]{2009ApJS..184...18G}. 
Ridges are massive, highly elongated counterparts to hubs  \citep{2018ARA&A..56...41M}; the 4000~M$_{\odot}$ ISF, which hosts the 2000 member ONC, is the nearest and best studied example of a ridge \citep[Fig~\ref{fig:isf},][]{1987ApJ...312L..45B,2016A&A...590A...2S,2019ApJ...882...45K,2016AJ....151....5M}.    The  column density PDFs of both massive hubs and ridges show shallow power-law tails extending to  N(H$_2$) $\approx 10^{23}$~cm$^{-2}$; these tails indicate the presence of high gas densities due to gravitational contraction  \citep{2011MNRAS.416.1436B,2016MNRAS.461...22P,2015A&A...577L...6S,2018MNRAS.473.2372K}. These tails suggests that hubs and ridges come from the collapse of cloud structures, with ridges resulting from the collapse of triaxial structures into massive filaments \citep[e.g.][]{1965ApJ...142.1431L}.   Millimeter line observations show evidence for flows of gas, both radially contracting onto ridges and hubs and along the filaments that radiate from these structures \citep{2010A&A...520A..49S,2013ApJ...766..115K,2017A&A...607A..22R}. These flows can supply gas to the hubs and ridges and sustain star formation. As is the case for the ISF, hubs and ridges are also the sites of high mass star formation \citep{2011A&A...533A..94H,2020A&A...642A..87K,2021MNRAS.508.2964A}.

Despite their elongation, ridges appear to be distinct from the narrow filaments that thread more diffuse star forming regions and more similar to hubs \citep{2011A&A...533A..94H,2012A&A...543L...3H}. Their mass-to-length ratios are  higher than those of filaments; the ISF has a mass to length to ratio of 390 M$_{\odot}$~pc$^{-1}$ within a radius of 1 pc from the spine of the ridge \citep{2016A&A...590A...2S}. Both ridges and hubs can have complex inner structures \citep{2017A&A...600A.141K,2018A&A...610A..77H}. Ridges also exhibit different radial density profiles than filaments. The density profiles of filaments show an inner flattening with a characteristic width of 0.1~pc and power-law volume density profiles of $q=-2$ ($\rho \propto r^q$) at larger radii \citep{2019A&A...621A..42A}.  In comparison the profiles of the ISF shows no flattening down to scales of 0.04~pc and a comparatively shallow $q = -1.6$  power-law density profile \citep{2016A&A...590A...2S}. Even for cross sections that include the density peak of the ONC cluster, \citet{2018MNRAS.473.4890S} show that the gas mass of the ISF dominates over the stellar mass of the ONC beyond 1~pc.  The high gas density and shallow density profile provides the conditions necessary for the formation of parsec scale clusters like the ONC.  

As discussed in Sec.~\ref{sec:structure}, the properties of clusters are inherited from the structure of the dense gas. The elongated clusters align with the morphologies of the hubs and ridges in which they form \citep[e.g. fig.~\ref{fig:isf}, also][]{2005ApJ...632..397G}.  \citet{2016A&A...586A..68P} show that the mass vs radius relationship for clusters is similar to that found for molecular clumps with massive stars by \citet{2014MNRAS.443.1555U}; these clumps are likely to be distant hubs or ridges in our galaxy. The increase in the peak stellar density with the number of members (Fig.~\ref{fig:cluster_prop}) reflects the increasingly high gas column densities present in successively more massive hubs and ridges. The increase in the highest gas column densities with cluster size can be seen in the gas column density PDFs for the cluster-forming hubs in Mon~R2. This can be seen by comparing the PDFs for regions 2 (smallest cluster), 4 (GG12-15, an intermediate mass cluster), and 8 (Mon~R2, the most massive cluster) in Fig.~13 of \citet{2016MNRAS.461...22P}.

%\begin{figure}[t!]
%\epsscale{1.}
%\plotone{figures/figure_pasp_proto_gas_v2.pdf}
%\caption{Distribution of protostars in the OMC2/3 region of the ISF. The greyscale shows the the APEX/SABOCA 350~$\mu$m map of the dense gas structure (Stanke, p. com.). The Class 0 (red), class 1 (green) and flat spectrum (yellow) protostars are overlaid  \citep{2016ApJS..224....5F}.}
%\label{fig:proto_gas}
%\end{figure}

\subsection{Tracing Fragmentation in Clusters}
\label{sec:frag}

The density of star formation in clusters depends on the spatial distribution of fragmentation sites. The distribution of sites is determined, in turn, by the internal structure of hubs and ridges.  %Evidence for internal structure is found within clusters, \citet{2009ApJS..184...18G} found that the distribution of YSOs in large cluster cores is inconsistent with a random distribution. 
In the Herschel dust continuum maps of the L1688 hub, \citet{2020A&A...638A..74L} find a network of 0.1 pc diameter filaments similar to those in more diffuse regions. Using velocity integrated C$^{18}$O line maps from CARMA, \citet{2019A&A...623A.142S} also resolved the ISF into a network of $\sim 0.1$~pc diameter filaments. 

An  alternative approach is to use the position and velocity information in  C$^{18}$O or N$_2$H$^+$ molecular line maps to map dense gas structure. \citet{2013A&A...554A..55H,2016A&A...589A..80H,2018A&A...610A..77H}  decomposed the L1495 hub, the NGC 1333 cluster hub in the Perseus cloud, and the ISF, respectively, into spatially and kinematically coherent, highly elongated structures called fibers.  The linewidths of the fibers are transonic, with the non-thermal components of the line profiles less than the sound speed \citep{2018A&A...610A..77H}. The fibers in L1495 and NGC 1333 have widths of $\sim 0.1$~pc; the ISF's fibers are narrower with widths of 0.02-0.05~pc \citep{2018A&A...610A..77H}. The narrow widths may be due to the use of a specific dense gas tracer, N$_2$H$^+$, to map the fibers in the ISF (also see Fig.~\ref{fig:proto_gas}).

The observed spatial distribution of pre-stellar cores in clouds, i.e. those that are unstable to collapse, shows that the fragmentation of the dense gas preferentially occurs in the filaments and fibers. Herschel dust continuum maps of the cluster forming Ophiuchus, Perseus and Serpens clouds (spatial resolution $\le 0.07$~pc) show  that within the hubs/ridges, $> 80$\% of the pre-stellar cores are found in filaments \citep{2020A&A...638A..74L,2021MNRAS.500.4257F,2021A&A...645A..55P}.  

Filaments and fibers containing cores typically have mass-to-length ratios $\ge  8$~M$_{\odot}$~pc$^{-1}$, similar to or in excess of the ratio for a stable, gravitationally bound isothermal filament (16 M$_{\odot}$~pc$^{-1}$). This is consistent with these structures being unstable to radial collapse and/or fragmentation \citep{2021A&A...645A..55P}. 

\citet{2018A&A...610A..77H} find that fibers mapped in the N$_2$H$^+$ line, a dense gas tracer, contain cores and conclude that these {\it fertile} fibers are unstable to fragmentation. They show that the number of fertile fibers per surface area scales with the overall mass-to-length ratio of their host structure; with the density of fibers increasing between the filaments of Taurus and the massive ISF ridge.  In this picture, the primary difference between cluster forming and diffuse regions is the surface density of fertile fibers.

The fragmentation scale, as measured by the separations between protostars or between cores, is comparable to the thermal Jeans length \citep{2006ApJ...636L..45T, 2016A&A...587A..47T}. It is, however, the {\it rate} a which the gas fragments into cores that primarily determines the rate of star formation in a cluster.
%In the ISF,  the offset of most protostars from  the fibers in the ISF \citep{2018A&A...610A..77H}, suggests that the time interval for fragmentation is significantly lower than the lifetime of protostars, 0.5~Myr; with rate $< N_{cores}$/0.5 Myr. 
An estimate of the rate can be obtained by extending the results of \citet{2018ApJ...853....5P}. They examined the fragmentation of the Perseus molecular cloud at each step of its structural hierarchy: from cloud to clumps, to cores, to envelopes and finally to protostars.  Here the protostars are identified by  VLA radio continuum surveys  (Sec.~\ref{sec:radio}), the envelopes are structures resolved inside the cores by SMA interferometric dust continuum maps, and the cores are found by single dish dust continuum maps of the cloud. The presence of multiple envelopes in a core or multiple protostars in an envelope imply multiplicity.  

At each scale, \citet{2018ApJ...853....5P} determined the number of Jeans masses, $N_J = M/m_J$, where $M$ is the total mass in a structure and $m_J$ is the Jeans mass calculated using the thermal sound speed. They then determined the efficiency of fragmentation, $\epsilon_f = N/N_J$ where $N$ is the observed number of fragments, on cloud, clump, core, and envelope scales. The clumps are parsec scale structures that include hub-filament systems. Fragmentation of the clumps into cores has an efficiency of $\epsilon_f \sim 0.2$. 

If the cores collapse over a free fall time, then the rate of fragmentation is $\approx 0.2 N_J /t_{ff}$, where $t_{ff} \approx 2 \times 10^5$~yrs for the average volume density of the clumps. For the clump forming the NGC 1333 cluster, this implies a fragmentation rate of $\sim 120$ per Myr. This is comparable to the rate of SFR from the YSO counts; about $\sim 70$ per Myr assuming a protostar lifetime of 0.5~Myr \citep{2009ApJS..184...18G}. The agreement is promising, particularly if a fraction of the cores do not collapse into stars or core lifetime exceeds a free fall time \citep{2017ApJ...846..144K,2019MNRAS.483..407S}. Future studies can refine this approach through more detailed examinations of the gas structure at each level of the hierarchy, and the application of the analysis to other star forming regions.   

\subsection{Age Spreads in Clusters}
\label{sec:age_spread}

\begin{figure}[t!]
\epsscale{1.2}
\plotone{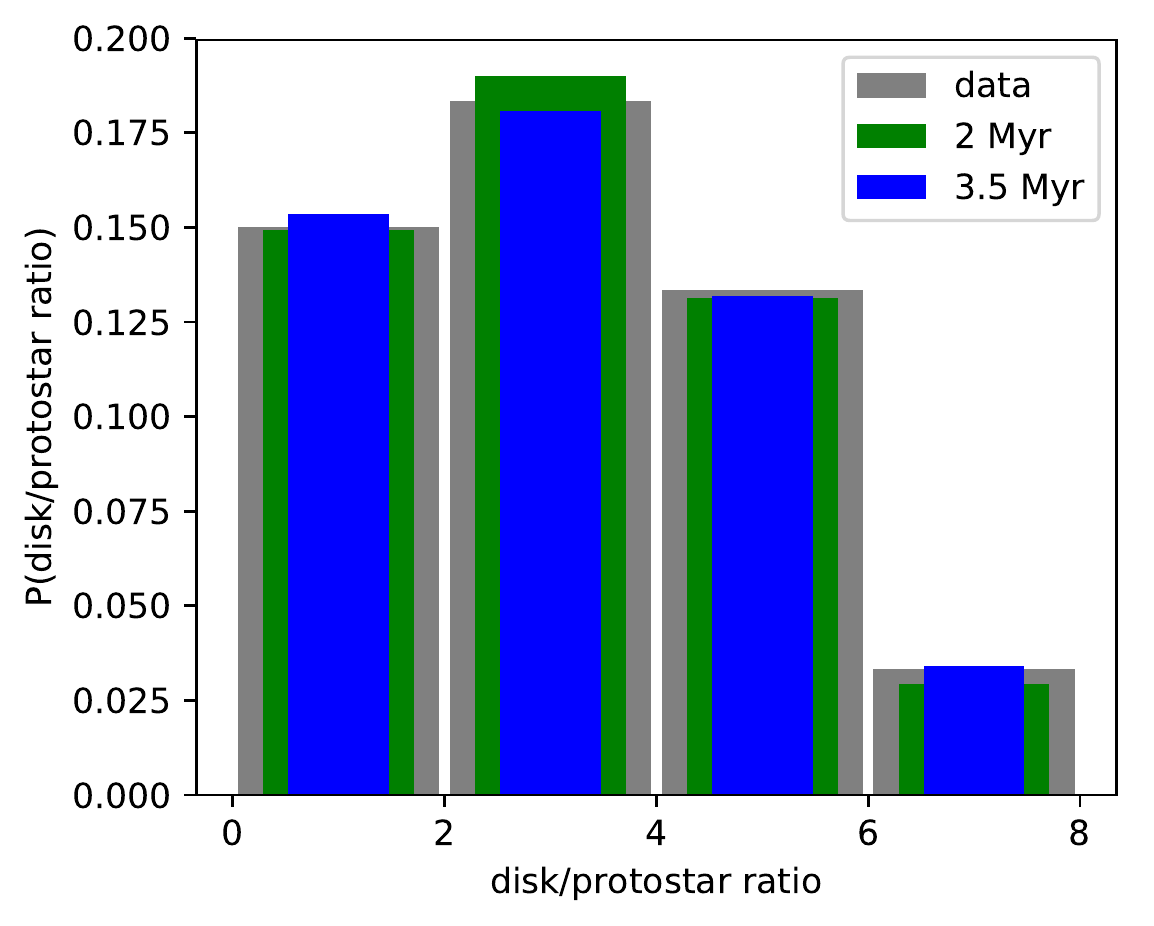}
\caption{Normalized histogram of the ratio of the number of pre-ms stars with disks to number of protostars in 36 cluster cores identified by \citet{2009ApJS..184...18G}. This is compared to simple models of cluster evolution with primary star formation episodes lasting 2 or 3.5~Myr. The histogram is reproduced if star formation is sustained over seven protostellar lifetimes in the cluster cores. See Appendix~\ref{sec:ratio}.} 
\label{fig:disk_protostar}
\end{figure}

Clusters sustain star formation over a few million years until they disperse their natal molecular gas and  star formation ceases. During this time, the number of protostars is proportional to the star formation rate smoothed over a protostellar lifetime. Over a few million years, the ratio of pre-ms stars with disks to protostars (also called the Class II/ Class I ratio) increases to values of 4-6 for a constant rate of star formation, and then higher as the star formation rate and the number of protostars decreases (Appendix~\ref{sec:ratio}). In Fig~\ref{fig:disk_protostar}, we show the distribution of the ratios for the cluster cores in \citet{2009ApJS..184...18G}. The ratios range from 1.4 for a very young, protostar rich cluster, to 31 for more evolved clusters in the later stages of gas dispersal. In a simple model where all protostars have a fixed lifetime and disks disappear exponentially with a half life of 2~Myr \citep{2009AIPC.1158....3M}, these ratios require a period of sustained star formation that lasts 2-3.5~Myr, or about seven protostellar lifetimes (Appendix~\ref{sec:ratio}). This conclusion depends on the adopted model for protostellar evolution. Assuming that protostars transition probabilistically into pre-ms stars with a fixed half life, \citet{2018A&A...618A.158K} find mean cloud ages of 1.2~Myr; these ages imply a shorter duration for cluster formation. 

%\footnote{The ratio is 4 achieved with for a constant star formation rate, a protostellar lifetime of 0.5~Myr, and a disk half life of 2~Myr for the pre-ms stars \citep{2014prpl.conf..195D}.}

For clusters where the members have effective temperatures and luminosities cataloged from a combination of photometric and spectroscopic observations, a comparison of the cluster HR diagram to theoretical pre-ms isochrones can determine ages and constrain star formation histories. The availability of accurate distances from Gaia has resolved one of the primary uncertainties in this method \citep[e.g.][]{2016ApJ...827...52L,2018ApJ...865...73O}. Substantial debate has focused on the interpretation of scatter within the HR diagrams of individual clusters. This scatter was interpreted by \citet{2000ApJ...540..255P} as evidence for accelerating star formation rates in clusters, with the oldest stars having ages approaching 10 Myr.  \citet{2001AJ....121.1030H} argued that much of the scatter is due to uncertainties in the stellar luminosities caused by variability and multiplicity. In this case, the implied acceleration and age spreads are artifacts caused by the scatter.
%\footnote{There is a discussion in the literature on whether the accretion history of the stars causes an inherent spread in the location of stars in the HR diagram - Baraffe Hosekawa}.
An analysis of the HR diagram of the ONC by \citet{2011A&A...534A..83R} showed that the uncertainties could not account for the full observed spread, and that the data are consistent with age spreads of 1.5 to 3.5~Myr, smaller than those origionally proposed by  \citet{2000ApJ...540..255P}. Further evidence for age spreads of this magnitude is found in the spread of surface gravities and radii
\citep{2007MNRAS.381.1169J,2011A&A...534A..83R,2016ApJ...818...59D}. In aggregrate, the disk to protostar ratios and the HR diagrams point to age spreads of a few million years.

Within these age spreads, there is evidence for spatial and temporal correlations. \citet{2017A&A...604A..22B} and \citet{2019A&A...627A..57J} found evidence for distinct episodes of star formation in the ONC. Using visible-light color-magnitude diagrams and Gaia DR2 distances, they found evidence for two, potentially three, episodes of star formation  occurring between 1.4 to 4.5 Myr ago. These episodes are extended along the ISF, with the youngest episode most concentrated toward the center of the ONC.

%Isochronal studies are often limited by extinction, large age spreads, inadequacies in pre-ms models, and uncertainties in the distance \citep[e.g.][]{2009AJ....137.4777W,2016ApJ...827...52L}. 

Spatial correlations may also be present in the form of age gradients.  On $> 1$~pc spatial scales, \citet{2013ApJ...768...99P} combined IR and X-ray surveys to measure the ratio of stars without disks to stars with disks, or Class III to Class II ratio. This ratio increases with age. They found that the Class III to Class II ratio increases in the outer regions of the ONC. Thus, the outer regions contain older stars that either formed in previous star forming episodes or migrated from the current region of active star formation.  

On $\le 1$~pc scales, \citet{2014ApJ...787..109G} found radial age gradients in the ONC and NGC~2024 clusters, with the age again increasing with distance from the cluster center. Combining X-ray data and near-IR photometry, they used an X-ray luminosity vs mass relationship and pre-ms tracks to convert L$_x$ and $M_J$, the X-ray luminosity and absolute $J$-band magnitude, of cluster members into an age, Age$_{jx}$. For the two largest clusters in Orion, the ONC and the NGC~2024 cluster, they found gradients in the median Age$_{jx}$ as a function of cluster radius, with the ages increasing with radius. Consistent with this age gradient, they also found the fraction of YSOs with K-band excesses decreasing with increasing median Age$_{jx}$.  In a study of NGC 1333 and Serpens Main cluster, \citet{2009AJ....137.4777W} found similar gradients in the isochronal ages, with the fainter, older stars concentrated in the outer regions of the clusters. They found, however, that the Class III/Class II ratio did not show a corresponding gradient. They argued that the gradients in isochronal ages were spurious and resulted from the high extinction values found in the centers of clusters hiding the faintest stars, thereby biasing the median age to younger values in the centers. Resolution of these contrary claims require studies that address incompleteness in clusters. 

In more evolved clusters,  star formation can continue on their edges. In the IC 348 cluster, \citet{2007AJ....134..411M} found protostars and a high rate of star formation on the edge of the cluster, while star formation in the center of the cluster has ceased due to gas dispersal. A similar configuration is found in the Cep~OB3b cluster, where star formation continues on the edge of the eastern sub-cluster \citep[Sec.~\ref{sec:gas_dispersal}, ][]{2009ApJ...699.1454G,2012ApJ...750..125A}.

\begin{figure}[t!]
\epsscale{1.15}
\plotone{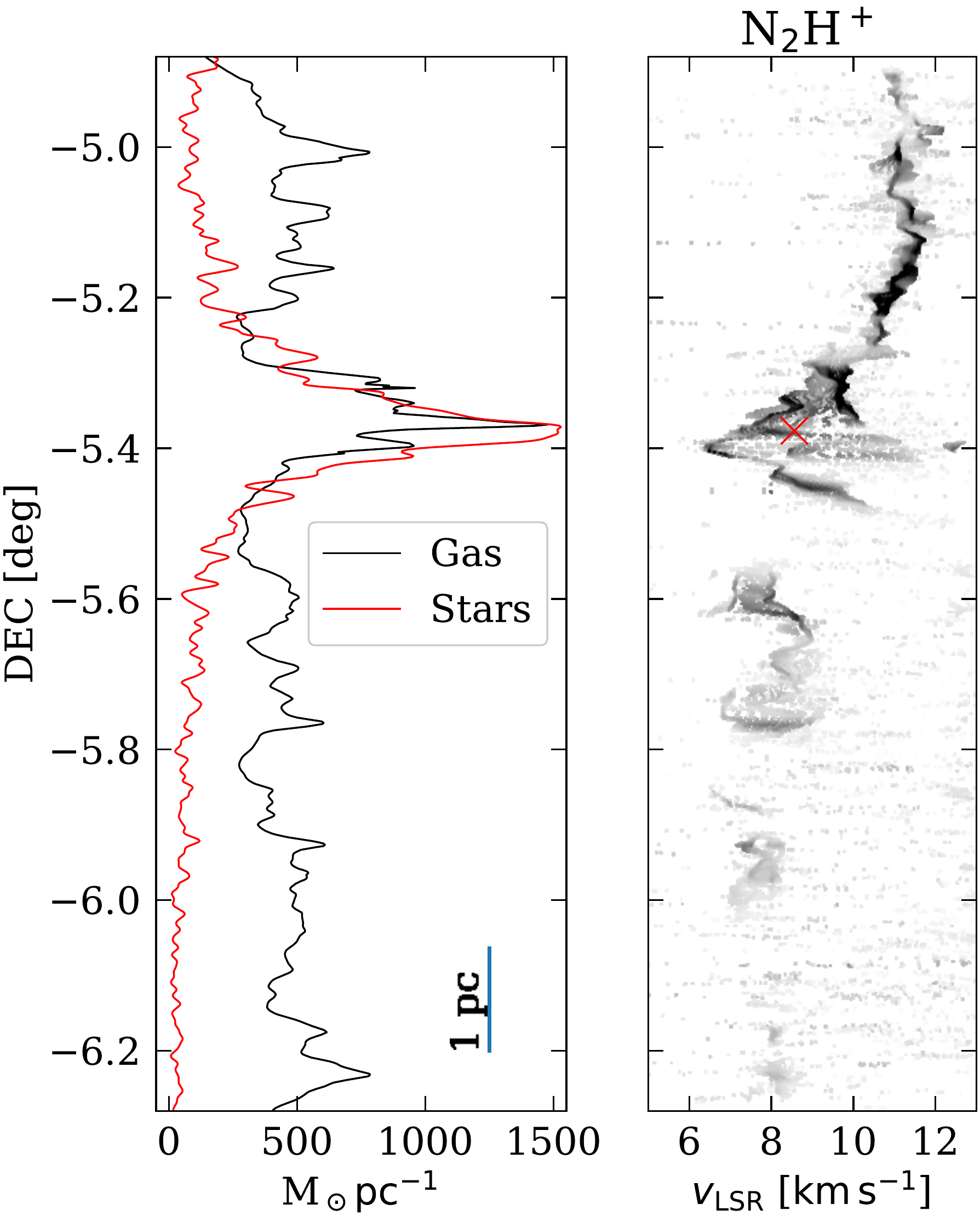}
\caption{{\bf Left:} mass-to-length ratios for stars and gas along the ISF; this shows that the stars are strong concentrated near the center of the ridge, but the gas mass-to-length ratio only increases by a factor of 2-3 at this position. {\bf Right:} intensity-weighted velocity centroids of the N$_2$H$^+$ (1-0) line along the ISF detailing the complex velocity structure of the filament. The x marks the center of mass and mean velocity of the ONC stars. Figure adapted from \citet{2019MNRAS.489.4771G}.} 
\label{fig:pv_isf}
\end{figure}

\subsection{The Kinematics of the ISF and ONC}
\label{sec:kinematics}

The incipient motions of the stars in an emerging cluster are inherited from the parental gas. Ridges such as the ISF show complicated motions over a range of scales (Fig.~\ref{fig:pv_isf}). These include turbulence \citep{2017ApJ...843...63F,2013ApJ...768L...5L},  oscillations \citep{2019MNRAS.489.4771G}, gravitational infall \citep{2017A&A...602L...2H} and rotational motions \citep{2021ApJ...908...86A,2021ApJ...908...92H}.

In the ISF, the  cores have motions that closely follow their parental dense gas \citep{2016A&A...590A...2S}, and these complex motions are important for understanding the kinematics of the nascent stars.  \citet{2019ApJ...882...45K} compared the radial velocities of the cores in the ISF (measured with the dense gas tracer NH$_3$) relative to the surrounding moderate density gas (traced by C$^{18}$O). They found a local velocity dispersion of 0.35~km~s$^{-1}$ \citep{2019ApJ...882...45K}; this value is comparable to the sound speed of the gas and less than the velocity dispersion of the moderate density gas of the ISF, $\sim 0.5$~km~s$^{-1}$.  This velocity dispersion is also less than the virial velocity estimated for the filament following the analysis of \citet{2000MNRAS.311...85F}, 0.8~km~s$^{-1}$. Other regions show similarly small core velocity dispersions relative to the surrounding gas \citep{2004ApJ...614..194W,2007ApJ...655..958W}.

A velocity dispersion can also be calculated from the observed core velocities. This core to core velocity dispersion is sub-virial in the NGC~1333 cluster of the Perseus cloud \citep{2007ApJ...655..958W,2015ApJ...799..136F} and the L1688 cluster in the Ophiuchus cloud  \citep{2007A&A...472..519A}. Due to the complex gas motions in the ISF, however, the global core to core velocity dispersion in the ISF is 2.9~km~s$^{-1}$, much higher than the estimated virial velocity. Thus, locally the motions of the cores are sub-virial relative to their natal gas, but on the scale of a cluster, the core to core velocity dispersion can be super-virial due to the complex motions of the parental ridge or hub inherited by the cores.

\begin{figure}[t!]
\epsscale{1.}
\plotone{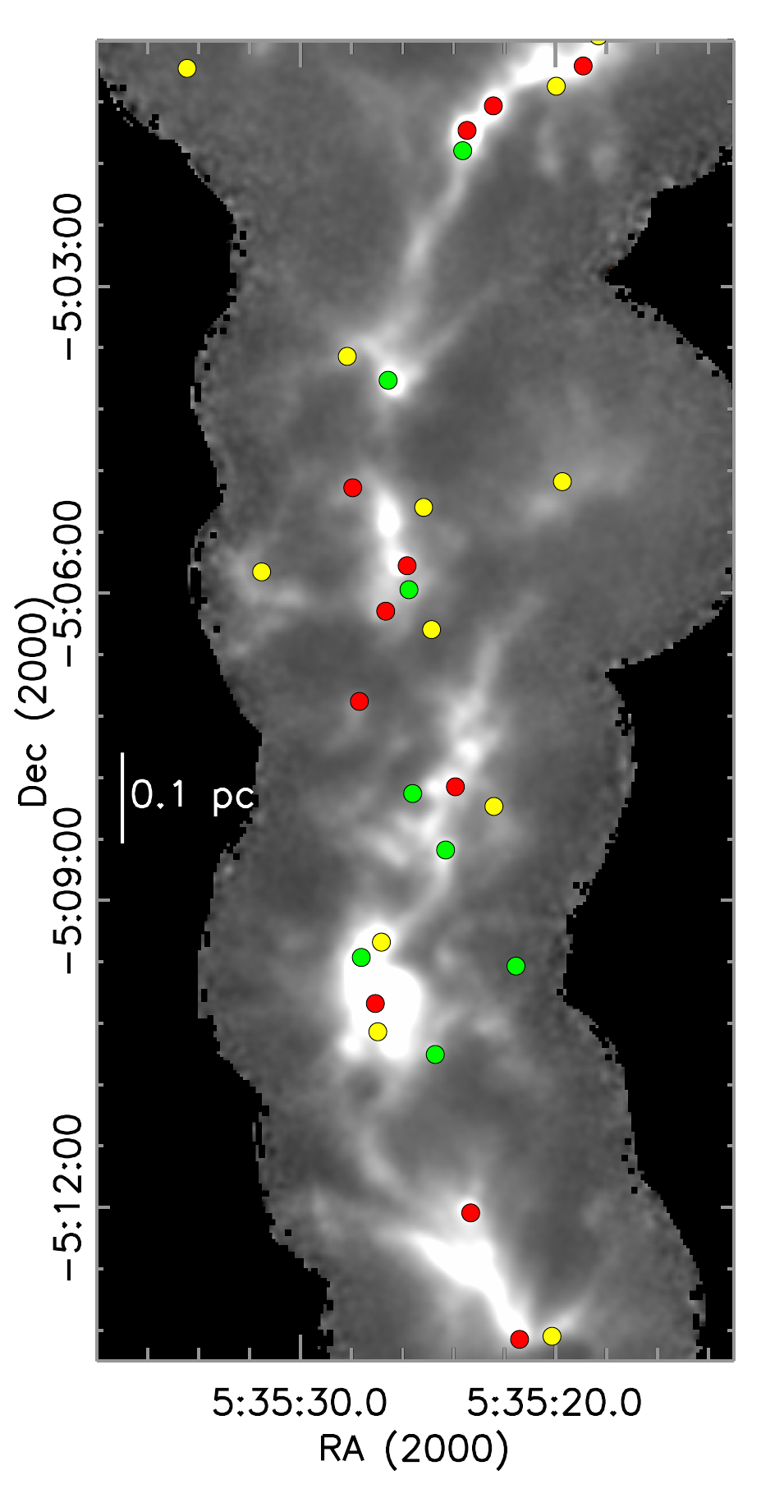}
\caption{Distribution of protostars in the OMC2/3 region of the ISF. The greyscale shows the APEX/SABOCA 350~$\mu$m map of the dense gas structure (Stanke, p. com.). The Class 0 (red), class 1 (green) and flat spectrum (yellow) protostars are overlaid  \citep{2016ApJS..224....5F}.}
\label{fig:proto_gas}
\end{figure}

As stars form, they decouple from the dense gas from which they collapsed. The 350~$\mu$m dust continuum map in Fig.~\ref{fig:proto_gas} shows the dense gas structures in the OMC2/3 region of the ISF; these structures closely follow the N$_2$H$^+$ emission used to identify fibers \citep[compare to Fig.~3 of][]{2018A&A...610A..77H}. Overlaid are the protostars separated into three different SED classes. These classes correspond approximately to a succession of evolutionary stages, although with some ambiguity due to the inclinations of the protostars \citep{2016ApJS..224....5F}. The youngest objects, the Class 0 protostars, are coincident with the dense gas structures. The intermediate aged Class I protostars show a mixture of sources on and off the structures. Most of the oldest protostars, the flat spectrum sources, are displaced from the structures. This successive detachment of the protostars from the dense gas may be driven by a combination of several factors: the accretion of the dense gas onto the protostars, the dissipation of the structures by feedback or turbulent motions in the surrounding gas, and the slingshotting of protostars out of the structures due to the acceleration of the gas \citep[e.g.][]{2016A&A...590A...2S,2017ApJ...840...36O}.  

After stars inherit the velocities of their natal cores, their motions are determined by the combined potential of the ISF and ONC and occasional flybys of other members.  For the kinematics of the ONC as a whole, the global, super-virial core to core velocity dispersion may be  more relevant. This high velocity dispersion may explain why pre-ms stars in the ONC are spread out in RA and are no longer concentrated in the dense backbone of the ISF along with the protostars (Fig.~\ref{fig:isf}).

One of the most distinctive features in the PV diagram of the ISF is a gradient in  radial velocity north of the center of the ONC (Fig.~\ref{fig:pv_isf}).
The velocities of the stars, cores and gas become increasingly blue shifted with decreasing projected distance to the center \citep{2009ApJ...697.1103T,2016ApJ...821....8K,2017A&A...602L...2H,2019ApJ...882...45K}. The gradient has been interpreted as the acceleration of the gas and stars in the ISF by the concentration of mass at the cluster center \citep{2008hsf1.book..590P,2009ApJ...697.1103T,2017A&A...602L...2H}, or as the result of wave-like oscillations in the ISF \citep{2016A&A...590A...2S,2018arXiv180711496S}.  

In the case of acceleration, the resulting motion would produce a mass infall onto the center of 50~M$_{\odot}$ in gas per million year \citep{2017A&A...602L...2H}. This interpretation requires that the ISF is inclined such that the center of the ONC is {\it closer} to the Sun than the filament to the north.  Gaia DR2 parallaxes for the infrared, visible, and X-ray selected YSOs in the ISF  suggest that the center of the ONC is {\it more} distant than the filament on either side \citep{2018arXiv180711496S,2019ApJ...870...32K,2019MNRAS.487.2977G}. 

For this inclination, the velocity gradient implies that gas is moving away from the center of ONC. The correlated variation of the gas radial velocity and  Gaia DR2 derived distance along the ISF has been interpreted as evidence for a wave-like oscillation \citep{2018arXiv180711496S,2019ApJ...882...45K}. Given the scatter in the parallaxes and the inability of Gaia to detect stars deeply embedded in the ISF, however, the inclination of the ISF remains uncertain.  

The density of stars is strongly peaked within 0.3 pc of the center of the ONC, which is located in the Orion Nebula where the most massive stars are found \citep[Fig.~\ref{fig:pv_isf}, \ref{fig:oncdetection},][]{1998ApJ...492..540H}. Although the ONC is highly elongated, the elongation decreases with increasing stellar density and disappears within 0.1 pc of the peak; this may be the result of dynamical relaxation \citep{2016AJ....151....5M}. Inside the central 0.3~pc, proper motions measured in the near-IR with HST and Keck \citep{2019AJ....157..109K} and in the radio with the VLA \citep{2017ApJ...834..139D} show no evidence of expansion, rotation or contraction while Gaia DR2 data \citep{2019ApJ...870...32K} shows a marginal detection of expansion.  The velocity dispersion is, to good approximation, gaussian and increases toward the center of the cluster \citep{2019AJ....157..109K,2021arXiv210505871T}. When including the gas mass inside the cluster, as estimated from the extinction to the member stars \citep{2014ApJ...795...55D}, the velocity dispersion is marginally super-virial  \citep{2017ApJ...845..105D,2019AJ....157..109K,2021arXiv210505871T}. 

Although this central cluster of the ONC may be dynamically evolved and bound, there are four reasons why it will not be in a stable equilibrium over timescales of 1-2 Myr, i.e. the relaxation time of the central 0.1 pc \citep{2016AJ....151....5M}. First,  toward the center of the ONC where the stellar mass is concentrated, the gravitational acceleration perpendicular to the ISF by the gas  exceeds that from stellar mass at distances greater than 0.36~pc \citep{2018MNRAS.473.4890S}.  The  O-star $\theta^1$C~Ori, which appears to be at the center of the cluster, is offset by 0.1-0.2 pc from the ionization front at the surface of the ISF \citep{1995ApJ...438..784W, 2017ApJ...837..151O}. Thus the ISF and central cluster are spatially offset. Accordingly, the cluster is likely being pulled and tidally stretched by the cloud \citep{2018MNRAS.473.4890S}. 

Second, the velocity dispersion along the filament is 25\% higher than that perpendicular to the filament: $\sigma_{v\alpha} = 1.64 \pm 0.12$~km~s$^{-1}$ and $\sigma_{v\delta} = 2.03 \pm 0.13$~km~s$^{-1}$ \citep{1988AJ.....95.1755J,2019AJ....157..109K,2019ApJ...870...32K,2021arXiv210505871T}. This higher velocity dispersion may result from stars falling into the cluster potential and contributing to the elongation of the cluster.  Third, the velocity dispersion in the line of sight, i.e.~toward the filament, is even higher: $2.56^{+0.16}_{-0.17}$~km~s$^{-1}$ \citep{2021arXiv210505871T}. Tidal acceleration by the filament may have inflated velocities in this direction \citep{2018MNRAS.473.4890S}. This velocity dispersion exceeds the virial velocity from \citet{2014ApJ...795...55D}. Furthermore, due to the tidal force of the filament, a higher mass may be required for virial equilibrium \citep{1969JKAS....2....1L}. The fourth and final reason is that the gas potential is changing on 1-2 Myr timescales due to gas dispersal and possible motions in the filament \citep[Sec.~\ref{sec:gas_dispersal}][]{2018MNRAS.473.4890S}.   

%Although the ONC is highly elongated, the elongation decreases with increasing stellar density and disappears within 0.1 pc of the peak \citep{2016AJ....151....5M}.  Numerous studies have shown that the inner parsec is close to virialized: i.) proper motions measured with HST and Keck data are consistent with a virialized cluster although with a higher velocity dispersion along the direction of elongation \citep{2019AJ....157..109K}, ii.) radial velocities from APOGEE-2 show a virialzed or slightly super-virial cluster \citep{2017ApJ...845..105D}, iii.) proper motions measured with the VLA show no evidence of rotation, contraction or expansion \citep{2017ApJ...834..139D}, and iv.) proper motions from Gaia DR2 show the cluster is in virial equilibrium and is not expanding \citep{kuhn2019}. The relaxation time in the center of the ONC is comparable to the age spreads discussed in Sec.~\ref{sec:age_spread} \citep{2016AJ....151....5M}. 

Another source of change  is the ejection of stars. In the ONC, $>$50 candidate members have been ejected within the last 4 Myr with velocities of $>$6~km~s$^{-1}$, most of which trace back to the innermost Trapezium region at the cluster center \citep{2019ApJ...884....6M,2020ApJ...900...14F,schoettler2020}. The most massive star in the cluster, $\theta^1$ Ori C has likely played a role in a large fraction of these ejections. 
%It is notable that there are significantly fewer ejected candidates in clusters that lack stars as massive as some of the ones found in the ONC (e.g., the NGC 2264, IC348, and NGC 1333 clusters, Schoettler et al., in prep). 
Almost all of these runaway and walkaway stars tend to be low mass; such stars are more abundant and it is easier to accelerate them. Only two O stars, $\mu$ Col and AE Aur, are known to have been ejected from the vicinity of the ONC in an event 2.5 Myr ago \citep{2001A&A...365...49H}. In general, the mass loss due to ejection is only a few percent of the total mass.  The ejection of OB stars, however, may alter cluster evolution by reducing feedback on the surrounding cloud \citep[e.g.][]{2018A&A...612A..74K}.

%{\color{black} In total, the ISF is the closest example of a massive ridge forming a large cluster, and it illustrates the complexities of cluster formation in a ridge. Despite the extensive observations of this region, we still lack a detailed understanding of the motions of the dense gas, how the cores become detached from the dense gas, and the subsequent dynamical evolution of the stars in the evolving gravitational potential.  A more complete description of the 3D structure and motions of the dense gas, cores and stars - as well models that adopt this structure and kinematics - are required to achieve a detailed physical model of this prototype ridge.  }  

\newpage
\subsection{The Effect of Gas Dispersal}
\label{sec:gas_dispersal}

{\color{black}
Feedback from the member stars is thought to be the primary means by which clusters lose their gas and star formation halts \citep[e.g.][]{2001ApJ...562..852H}.
Since the molecular gas dominates the  masses of embedded clusters (Sec~\ref{sec:SFE}), the dispersal of the gas should have a large impact on the evolution of the clusters \citep[e.g.][]{1980ApJ...235..986H}.  Understanding the evolution of clusters through gas dispersal has implications for a range of astrophysical inquiry, from the the evolution of planetary systems \citep[][]{2010ARA&A..48...47A} to the use of clusters to trace star formation in galaxies \citep[e.g.][]{2019ARA&A..57..227K}.

By comparing the rate of formation of embedded and bound open clusters, \citet{2003ARA&A..41...57L} estimated that only 7\% of embedded clusters survive gas dispersal to become bound clusters. With censuses of open clusters and YSOs around the Sun provided by Gaia and Spitzer, respectively, we can estimate the fraction of all stars formed that remain  bound in clusters after gas dispersal \citep{2012MNRAS.426.3008K}.  \citet{2021A&A...645L...2A} identified all open clusters in a 2 kpc radius cylinder centered on the Sun.  From the masses and ages of the clusters, they estimated the rate of open cluster formation in units of mass is $250^{+190}_{-130}$~M$_{\odot}$~Myr$^{-1}$~kpc$^{-2}$.  From the total number of YSOs in the Orion clouds plus all clouds within 500 pc, the local star forming rate is $1700\pm22$~M$_{\odot}$~Myr$^{-1}$~kpc$^{-2}$. Thus, in the local region of our galaxy, $15^{+11}_{-08}\%$ of stars emerge from their birth clouds in bound open clusters. If open clusters come from the 65\% of stars that form in $\ge 100$ member embedded clusters \citep{2001ApJ...553..744A,2016AJ....151....5M}, $23^{+17}_{-12}\%$ of all stars that form in such embedded clusters emerge in open clusters. See Appendix~\ref{sec:sfr_500pc} for details.

%The high incidence of infant mortality was considered to be the result of the rapid gas dispersal (ref?). However, the high rate of infant mortality came into question, after discovery of large populations of massive, bound clusters in other galaxies. 

N-body simulations of cluster evolution during gas dispersal initially assumed that the stars and gas have similar spatial distributions and are initially virialized \citep[e.g.][]{1984ApJ...285..141L}.  These models found that the fraction of stars that remain in bound clusters depend on the integrated SFE and the rate of gas dispersal.  In the case of rapid gas dispersal, \citet{2007MNRAS.380.1589B} found that integrated SFEs $\ge 33\%$ are needed for clusters to survive. Alternatively, when the removal of the gas occurs over multiple crossing times, clusters with efficiencies as low as 10\% survive. 

Motivated by simulations of cluster formation in turbulent clouds that predicted  sub-clusters dominated by stellar mass with sub-virial stellar velocities, subsequent work predicted that most clusters survive rapid gas dispersal  \citep[e.g.][]{2009Ap&SS.324..259G,2012MNRAS.419..841K,2018MNRAS.476.5341F}. Furthermore, based on comparisons of clump masses to the properties of young massive clusters, conveyor belt models were proposed in which gas steadily flows into the central cluster as the stellar population grows \citep{2014prpl.conf..291L,2016MNRAS.457.4536W}.  The hierarchical structure, sub-virial velocities and conveyor belt flows increases the fraction of stars that remain bound and the chances that clusters with low integrated SFEs survive.

In contrast, models incorporating feedback show that irregular shells of gas swept up by winds and radiation from massive stars can accelerate young stars; this process can help disperse clusters \citep{2019MNRAS.488.3406Z}. These theoretical developments motivate a detailed examination of the process of gas dispersal. While the nearest 1~kpc lacks the sample size of studies of more distant regions in our galaxy and others, it does host specific examples that can be studied in detail. The remainder of this section concentrates on these examples. 
}

\begin{figure}[t!]
\epsscale{1.2}
\plotone{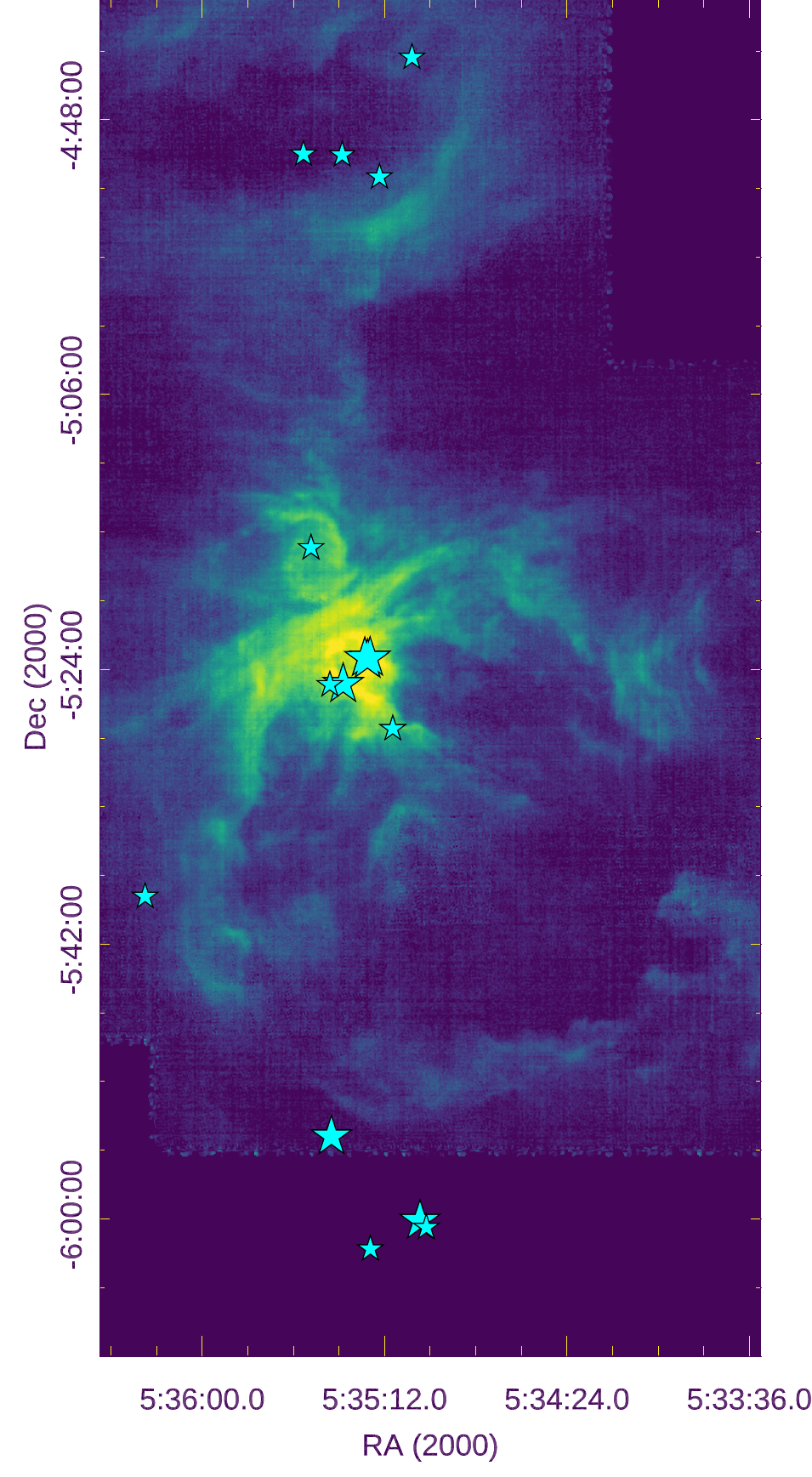}
\caption{Integrated 158~$\mu$m [CII] map of the ISF from the upGREAT instrument on SOFIA \citep{2019Natur.565..618P}. The O and B stars from \citet{1994A&A...289..101B} are the large and small star markers, respectively. The Orion nebula and Trapezium are at the center. This image shows the massive 1500~M$_{\odot}$ bubble extending southwest from the Trapezium, the bubble around the B-stars in NGC~1977 to the north, and the small bubble around a single B-star in M43 just northeast of the center.}
\label{fig:sofia_cii}
\end{figure}

Observations are now directly measuring the effect of the OB stars in the ONC on the ISF (Fig.~\ref{fig:sofia_cii}).  SOFIA [CII] maps trace a 1500~M$_{\odot}$ neutral hydrogen shell expanding at 13~km~s$^{-1}$  \citep{2019Natur.565..618P}. This shell is part of a 0.2 Myr old bubble  driven by the wind of the O7 star $\theta^1$~Ori~C, which is in center of the Orion Nebula. The shell is apparent in Fig.~\ref{fig:isf} as a large arc of nebulosity that extends to the southwest of the nebula.  The other two bubbles surround B-stars and appear to be driven by the expansion of the hot, ionized gas \citep{2020A&A...639A...2P}. One of these  bubbles surrounds the NGC~1977 sub-cluster at the northern tip of the ISF \citep[Fig.~\ref{fig:isf},][]{2008hsf1.book..590P}. The age of this bubble is 0.4 Myr.  
%Interestingly, the ISF bends sharply to the west as it meets this sub-cluster; this bend may result from the the feedback by the B-stars in NGC~1977. Alternatively, rotation or magnetically driven oscillations of the ISF may have alternatively accelerated the dense gas away from the cluster \citep{2007ApJ...654..988H,2016A&A...590A...2S}; such accelerations may also play a role in clearing dense gas from clusters. 
These observations show that gas dispersal starts at multiple locations along a ridge driven by either winds or HII regions.    

Is the ONC forming a bound cluster? The most likely part of the ONC to survive gas dispersal is the central cluster, i.e. the stars within 0.3~pc of the center. This high stellar density region appears to be undergoing relaxation \citep{2016AJ....151....5M}.  The massive stars in this central region drive the wind bubble detected in [CII] observations. The age of this bubble, 0.2 Myr, is similar to the crossing time of the central 0.3 pc of the cluster, indicating that the  dispersal of the gas is rapid \citep{2007MNRAS.380.1589B,2020A&A...639A...2P}. 

Although the  velocity dispersion of  stars in the cluster slightly exceeds the virial velocity, the cluster shows no signature of contraction and only a weak signature of expansion (Sec.~\ref{sec:kinematics}). The stellar and gas masses in the central 0.3 pc are  approximately equal, suggesting a $SFE \sim 50\%$ (Fig.~\ref{fig:pv_isf}). This does not include, however, the gas in the  1500~M$_{\odot}$ wind bubble \citep{2020A&A...639A...2P}. If the swept up gas started in the inner parsec, then the SFE at the onset of gas dispersal may have been closer to 30\%.   

The large bubble driven by $\theta^1$~Ori~C is filled with YSOs, including protostars, that are displaced from the spine of the ISF (Fig.~\ref{fig:isf}).  These stars may have formed in the gas as it was accelerated leaving a trail of young stars behind, or they may have been accelerated by the gravitational pull of the expanding, inhomogenous bubble \citep{2019MNRAS.488.3406Z}. There are hints that these stars are expanding away from the ONC in the kinematic data  \citep{2008ApJ...676.1109F,2019MNRAS.487.2977G}. These data suggest that the expansion of some of the stars are driven by feedback sweeping up the molecular gas. The bulk of the stars in the core of the cluster, however, remain unaffected. 

In aggregate, these observations suggest the central 0.3~pc of the ONC is on the cusp of forming a few hundred solar mass bound cluster \citep{2007MNRAS.380.1589B,2018MNRAS.476.5341F}, although the effect of the filament on the cluster remains unclear (Sec.~\ref{sec:kinematics}). Continued star formation in the filament will increase the size of the eventual bound cluster \citep{2021ApJ...923..221O}. This star formation may potentially be fed by molecular gas falling into the central 0.3 pc (Sec~\ref{sec:kinematics}), similar to the flows invoked in conveyor belt models. 

\begin{figure}
    \centering
    \includegraphics[width=0.46\textwidth]{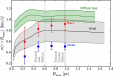}
    \caption{Velocity dispersions of cores, stars and gas within a maximum radius from the center of the NGC~1333 cluster. These are compared to the virial mass calculated from the total stellar and gas mass enclosed within $R_{max}$; this virial mass does not include terms for magnetic fields or external pressure \citep{2015ApJ...799..136F}. The plot shows that cores have a sub-virial distribution while the stars appear virialized. Figure from \citet{2015ApJ...799..136F}.  }
    \label{fig:foster}
\end{figure}

\begin{figure*}[t!]
%\epsscale{2}
\center{\plotone{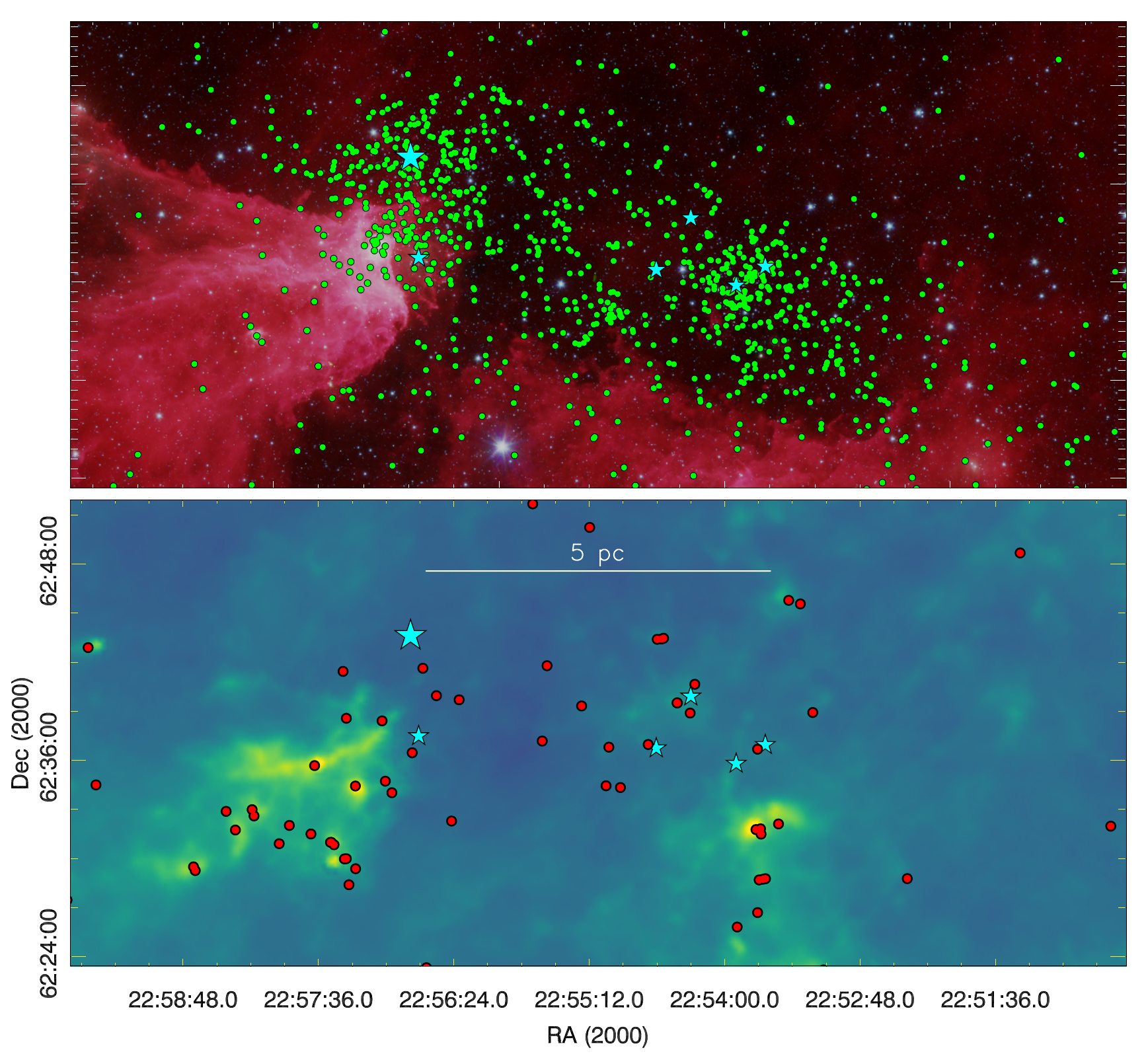}}
\caption{The spatial distribution of YSOs in the Cep~OB3b cluster. The top panel shows the Spitzer 3.6, 4.5 and 8~$\mu$m combined image with the locations of the pre-main sequence stars with disks displayed \citep[][data from SESNA]{2012ApJ...750..125A}. The bottom image is a N(H$_2$) column density map of the remnant molecular cloud (Pokhrel et al. 2020) with the location of protostars overlaid (also from SESNA). The protostars that are not coincident with gas structures are likely to be edge-on disks or extragalactic contamination \citep{2012AJ....144...31K}. The location O and B stars are shown as large and small star markers, respectively \citep{2012ApJ...750..125A}. The 5 pc bar shows the approximate length of the ISF shown in Fig.~\ref{fig:isf}; this cluster has a similar  number of members to the ONC but is in the late stages of gas disruption.}
\label{fig:cepob3}
\end{figure*}

Other embedded clusters show  approximately virial velocity dispersions for stars. \citet{2015ApJ...799..136F} found that while the cores in the NGC 1333 cluster have a sub-virial velocity dispersion, the stars and the gas have velocity dispersions that are comparable to the virial velocity for the total star and gas potential (Fig.~\ref{fig:foster}).  In addition, the gas density exceeds the stellar density on all spatial scales \citep{2015ApJ...799..136F}.  Gaia proper motiones of  NGC 1333 show no evidence of rotation or expansion \citep{2018ApJ...865...73O}. Similarly, the pre-stellar cores in L1688 also have a sub-virial velocity dispersion \citep{2007A&A...472..519A}.  The stars in this cluster once again have a virial velocity dispersion for the combined  star and gas potential  \citep{2016A&A...588A.123R}. These examples demonstrate that the young stars in clusters can achieve virial velocity dispersions before gas dispersal. Since both of these clusters have instantaneous SFE of 30\% \citep{2009ApJS..184...18G}, it is likely both will survive to form bound clusters \citep{2007MNRAS.380.1589B,2018MNRAS.476.5341F}.

If the stars in embedded clusters have approximately virial velocity dispersions in a potential that is dominated by the gas, then the dispersal of the gas should lead to the expansion of the clusters. Indirect evidence for expansion is found in the densities and structures of clusters. In a study of three clusters, \citet{2005ApJ...632..397G} show that two embedded clusters are elongated and aligned with their parental hubs and ridges, while the third cluster, which has dispersed much of its gas, has a lower stellar density and no elongation. They propose that the lower density and circular symmetry of the third cluster is due to its expansion after gas dispersal from a denser, elongated initial configuration.

Kinematic evidence for expansion is less clear. The IC~348 cluster in Perseus is in the late stages of gas dispersal, with the central cluster largely cleared of gas \citep{2007AJ....134..411M}.  IC~348 shows a  super-virial velocity dispersion relative to the stellar mass \citep{2015ApJ...807...27C}. It also shows, surprisingly, evidence that the stars on the near and far side of the cluster are converging; this is interpreted as either two converging sub-clusters or the cluster oscillating around a new equilibrium radius in response to gas dispersal  \citep{2015ApJ...807...27C}. Yet, Gaia DR2 data of IC~348 show no evidence of rotation and expansion  \citep{2018ApJ...865...73O}. 

%It is possible that IC~348 will form a bound cluster as it relaxes into a new equilibrium, although potentially with a significant loss of stars.  

%Maybe see \citep{2016MNRAS.457.3430P}. 

In the nearest 1~kpc, the strongest evidence for expansion due to gas dispersal is found in Cep OB3b, a massive, ONC sized cluster in the later stages of gas dispersal which contains two distinct sub-clusters \citep[Fig.~\ref{fig:cepob3},][]{2012ApJ...750..125A}.  The smaller western sub-cluster contains B-stars.  Its relatively low density indicate that this sub-cluster has expanded due to gas dispersal. Despite this expansion, the radial velocities suggest that the emerging cluster is bound  \citep{2019ApJ...871...46K}.

In contrast, the  more massive eastern sub-cluster \citep[also referred to as Cep~B, e.g.][]{2006ApJS..163..306G} hosts an O7 star.  Its radial velocity dispersion suggests that this sub-cluster is on the whole unbound, although there are large uncertainties due to multiplicity \citep{2019ApJ...871...46K}.  Gaia~DR2 astrometry of this sub-cluster shows  expansion with a magnitude consistent with the radial velocity dispersion \citep{2019ApJ...870...32K}. The relatively low surface density and circular symmetry of the sub-cluster are also indicative of expansion \citep{2019ApJ...871...46K}.  

The most massive stars in the Cep~OB3b sub-clusters appear to influence the fraction of stars that survive gas dispersal in bound clusters.  Comparing the virial ratios of the sub-clusters to the results of N-body codes, \citet{2019ApJ...871...46K} argue that both sub-clusters will form  bound clusters with $\sim300$ members. In the larger eastern sub-cluster, which hosts an O-star, only a relatively small fraction of  members remain bound  \citep{2018MNRAS.476.5341F}. In contrast, in the western sub-cluster, which is undergoing slower gas dispersal by B-stars, a large fraction of members remain bound. Averaging Gaia DR2 proper motions, \citet{2019ApJ...871...46K} find that the two sub-clusters are moving apart to form a double cluster. 

\citet{2020ApJ...900L...4P} find a possible 25~Myr analog to Cep~OB3b using Gaia DR2: the double cluster NGC~2232 and LP~2439. While both are expanding, the smaller NGC~2232 appears to be bound and undergoing re-virialization, while only half the members of the larger LP~2439 are bound. Like the eastern sub-cluster of Cep~OB3b, LP~2439 probably underwent rapid gas dispersal and may ultimately dissolve while the lower mass cluster survives. 

Although the current number of examples are small in number, they point to a picture where clusters become virialized before the gas is dispersed. Gas dispersal then leads to super-virial velocity dispersions and expansion, but the clusters can survive the loss of gas in a bound state. This is often accompanied with a significant loss of stars \citep[e.g.][]{2007MNRAS.376.1879W}.  The ONC will likely lose the lower density regions outside the central 0.3~pc (Fig.~\ref{fig:isf}), while the eastern sub-cluster of Cep~OB3b will lose more than half its stars due to its globally unbound state \citep{2019ApJ...871...46K}. Clusters with slower gas dispersal, such as the western sub-cluster of Cep~OB3b, lose a smaller fraction of stars. Although these (sub)-clusters survive gas dispersal, subsequent dynamical evolution, stellar evolution, or tides may disrupt them \citep[e.g.][]{2012MNRAS.426.3008K, 2012MNRAS.425..450M,2021MNRAS.508.5410D}.

Gas dispersal will also affect the merging of sub-clusters.
{\color{black} In a study of 28 clusters and associations in our galaxy, \citet{2019ApJ...870...32K} use Gaia~DR2 to show that in cases where there are multiple sub-clusters in a region, they are moving away from each other and  will not merge. As is the case for Cep~OB3b, the sub-clusters have dispersed much of their parental gas. \citet{2019ApJ...870...32K} argue that if merging occurs, it must happen during an earlier embedded phase when gas dominates the mass.}

\section{Clusters and Associations}
\label{sec:assoc}

\begin{figure*}[t!]
\includegraphics[width=1\textwidth]{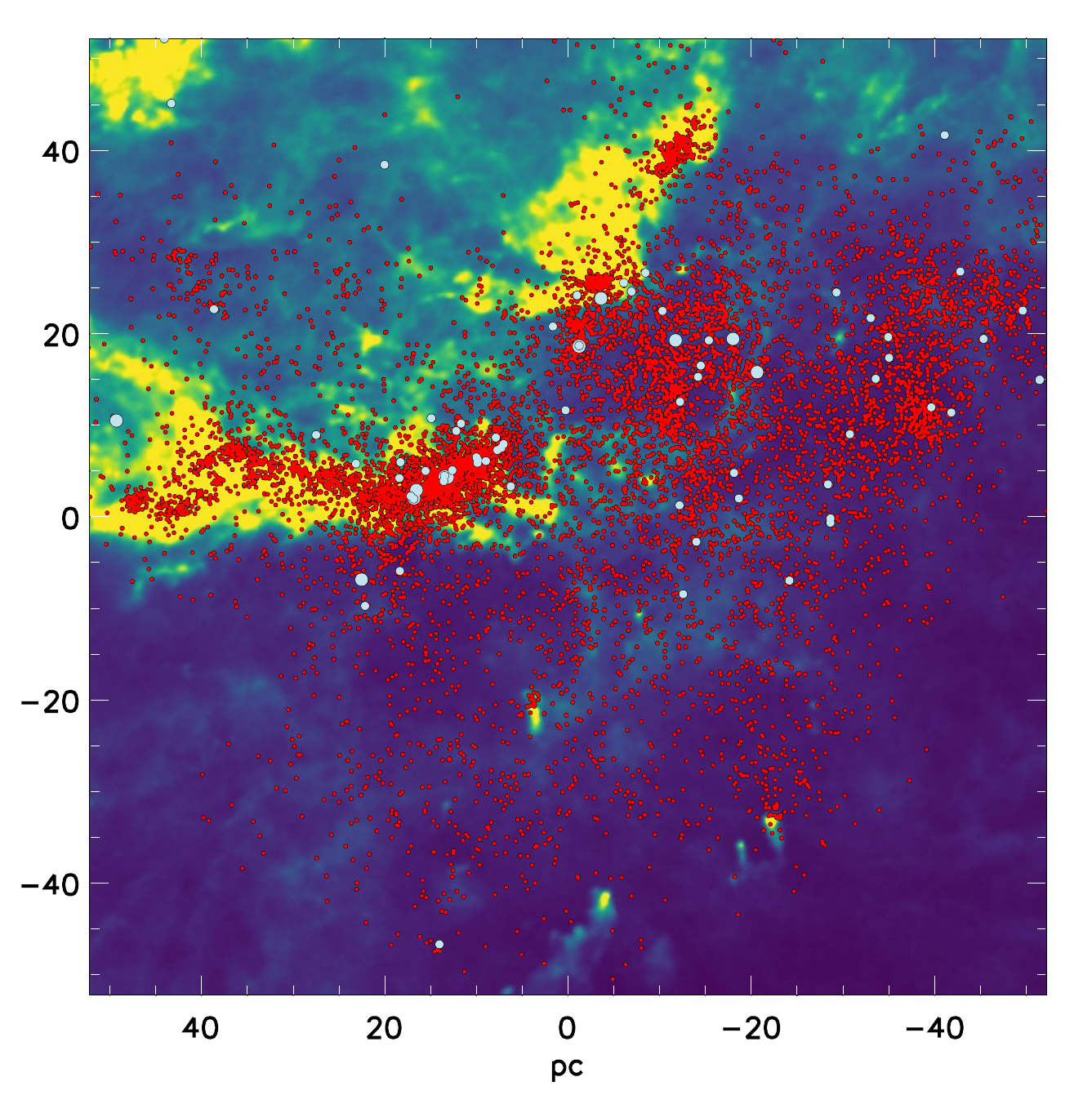}
\vskip -0.1 in
\caption{The young stars in the Orion molecular cloud complex and OB1 association plotted over an A$_v$ map from the Planck Legacy Archive \citep{2014A&A...571A..11P}.  The red dots are the young stars either from the Spitzer survey \citep{2012AJ....144..192M,2016AJ....151....5M} or from  Gaia DR2 \citep{2018AJ....156...84K}. Also shown are the O stars (large blue circles) or B stars (small blue circles) of the Orion OB1 association \citep{1994A&A...289..101B}. The coordinates are  parsecs at a distance of 400~pc. The map is oriented in galactic coordinates: the ONC is centered near X:~10~pc and Y:~5~pc.}
\label{fig:ob1}
\end{figure*}

%\citep{2010ARA&A..48...47A}

The clusters and molecular clouds discussed in previous sections are often parts of larger complexes of stars and clouds. The ISF, which contains the ONC, is part of the Orion A molecular cloud, a highly elongated, filamentary cloud that extends 90 pc in length \citep[Figure~\ref{fig:oriona_cluster},][]{1987ApJ...312L..45B,2018A&A...619A.106G}. Orion A, in turn, is part of the Orion Complex which extends for almost 200 pc and contains multiple generations of stars that have formed over the last 12 Myr in a succession of epochs  \citep{2008hsf1.book..838B,2018AJ....156...84K}. This complex contains the Orion OB1 association, whose massive stars power a superbubble 400 pc in diameter \citep{2008hsf1.book..459B,2015ApJ...808..111O}.

Complexes contain a mixture of low and high mass stars located both in molecular clouds and in large populations that have dispersed their natal clouds. While low mass stars in the molecular clouds can be identified by the presence of dusty disks and envelopes, coronal X-ray emission, or compact clusters, identifying the low mass stars outside of clouds is more difficult. These older stars have low disk fractions, are dispersed over areas too large to efficiently survey at X-ray wavelengths, and often require time consuming spectroscopic measurements \citep{2007ApJ...671.1784H, 2007ApJ...661.1119B}. For these reasons, the full extent of many complexes have historically been traced by their massive stars and referred to as OB associations \citep{1999AJ....117..354D}. 

With the release of Gaia DR2 and EDR3 and systematic spectroscopic surveys such as APOGEE-2, it is now possible to identify and characterize the populations of low mass stars outside of the clouds using the 6D data obtained by combining their positions in the sky with parallaxes and 3D motions. In the Orion complex, which includes the Ori~OB1 association, \citet{2018AJ....156...84K} used 6D data to identify 10,000 low mass members in the complex. They find that the entire complex is co-moving, with a velocity dispersion of a few km s$^{-1}$. This indicates a common origin. The densest regions of the association are the clusters embedded in the molecular clouds; yet the lower density off-cloud population shows considerable spatial and kinematic structure \citep[Fig.~\ref{fig:ob1},][]{2018AJ....156...84K}.  

{\color{black}

The nearby Sco OB2 association also shows significant spatial structure as well as a diffuse population that contains most of the stars  \citep{2008hsf2.book..235P,2018MNRAS.476..381W}. The densest regions of the association are the embedded cluster in the Ophiuchus cloud (Fig.~\ref{fig:oph_taurus}) and an isolated off-cloud peak; it is not known if this peak is a bound cluster \citep{2018A&A...614A..81R,damiani2018}. Proper motions show most  of the association is unbound, with asymmetric velocities dispersions and complex kinematic sub-structure \citep{2018MNRAS.476..381W,damiani2018}. 

%\citep{2018MNRAS.475.5659W,2020MNRAS.495..663W}

In a Gaia DR2 study of  109 galactic OB associations,\citet{2020MNRAS.495..663W} find substantial kinematic sub-structure.  They rule out models where the associations arises from the expansion from one or several clusters.  Their model of the observed motions requires several distinct components including one with random velocities,  sub-structures with correlated velocities, and expansions from multiple clusters \citep[also see][]{2016MNRAS.460.2593W}. Looking back at the stellar distribution of the Orion~A cloud (Fig.~\ref{fig:oriona_cluster}), the spatially correlated components may arise from the small groups/clusters along the filament of the cloud; as shown by \citet{2018AJ....156...84K}, these stars exhibit complex kinematics that include a large velocity gradient along the filament. The expanding cluster could come from the dense center of the ONC. These components will merge into the association as the Orion~A cloud disperses.  The Gaia DR2 analysis of  \citet{2019ApJ...870...32K} finds expanding clusters in several associations. }

With  Gaia DR2 and EDR3, it is possible to identify distinct populations of stars in the Orion complex \citep{2018AJ....156...84K,2020A&A...643A.114C}, as well as use the observed trajectories of the stars to infer dynamical evolution. Many of the observed trajectories indicate infall. Several clouds, as well as a number of small populations of young stars that have already dispersed their gas, are moving inward toward the complex's center of mass.  
%On large scales, there is some evidence for the importance of the self-gravity of the complexes. The large mass of the Orion Complex in particular results in an infall of a number of clouds (or small populations of young stars that have already dispersed the gas) towards the center of Orion. 
The most massive of these clouds is the Orion~B cloud. On smaller scales, the $\sigma$ Ori cluster, the densest, post-gas dispersal cluster in the complex, is moving toward the Orion~B cloud \citep{2018AJ....156...84K}.
%On  smaller scales, it is possible to observe the gravitational attraction of the $\sigma$ Ori cluster by Orion~B \citep{kounkel2018a}. 

%On the other hand, substructures identified within individual clusters (or even arguably multiple clusters in a single molecular cloud), in large part do not have convergent motion towards each other, and they are unlikely to further grow clusters via hierarchical assembly \citep{kuhn2019}. 

%Taken as a whole, individual clusters show little evidence for bulk rotational motion, and there is a preference for them to expand, with the typical expansion velocity of $\sim$0.5~km~s$^{-1}$. 

Extreme expansions are also observed in the Orion Complex. The observed expansions are ballistic, and given their magnitude ($ > 6$~\kms), may be attributed to the shockwave of two supernovae sweeping away the molecular gas and triggering star formation in the process \citep{kounkel2020a,2021A&A...647A..91G}. One of them dates back $\sim$4 Myr within the $\lambda$ Ori cluster. The other one dates back $\sim$6 Myr, likely originating along the sightline near $\eta$ Ori at the geometrical center of Barnard's loop. This supernova has been responsible for shaping much of the morphology of the Orion Complex. Given its location, the gas swept up along its shock front may have led to the formation of the ONC, the most massive cluster in the solar neighborhood \citep[for a different view on these motions, see][]{2021ApJ...917...21S}. 

The  properties of associations result from several factors.  First, associations form in complexes of elongated and filamentary clouds that can extend over 100 parsecs in length: the Orion A and B clouds span the diameter of the Orion complex \citep[Fig.~\ref{fig:ob1},][]{2016AJ....151....5M}. The overall size of the association is mostly determined by the size of the complex and not by expansion. 

{\color{black}
Second, associations are assembled primarily from the embedded groups and clusters that form in the clouds, and they inherit the velocity structure of the gas. The stars from the groups and clusters expand and/or disperse into the association creating complex kinematic and positional sub-structure. Infall motions and expansions driven by massive feedback add to the complex kinematic structure.

Third, star formation is sustained in associations for $\sim10$~Myr, longer than the ages of individual clusters or clouds \citep[Sec.~\ref{sec:age_spread},][]{2001ApJ...562..852H}, in a patchwork of star-forming regions that come and go through time \citep{2019A&A...628A.123Z,2020NewAR..9001549W}. In nearby associations such as Orion, massive star formation appears to have occured throughout this $\sim 10$~Myr interval \citep{2020NewAR..9001549W}. This is in tension with results from extragalatic studies that find cloud lifetimes of 10-30~Myr with the  massive stars and HII regions  appearing in only the last 1-5~Myr \citep[][]{2019Natur.569..519K,2022MNRAS.509..272C}.

%In contrast, the diffuse nature of T associations of low mass stars, such as Taurus, is primarily due to the first process \citep{2018AJ....156..271L}.

Finally, due to the large spatial extent of the molecular cloud complexes, associations  are assembled over regions where the crossing time, as calculated with the stellar velocity dispersions in localized sub-regions such as the ONC ($\sim 50$~Myr for a $\sim$~2~km~s$^{-1}$ velocity dispersion), is much longer than the star forming time in a cloud ($\sim 3.5$~Myr) or association ($\sim 10$~Myr) \citep{2001ApJ...562..852H,2020NewAR..9001549W}.  Within an association, embedded clusters form with crossing times less than the their lifetimes; these may survive and remain bound after the association disperses  \citep{2011MNRAS.410L...6G}.  }

\begin{figure*}
\epsscale{1.0}
\plotone{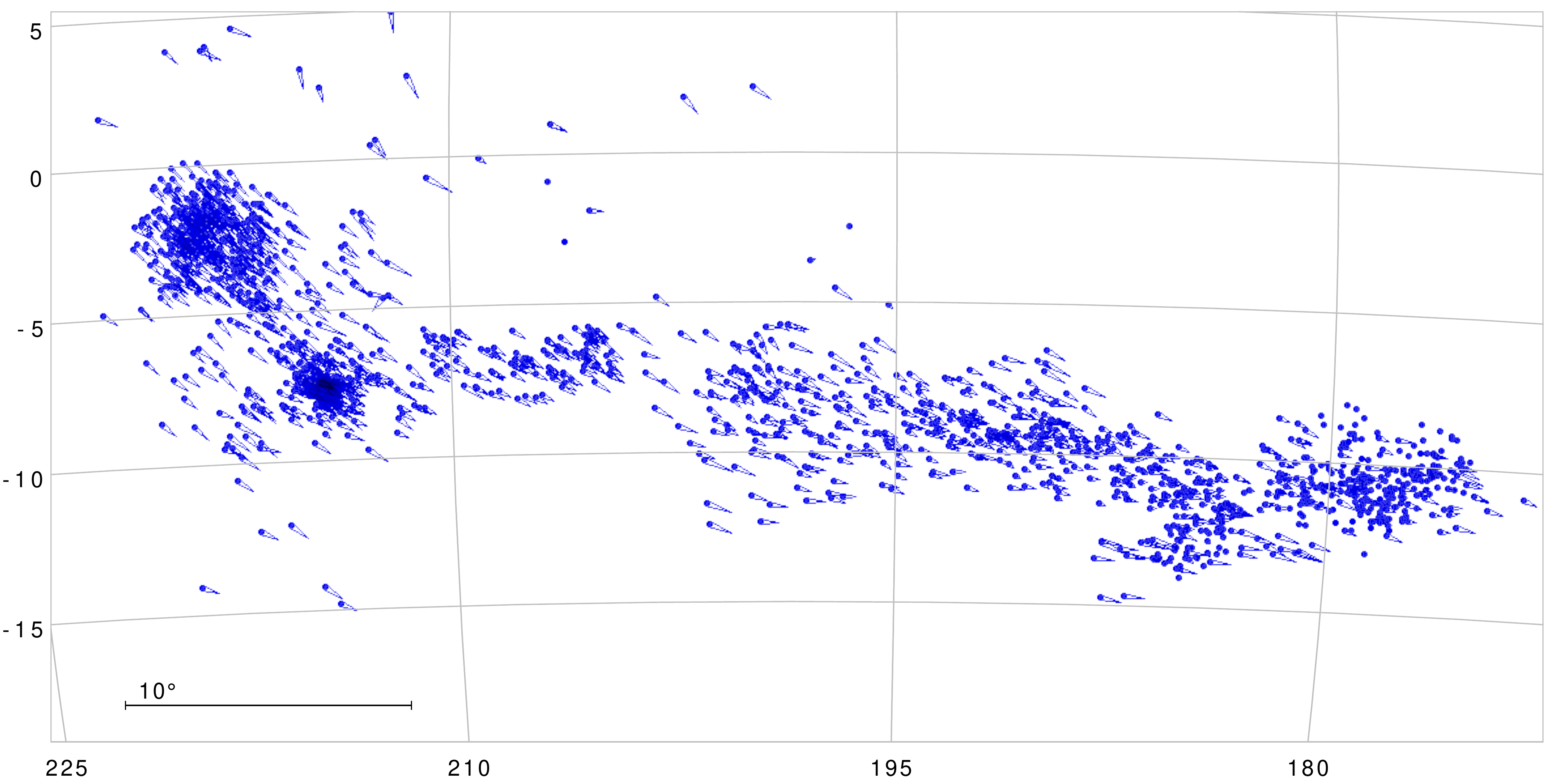}
\caption{Distribution of members of a $\sim$20 Myr old stellar string associated with the cluster NGC 2232. The plot is in the galactic coordinates. Arrows how the proper motions of the stars in the local standard of rest. Members were identified in \citet{kounkel2020}, see also \citet{tian2020}.
\label{fig:ngc2232}}
\end{figure*}

All young populations are slowly dissolving into the field as they age. However, because an individual molecular cloud tends to form several hundred to a few thousand stars, and because their original dispersion velocity mirrors that of the clouds and is much lower than the field distribution, they  persist as comoving groups. In recent years, with the improved ability to identify such comoving stellar populations afforded by Gaia, a number of older structures have been found \citep{2019AJ....158..122K,2019MNRAS.489.4418J,beccari2020}. %The ages of the stars inside them can be determined from the isochrone fitting. Although they also have a wide complexity in morphologies, 
{\color{black} Many of these populations are string-like configurations that extend in length on average 200 pc and have widths of only $\sim$10 pc (Fig.~\ref{fig:ngc2232}). They are unlikely to have formed by tidal stretching of clusters, or associations significantly more compact than their current dimensions, as such lengths are observed in strings of all ages from $<10$ Myr to $>>100$ Myr. }Rather, they resemble the shapes of filamentary  molecular clouds from which they likely formed \citep{2018ApJ...864..153Z}, {\color{black} and they are well-matched by the typical dimensions observed in the molecular clouds \citep{2018ApJ...864..153Z}}. Some, but not all, strings contain star clusters. Only a few young star clusters exist in isolation without extended stellar populations surrounding them. This shows that most clusters formed as parts of filamentary clouds. 

%This shows that cluster formation tends to be a biproduct of larger scale collapse, although more observational constraints are needed in order to determine the conditions under which clusters arise.

As more and more comoving populations with different ages have been identified, it is now possible to determine statistics about their survivabillity.  \citet{2019AJ....158..122K} find that most strings persist up to ages of $\sim$200--300 Myr, comparable to the orbital time of the Sun around the Galaxy. Beyond that age, the remaining comoving groups are typically compact and isolated - potentially remnants of the densest cluster-like environments that have formed inside of a cloud, while the rest of the diffuse population has largely been dispersed. Nonetheless, it is also possible to find a few stellar strings up to the ages of a few Gyr. These strings tend to be much more diffuse than their younger counterparts, with only a few dozen stars remaining. They are likely the surviving remnants of the populations that were the most massive at the time of their formation. Overall, the number of stars in a population that remains comoving as a function of age, regardless of size or mass, can be characterized by the relation  

\begin{equation}
N_t=N_o\times10^{(7-\log{t/\mathrm{yr}})/1.5},
\end{equation}

\noindent
where $N_o$ is the typical number of stars found in a population with an age of 10 Myr, {\color{black} and $t$ is the age of the population in years.} \citep{2019AJ....158..122K}.

{\color{black}
\section{Summary}
\label{sec:summary}}

%Young low mass stars and protostars, or YSOs, are a powerful probe of star formation in molecular clouds.  Populations of YSOs in molecular clouds can be studied using observations that span much of the electromagnetic spectrum, from X-rays to centimeter radio waves. Near to far-IR imaging and spectroscopy have proven essential for characterizing YSOs. Visible light observations are also powerful for less embedded population, providing the photometry and spectroscopy needed to identify and study young stars as well as measure their motions and distances. Gaia DR2, in particular, has enabled the study of the populations of the associations of young stars in which molecular clouds are found. We summarize the main points of this review as the following:

In Sec.~\ref{sec:scope}, we introduced four questions in the current study of star formation.  In this review, we addressed these questions using surveys of low mass YSOs combined with studies of the associated molecular gas. Here we summarize  our current progress in answering these questions, and suggest future lines of inquiry. 

\vskip 0.1 in
\noindent
{\it What are the rates, densities and efficiencies of star formation in molecular clouds, and how much do they vary with environment?}
\vskip 0.1 in

1.~Young low mass stars and protostars, or YSOs, are important tracers of star formation  on cluster and cloud scales. Over the last 20 years, surveys at X-ray, visible, IR and radio wavelengths have detected, identified, and characterized  populations of YSOs in many clouds in the nearest 1.5 kpc. The resulting catalogs of YSOs are the basis of wide ranging investigations into star and cluster formation. (Sec.~\ref{sec:techniques}) 

2.~The surface densities of YSOs measured in clouds within $\le  500$~pc of the Sun range from 0.1 to 10$^4$~pc$^{-2}$. The combined, YSO sampled PDF of these clouds is approximately lognormal for low to moderate densities, peaks near 20~pc$^{-2}$, but has a distinct wing at higher densities primarily due to the rich clusters found in Orion.  (Sec.~\ref{sec:PDF})

3. Within individual molecular clouds, comparisons of the surface densities of YSOs to the column densities of gas show power-law, star-gas correlations in regions without significant gas dispersal, where $\Sigma_{\scalebox{.5}{YSO}} \propto \Sigma_{gas}^2$. (Sec.~\ref{sec:sf_law})

4. The  spatially averaged, instantaneous SFE varies from $\sim 4\%$ for entire molecuar clouds to $\sim 13\%$ for clusters, although with large scatter.  This is consistent with the star-gas correlations, which imply that the instantaneous star formation efficiency (SFE) within a cloud varies linearly with the surface density of gas until a significant fraction of the gas is expelled by feedback.  (Sec.~\ref{sec:SFE} and \ref{sec:sf_law})

5. The star formation rate (SFR) can be measured from the number of dusty YSOs and  adopted lifetimes of disks and protostars, or from the total luminosities of protostars.  Inconsistencies between these approaches remain unresolved. (Sec.~\ref{sec:sfr})

6. The star-gas correlations can  be interpreted as a sf-relation where the SFR per area scales as the square of the gas column density. These correlations can also be reproduced by a sf-relation with a constant SFE per free fall time of $\approx 0.01 - 0.03$. (Sec.~\ref{sec:sf_law} and \ref{sec:sf_ff})

\vskip 0.1 in

\noindent
{\it What are the differences between diffuse and clustered star formation, and how do they depend on cloud properties?}

\vskip 0.1 in

7. Several techniques are used to extract clusters from 2D maps of YSOs, with notable differences between the results.  Cluster extraction applied to cloud surveys shows that clouds contain multiple assemblages, ranging in size from small groups to massive clusters. In clouds with clusters, the most massive clusters contain approximately half the stars while $\sim 20\%$ of the YSOs are relatively isolated. (Sec.~\ref{sec:clusters} and \ref{sec:demo}) 

8. Due to the sf-relations in clouds, clusters form in regions with high gas column densities and are observed as peaks in the stellar density. Thus, clustered and diffuse star formation are a continuum that is a function of the density of gas. The SFE per free fall time sf-relation varies little between clouds forming massive stars and low mass stars, showing that it does not depend on the feedback of massive stars.  (Sec.~\ref{sec:sf_law} and \ref{sec:sf_ff}) 

9. The PDFs of individual clouds vary significantly. Clouds without clusters have median densities $< 10$~pc$^{-2}$ while those with clusters have median densities $> 20$~pc$^{-2}$. The presence of clusters is determined by the structure of the dense gas in a cloud. (Sec.~\ref{sec:PDF})

10.  Clusters form in hubs and ridges while diffuse star formation occurs primarily in networks of filaments with 0.1 pc diameters. Hubs are massive concentrations of molecular gas that appear at the intersections of multiple filaments. The filaments can supply hubs with flows of gas as the hubs form stars.  Ridges are much more elongated than hubs, and have higher mass to length ratios and different radial density profiles than filaments. (Sec.~\ref{sec:gas_env})

11. Hubs and ridges contain significant substructure that have been characterized as filaments (using 2D intensity maps) or as fibers (using 3D position-velocity data); these structures appear to be the sites of fragmentation. The density of unstable fibers/filaments may correlate with the density of star formation, with the highest densities found in clusters. (Sec.~\ref{sec:frag}) 

\vskip 0.1 in
\noindent
{\it What is the duration of cluster formation, and during this time, what processes shape structure and kinematics of the clusters?  }
\vskip 0.1 in

12. Based on the observed ratios of the number of pre-ms stars with disks to the number of protostars, star formation in clusters is sustained for approximately 2-3.5 million years, or about seven protostellar lifetimes. This time interval is also reflected in the age spreads in HR diagrams. (Sec.~\ref{sec:age_spread}) 

13. There is evidence for spatial and temporal correlations of the SFR within clusters, both in the form of age gradients between the edges and interiors of clusters, and the presence of multiple, distinct episodes of star formation extending across a cluster. (Sec.~\ref{sec:age_spread}) 

14. Structural analyses of the extracted clusters show a mass to radius relationship and a correlation between the peak YSO surface density and the total number of members. Clusters are also typically elongated.  These correlations appear to be inherited from their formation in hubs and ridges. (Sec.~\ref{sec:structure} and Sec.~\ref{sec:gas_env})

15. YSOs inherit their velocities from the complex gas motions observed in hubs and ridges. Locally, the velocities of cores relative their natal gas appears sub-virial. The global core-to-core velocity dispersion, however, can be  sub-virial or supervirial, depending on the region. In the ISF the global dispersion is supervirial due to the motions over the length of the ISF. (Sec.~\ref{sec:kinematics})

16. The velocities of cluster members show  that clusters are often dynamically evolved, neither expanding or rotating, and often have velocity dispersions close to the virial velocity. The presence of massive, nearby gas structures, asymmetries in the velocity dispersion, and rapid gas dispersal suggest, however, that clusters are not in equilibrium.  (Sec.~\ref{sec:kinematics})  

17. Embedded clusters can lose mass through ejections. In the case of the ONC, the observed rate is small and will only reduce the cluster mass by a few percent.  The rare ejections of massive stars, however, may have a large effect by changing the amount of feedback. (Sec.~\ref{sec:kinematics})

\vskip 0.1 in
\noindent
{\it How does star formation in cloud complexes produce a mixture of bound clusters and unbound associations, and how long do these assemblages persist in the galactic disk after star formation ends?}
\vskip 0.1 in

18. The SFR within 500 pc of the Sun is 1700~M$_{\odot}$~Myr$^{-1}$~kpc$^{-2}$. A comparison of the rate at which open clusters form with this SFR finds that $15^{+11}_{-08}\%$  of stars emerge from their birth clouds in open clusters. If clusters are primarily formed from $\ge 100$ member embedded clusters, then  $23^{+17}_{-12}\%$ of the stars in embedded clusters emerge in open clusters. (Sec.~\ref{sec:gas_dispersal})

19. The cluster forming gas is removed by the expansion of HII regions around early B stars or winds of O stars; these can occur at multiple locations along a ridge or hub. The gas dispersal is  being mapped by observations of emission lines from PDRs with SOFIA. (Sec.~\ref{sec:gas_dispersal})

20. Since the velocity dispersions of the stars in embedded clusters are often close to the virial velocity for the star and gas potential, the loss of gas can have a large effect. Measurements of cluster surface densities and kinematics have found clusters expanding in response to the removal of their parental gas. Sub-clusters are observed to be moving apart and do not appear to merge to form larger clusters, in part due to the dispersal of the gas that binds them. (Sec.~\ref{sec:gas_dispersal})

21. There is some evidence that the rate of gas dispersal is important in the production of clusters, with clusters undergoing rapid dispersal by O-stars showing more rapid expansion and retaining only a relatively small fraction of their stars in a bound cluster. Clusters undergoing slower gas dispersal by B-stars may retain a much larger fraction of their stars. (Sec.~\ref{sec:gas_dispersal})

22. Clusters are typically found in large, filamentary molecular clouds, and these clouds are often parts of large complexes that can span 200 pc and sustain star formation for 10 Myr. Using Gaia and radial velocities, 6D observations of star formation complexes now trace the older, lower density stellar populations that have dispersed their natal gas and expanded. (Sec.~\ref{sec:assoc})

23. The kinematics of the associations of stars outside the clouds require that they are formed from multiple components. These components likely originate from a mixture of  multiple  clusters, smaller groups of stars, and more isolated stars that form in clouds.  The stars within complexes also show both expansive motions due to supernovae, and inward motions which may be due to gravity. (Sec.~\ref{sec:assoc})

24. The stellar associations of embedded and post-embedded stars that span these complexes have crossing times, determined with local stellar velocity dispersions, that are much longer than the 10~Myr lifespans of the complexes. These associations, however, contain clusters of stars whose crossing times are less than their ages and can form bound clusters. (Sec.~\ref{sec:assoc})

25. Despite the complex motions, remnants of complexes can persist for up to 200-300 Myr. These remnants appear in Gaia observations of the local galaxy as string-like structures, many containing clusters. (Sec.~\ref{sec:assoc})

{\bf Future directions:} The next few decades will bring new opportunities for extending surveys of low mass YSOs.  JWST will survey dusty YSOs in more distant regions of our Milky Way, as well as the LMC and SMC, providing a more representative sample of star formation in our Galaxy and others. SDSS-V and the SKA will find young stars without requiring the presence of a disk or envelope. This capability will be further expanded by the ngVLA and future X-ray space telescopes.  

A deeper understanding of protostellar and pre-ms evolution will improve measurements of the star formation rates and histories in clouds.  Spectroscopic studies of protostars with JWST and near-IR spectrographs on large telescopes can measure accretion onto protostars and constrain their effective temperatures and radii, 
while ALMA can determine their masses. These data will provide strong observational constraints on protostellar accretion and evolution \citep[e.g.][]{2016ARA&A..54..135H,2020ApJ...905..162T}. Improved surveys for multiplicity, observational constraints on the initial conditions of pre-ms contraction, and a deeper understanding of the role of magnetic fields, rotation, and star spots in pre-ms evolution are needed to obtain robust ages and star formation histories in clouds and clusters \citep[e.g.][]{2021MNRAS.505.1280B,2021ApJ...923..177S,2022ApJ...924...84C}. 

Surveys of dense cores with both single-dish telescopes in nearby regions and ALMA in more distant regions will provide an independent means to study star formation rates and efficiencies \citep[e.g.][]{2019MNRAS.483..407S,2021MNRAS.508.2964A}. This includes using observations of cores to measure the rate of fragmentation or the efficiency of fragmentation per free fall time over the full range of cloud environments (Sec.~\ref{sec:frag}). Such studies coupled with  observations of turbulent motions, magnetic fields and outflows can guide models of how star formation is regulated in clouds \citep[e.g][]{2015ApJ...808...48B,2020NatAs...4.1195P,2021arXiv211107995X}.

Even in the ISF, our understanding of the evolution of clusters during gas dispersal remains incomplete. Kinematic studies can be expanded to a larger number of clusters within the nearest 1-2~kpc that sample the full range of observed SFEs  (Fig.~\ref{fig:sfe}). Such studies will be performed with a combination of SDSS-V, Gaia, and potentially through proper motion measurements of embedded stars with the Roman space telescope.  Observations of stars outside of clusters, many using Gaia, can estimate the fraction of stars in embedded clusters that are lost during gas dispersal  \citep[e.g.][]{2007MNRAS.376.1879W,2020ApJ...890..129K,2021arXiv211004296H}. Models of cluster evolution need to account for the continual formation of stars and feedback from those stars over multiple generations \citep{2021ApJ...911..128K,2021MNRAS.506.3239G,2021MNRAS.502.3646G} and the gas motions in ridges that can inflate the velocity dispersion in clusters \citep[e.g.][]{2016A&A...590A...2S,2019MNRAS.489.4771G}. 

\bibliography{refs}

\acknowledgments

We thank Dr. Amelia Stutz for a careful reading and insightful comments on cluster kinematics. Dr. Riwaj Pokhrel supplied plots and expertise into star-formation relations. Dr. Michael Fellhauer provided insights in cluster evolution during gas dispersal. Dr. Thomas Stanke produced the beautiful APEX/SABOCA 350~$\mu$m map of the OMC2/3 region.  Sam Federman carefully read the text and gave numerous useful comments. Drs. Cornelia Pabst and Xander Tielens provided the [CII] data cube from SOFIA. STM received funding from the NASA ADAP grant 80NSSC18K1564 and STM, RAG and MAK received funding from the NASA ADAP grant 80NSSC19K0591. RAG also acknowledges funding support from: NASA ADAP awards NNX11AD14G, NNX13AF08G, NNX15AF05G, and NNX17AF24G; NSF AST grants 1636621, 1812747, and 2107705; NASA-USRA SOFIA grants 05-0181, 07-0225, and 08-0181; NASA-JPL/Caltech Spitzer grants 1373081, 1424329, and 1440160 and Herschel grant 1489384. During the completion of this review, STM was a Fulbright Scholar at the Universidad de Concepc\'ion, Chile; he thanks Dr. Amelia Stutz and UdeC for hosting him. Finally we thank an anonymous referee for their insightful comments and valuable suggestions, and the editorial staff of the PASP for their infinite patience. 

\appendix

\section{Calculating the SFR from the Integrated Protostellar Luminosity}
\label{sec:appendix_SFR}

In Sec.~\ref{sec:sfr}, we show a calculation of the SFR in bins of declination along the ISF (Fig.~\ref{fig:sfr}). One of the calculations used the integrated protostellar luminosity to get a total mass accretion rate, which is equal to the current SFR (Eqn.~\ref{eqn:sfr_lum}).  The displayed luminosity histograms were calculated using the bolometric luminosities of the HOPS protostars given by  \citet{2016ApJS..224....5F}; due to extinction and inclination effects, the actual luminosities may be as much as a factor of two times higher. As protostars evolve, their accretion luminosities drop and their luminosities begin to be dominated by the intrinsic luminosities of the central protostars \citep{2017ApJ...840...69F}.  To limit the sample to protostars with accretion dominated luminosities, we require that they have $T_{bol} < 100$~K. Finally, we adopted $\eta = 0.8$ (this is the fraction of the accretion-generated energy that is radiated away as luminosity) and $m_{\star}/r_{\star} = 0.15$ \citep[see][]{2017ApJ...840...69F}. The HOPS survey did not include the Orion Nebula region due to bright nebulosity. To correct for this bias, we used the protostars detected at shorter wavelengths in \citet{2012AJ....144..192M,2016AJ....151....5M}. To add in this contribution, we included all of the protostars in the nebula that were not included in HOPS. To determine the total accretion luminosity of these protostars, we first multiplied the number of non-HOPS protostars in each of the declination bins coincident with the nebula by the fraction of  HOPS protostars with $T_{bol} < 100$~K. We then multiplied the resulting number of protostars by the mean luminosity of the HOPS protostars.  In Fig.~\ref{fig:sfr}, we show the accretion rates with and without the correction for the missing protostars in the nebula.

\section{Calculating Cluster Properties}

In Fig.~\ref{fig:cluster_prop}, we combine the number of members, mean densities, peak densities and radii from three studies that have systematically extracted and characterized clusters in Spitzer and/or Chandra  data. In this appendix, we summarize how these values were calculated. \citet{2014ApJ...787..107K,2015ApJ...802...60K} fit isothermal ellipsoids plus a constant background to their distributions of X-ray and IR selected YSOs. As discussed in Sec.~\ref{sec:clusters}, multiple ellipsoids can be fit to a cluster like the ONC, and they are referred to in \citet{2014ApJ...787..107K} as sub-clusters.  To determine the properties, we used the $r_{4,major}$, $r_{4,minor}$ and $n_{xlf}$ values in Table 2 of \citet{2015ApJ...802...60K}. The first two values are the radii of the semi-major and semi-minor axis for an ellipsoid with a radius four times the core radius, and $n_{xlf}$ is the number of members corrected for incompleteness using the XLF of each region. From these values, we calculated 

\begin{equation}
    r = (r_{4,major} \cdot r_{4,minor})^{0.5},~N_{mean} = \frac{n_{xlf}}{\pi r^2},~ N_{peak} =  \frac{16}{ln(17)} N_{mean},~{\rm and}~{\alpha} = \frac{r_{4,major}}{r_{4,minor}},
\end{equation}

\noindent
where $r$, $N_{mean}$, $N_{peak}$ and $\alpha$ are the radius, mean density, peak density and aspect ratio, respectively.  In our analysis, we only use the clusters at distances $\le 1$~kpc; i.e. those sub-clusters associated with the ONC, W40, the Flame Nebula and the NGC~2264 regions.

\citet{2009ApJS..184...18G} and \citet{2016AJ....151....5M} use different methods for extracting clusters from the distribution of IR selected YSOs, but once extracted, they use the same equations for determining the cluster properties. For each cluster, they determine a convex hull that contains the cluster, with each vertex being a cluster member; the area of this convex hull is $A_{hull}$.  They also measure the minimum radius of a circle that contains the entire cluster, $r_{circ}$.  Finally, they calculate the nearest neighbor density $N_n$ for each member.  The cluster properties are then given by

\begin{equation}
   r = \sqrt{\frac{A_{hull}}{\pi}},~N_{mean} = \frac{n_{\scalebox{.5}{YSO}}}{\pi r^2},~N_{peak} = max(N_n),~{\rm and}~{\rm asp}=\frac{r_{circ}}{{r_{hull}}}.
\end{equation}

\noindent
\citet{2016AJ....151....5M} use $N_{10}$ to calculate the peak density while \citet{2009ApJS..184...18G} use $N_6$. In contrast to \citet{2015ApJ...802...60K}, these surveys only count YSOs with dusty disks and envelopes that can be identified in the Spitzer data. Furthermore, \citep{2009ApJS..184...18G} do not apply any correction for incompleteness, while \citet{2016AJ....151....5M} apply a correction for incompleteness due to the bright nebulosity in Orion.

To extract the clusters, \citet{2009ApJS..184...18G} construct a minimum spanning tree connecting the YSOs for each Spitzer field \citep[e.g.][]{2004MNRAS.348..589C}. They rank the branches by their length and cut the branches longer than a critical branch length. This critical length corresponds to the transition from a rapid to slow increase in length with rank. They refer to the extracted structures as cluster cores.  The cluster core properties are taken from their Tables 6 and 8. In contrast, \citet{2016AJ....151....5M} use a density threshold. They identify all sources with a $N_n \ge 10$~pc$^{-2}$; for all YSOs that satisfy the criteria, the 10 nearest YSOs that also meet this criteria are considered friends. Using a friends of friends analysis, they identify clusters of friends (see Fig.~\ref{fig:oriona_cluster}). The cluster properties are given in their Table 2; we used the completeness corrected number, radius, aspect ratio and peak density from that table.  

%\section{Calculating the Infall Rate of YSOs in the ISF}
%\label{sec:infall_YSO}

%To determine the rate of stellar mass infall, we modified the equation  that \citet{2017A&A...602L...2H} used to calculate the infall of gas.  We used
%\begin{equation}
%\dot M = m_{\star} \frac{n_{\scalebox{.5}{YSO}}}{\Delta L} V_{infall}
%\end{equation}

%\noindent
%where $n_{\scalebox{.5}{YSO}}$ is the number of IR identified YSOs in a rectangular region that extends over a length $\Delta L$.  We used a rectangle spanning from 400'' to 600'' in declination and from -74'' to 74'' in R.A. relative to our adopted center (${\rm R.A.} = 83.83279^{\circ}$, ${\rm Dec.} = -5.3725^{\circ}$); this rectangle is just north of the section of the ISF that shows the steep velocity gradient in \citet{2017A&A...602L...2H}, and it therefore gives a measure of the density of stars before acceleration. Dividing  $n_{\scalebox{.5}{YSO}}$ by $\Delta L$ gives a line density for the adopted a distance of 388~pc.  Adopting $V_{infall} = 1$~km$^{-1}$ and an average mass $m_{\star} = 0.5$~M$_{\odot}$, the infall rate is 30~M$_{\odot}$~Myr$^{-1}$.

\begin{figure}[t!]
\epsscale{1.}
\plotone{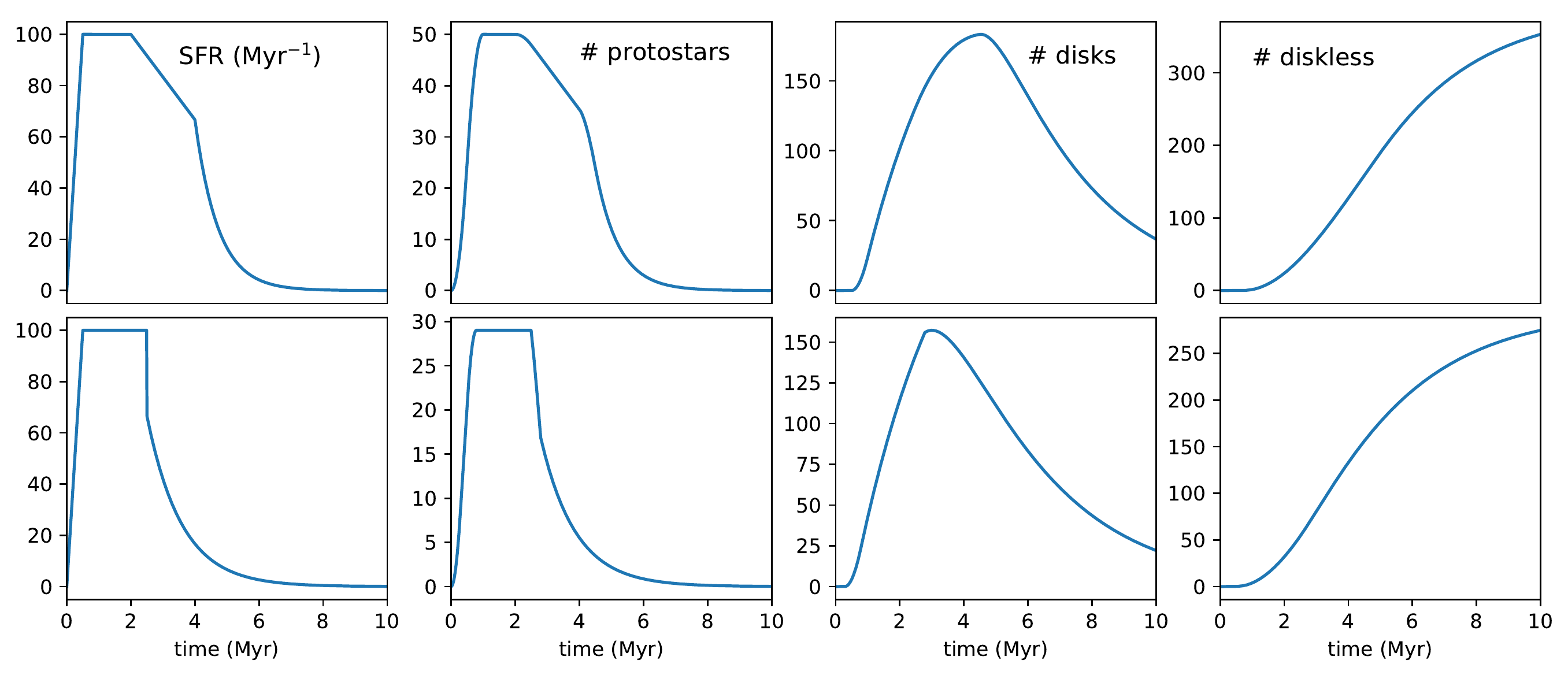}
\caption{The two models used to calculate the distribution of ratio of pre-ms stars with disks to protostars in Fig.~\ref{fig:disk_protostar}. Each row shows one of our two model, with the adopted SFR and the resulting number of protostars, pre-ms stars with disks, and pre-ms stars without disks as a function of time.} 
\label{fig:ci_ii_iii_model}
\end{figure}

\section{The Ratio of Disks to Protostars and the Star Formation History}
\label{sec:ratio}

To model the histogram of disk to protostar ratios for an ensemble of young clusters, we adopted a model SFR, $\phi(t)$, which starting at time 0 shows a linear rise to a maximum star formation rate at $t_0$.  The SFR plateaus at $\phi_{max}$ between $t_0$ and $t_1$.  Between $t_1$ and $t_2$ there is a linear decrease in the SFR to $(2/3) \phi_{max}$. After $t_2$, the $\phi(t)$ drops exponentially with an e-folding time of $t_{3}$. This is described  by the equations

\begin{eqnarray}
\phi(t) & = & \phi_{\rm max}(t/t_0)~{\rm when}~t \le t_0 \\
& = & \phi_{\rm max} ~{\rm when}~ t_0 < t \le t_1 \\
& = & \phi_{\rm max}(1-(1/3)(t-t_1)/(t_2-t_1))~{\rm when}~t_1 < t \le t_2 \\
& = & (2/3)\phi_{\rm max}e^{-(t-t_2)/\tau_{3}}~{\rm when}~t > t_2.
\end{eqnarray}

\noindent
The protostellar stage is assumed to last for $t_{\rm proto}$, hence the number of protostars for a cluster with age $t_{\rm cluster}$ is given by

\begin{equation}
    N_{\rm proto}(t_{\rm cluster}) = \int^{t_{\rm cluster}}_{t_{\rm cluster}-t_{\rm proto}} \phi(t) dt.
\end{equation}

\noindent
After the protostellar stage, a star immediately transitions into a pre-ms star with disk.  The optically thick disks detected by Spitzer  disappear exponentially with time with a half-life of $t_{half}$ \citep{2009AIPC.1158....3M}.  Thus, the number of disks (i.e. Class II objects) at time $t_{cluster}$ is 

\begin{equation}
    N_{\rm  disk}(t_{\rm cluster}) = \int^{t_{\rm cluster}}_{t_{\rm proto}} \phi(t-t_{\rm proto}) e^{-ln(2)(t_{\rm cluster}-t)/\tau_{\rm half}} dt
\end{equation}

\noindent
Finally, after the  optically thick disks disappear, the number of diskless stars (Class III objects) is  

\begin{equation}
    N_{\rm diskless}(t_{\rm cluster}) = \int^{t_{\rm cluster}}_0 \phi(t) dt - N_{\rm proto} - N_{\rm disk}
\end{equation}

The SFR, $N_{\rm proto}$, $N_{\rm disk}$, and $N_{\rm diskless}$ are plotted as a function of $t$ in Fig.~\ref{fig:ci_ii_iii_model}. To obtain the distribution of disk to protostar ratios, we assume that the observed cluster cores are uniformly distributed in $t_{\rm cluster}$ between 0 and 10~Myr and that all the observed clusters have $N_{disk} \ge 0.25 N_{proto}$; in other words, the youngest clusters are not included in our sample. We perform the calculations shown in Fig.~\ref{fig:ci_ii_iii_model} for discrete values of $t_{\rm cluster}$ separated by a constant time interval. We then plot histograms of the ratios of $N_{\rm disk}$ to $N_{\rm proto}$ for all values of $t_{\rm cluster}$ where $N_{\rm \rm disk} \ge 0.25 N_{\rm proto}$. 

We find two solutions that reproduce the observed distribution.  The first is where $t_0 = 0.5$~Myr, $t_1 = 2$~Myr, $t_2 = 4$~Myr, and the exponential decay time is $\tau_3 = 0.5$~Myr (top row of  in Fig.~\ref{fig:ci_ii_iii_model}).  In this solution the protostellar lifetime, $t_{\rm proto}$ is 0.5~Myr and the disk half life is $t_{\rm disk} = 2$~Myr. In the second solution, $t_0 = 0.5$~Myr, $t_1 = 2.5$~Myr, $t_2 = 2.5$~Myr, and the exponential decay time is $\tau_3 = 0.75$~Myr (bottom row in Fig.~\ref{fig:ci_ii_iii_model}).  In this solution, we reduced the protostellar lifetime to $t_{\rm  proto} =0.285$~Myr and kept the same disk half life of $t_{\rm  disk} = 2$~Myr.  In both cases, the primary star formation time, between the onset of the plateau and the end of the 33\% drop in SFR, is seven times the protostellar lifetime. 

\section{The Star Formation Rate in the Nearest 500pc}
\label{sec:sfr_500pc}

To compare the SFR in open clusters within a 2 kpc cylinder centered on the Sun, as estimated by \citep{2021A&A...645L...2A}, to the current rate of SFR, we use the  census of all dusty YSOs, i.e. protostars and pre-ms stars with disks, within 500 pc of the Sun. We restrict our estimate to this distance since our census of YSOs is much more complete in the nearest 500 pc.  We assume that the SFR per galactic surface area in the nearest 500 pc is similar to that in the nearest 2 kpc.

We use the entire SESNA processing for the clouds within 500 pc to determine the number of YSOs in the Auriga, Cepheus Flare, Chameleon, Coronus Australis, Lupus, Musca, Ophiuchus, Perseus, and Scorpius clouds  \citep[Gutermuth et al. in prep.,][]{2009ApJS..181..321E,2015ApJS..220...11D,2020ApJ...896...60P}. The total number of dusty YSOs is 2953; after correcting for contamination of 9 objects per sq.~deg.\citep{2020ApJ...896...60P}, this value is 2624.  We augment these using the Orion YSO catalog supplemented by Chandra X-say sources in the ONC and NGC~2024 described by \citet{2016AJ....151....5M}; this is 3889 YSOs which drops to 3810 YSOs after correcting for contamination. The number of dusty YSOs in Taurus is 226 \citep{2010ApJS..186..111L}.  Finally, the number of dusty YSOs in the Pipe molecular cloud is 18 \citep{2009ApJ...704..292F}. Using these numbers and the equation

\begin{equation}
    {\rm SFR}~({\rm M}_{\odot}~{\rm Myr}^{-1}~{\rm kpc}^{-2}) = \frac{0.5~{\rm M}_{\odot}~n_{\rm YSO}}{2.5~{\rm Myr}~\pi~ (0.5~{\rm kpc})^2},
\end{equation}

\noindent
(see Sec.~\ref{sec:sfr}), we find that the SFR = $1700 \pm 22$~M$_{\odot}$~Myr$^{-1}$~kpc$^{-2}$, assuming Poisson uncertainties in the numbers of sources and contamination. The value of the SFR in open star clusters ${\rm SFR}_{\rm cluster} = 250^{+190}_{-130}~$M$_{\odot}$~Myr$^{-1}$~kpc$^{-2}$. Propagating the uncertainties we find that the percentage of stars formed that emerge in clusters is $15^{+11}_{-08}\%$.  If we assume that only the 65\% of stars that form in embedded clusters with 100 members or more can become open clusters  \citep{2016AJ....151....5M}, then the fraction of those stars that emerge in open clusters is $23^{+17}_{-12}\%$.

\end{document}